# Using Wave-Packet Interferometry to Monitor the External Vibrational Control of Electronic Excitation Transfer

Jason D. Biggs and Jeffrey A. Cina*

Department of Chemistry and Oregon Center for Optics,

University of Oregon, Eugene, Oregon 97403

**Abstract.** We investigate the control of electronic energy transfer in molecular dimers through the preparation of specific vibrational coherences prior to electronic excitation, and its observation by nonlinear wave-packet interferometry. Laser-driven coherent nuclear motion can affect the instantaneous resonance between site-excited electronic states and thereby influence short-time electronic excitation transfer (EET). We first illustrate this control mechanism with calculations on a dimer whose constituent monomers undergo harmonic vibrations. We then consider the use of nonlinear wave-packet interferometry (nl-WPI) experiments to monitor the nuclear dynamics accompanying EET in general dimer complexes following impulsive vibrational excitation by a sub-resonant control pulse (or control pulse sequence). In measurements of this kind, two pairs of polarized phase-related femtosecond pulses following the control pulse generate superpositions of coherent nuclear wave packets in optically accessible electronic states. Interference contributions to the time- and frequency-integrated fluorescence signal due to overlaps among the superposed wave packets provide amplitude-level information on the nuclear and electronic dynamics. We derive the basic expression for a control-pulse-dependent nl-WPI signal. The electronic transition moments of the constituent monomers are assumed to have a fixed relative orientation, while the overall orientation of the complex is distributed isotropically. We include the limiting case of coincident arrival by pulses within each phase-related pair in which control-influenced nl-WPI reduces to a fluorescence-detected pump-probe difference experiment. Numerical calculations of pump-probe signals based on these theoretical expressions are presented in the following paper [J. D. Biggs and J. A. Cina, J. Chem. Phys. xxx].

* Electronic mail: cina@uoregon.edu

## 1. INTRODUCTION

Careful selection of the initial vibrational wave function can dramatically influence excited state dynamics when nuclear motion takes place either on multiple potential surfaces[1,2,3,4,5,6] or on a single adiabatic potential.[7,8] By controlling the initial state, and thus the subsequent evolution, surface-crossing transitions can, in particular, be hastened or curtailed. Here we propose a method to control the flow of electronic excitation energy (high energy quanta) between coupled chromophores using vibrational motion (low energy quanta). Coherent nuclear motion, imparted by infrared absorption or impulsive stimulated Raman scattering (ISRS)[9,10,11,12,13,14,15,16,17,18] in the electronic ground state, acts to influence the instantaneous local resonance between subsequently populated site-excited states and hence the time-course of the surface-crossing transitions corresponding to energy transfer.

To set the stage for theoretical and experimental investigations of this general control strategy, we derive expressions for the fluorescence-detected nonlinear wave-packet interferometry (nl-WPI) signal from a vibrationally excited sample of isotropically oriented energy-transfer dimers of fixed internal geometry. The signal expressions are parametrized by the relevant nuclear Hamiltonians, and should be valid for systems with weak to moderate coupling between site-excited states.

Our basic approach is the common one of describing the molecular system and its environment quantum mechanically, while the laser fields are taken to be classical electromagnetic waves whose interaction with the active chromophores is treated perturbatively. The control processes of interest are most easily visualized in terms of externally driven or freely evolving nuclear wave packets undergoing transfer between different electronic potential energy surfaces under the influence of resonant light pulses or spatially localized EET surface-crossings. The measured quantities are naturally expressed as quantum mechanical overlaps between superposed wave packets that have undergone different sequences of transitions in state space. For these reasons, we find it convenient to organize our calculations in terms of sums over the contributing wave-packet overlaps, each one of which involves a unique sequence of pulse-induced propagation events, free-evolution intervals on a single electronic potential-energy surface, and surface-crossing electronic transitions. The results of our analysis could, however, readily be transcribed to the widely used and formally equivalent notation of time-dependent nonlinear optical response functions, the various contributions to which are often pictured in terms of bra- and ket-sided time diagrams.[19,20,21,22]

## A. Control Scheme

Electronic excitation transfer (EET) involves the intermolecular or intramolecular redistribution of electronic excitation and competes with other relaxation mechanisms available to the system, such as fluorescence or radiationless decay.[23] In natural photosynthetic systems, the efficiency of EET ensures that the absorbed photon energy is preferentially funnelled towards a reaction center, where it is converted to chemical energy that can be stored for future use.[24,25,26] The roles of nuclear motion and molecular structure in these large EET complexes have been the subject of several recent investigations.[27,28,29] In this subsection, we discuss how the interplay between vibrational and energy-transfer dynamics brought to light by recent work suggests the use of low-frequency vibrational dynamics to influence higher-frequency electronic excitation-transfer dynamics.

A recent theoretical study viewing EET in molecular dimers as an electronic potential energy surface-crossing process,[30, see also 31] showed that the vibrational coherence transfer that naturally accompanies electronic excitation transfer can give rise to quantum beats in time-resolved polarized emission spectra that are in-phase between parallel and perpendicular emission channels, as had been observed in the LH-1 antenna of photosynthetic bacteria.[24,32,33] Those calculations on a model complex composed of two identical chromophores, each supporting a single harmonic vibrational mode, demonstrated that vibrational coherence transfer is a consequence of the nuclear motion resulting from short-pulse excitation to the Franck-Condon point of a site-excited state (the donor-excited state). The periodic return of this localized wave packet to the surface-crossing seam ($q_a = q_b$), where amplitude can be transferred to the acceptor-excited state, results in a stepwise decrease in donor population (see Fig. 1 – solid line) and a concomitant transfer of vibrational coherence to the acceptor-excited state. It was shown in Ref. 30 that the resulting coordinated nuclear motion in donor- and acceptor-excited states is manifested by in-phase quantum beats in the polarized pump-probe signals similar to those observed experimentally.

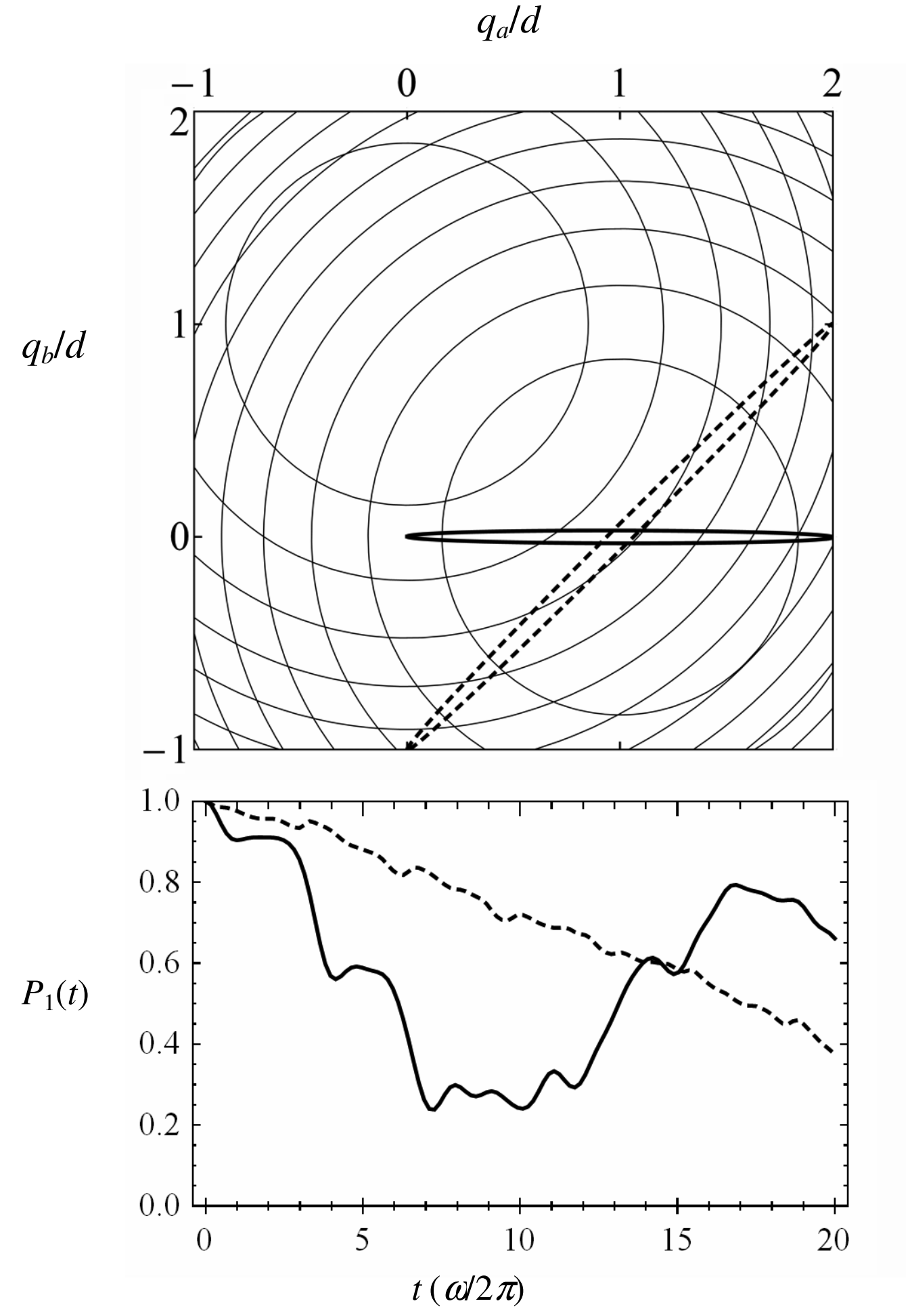

FIG. 1. Top Panel: Initial trajectories of wave packets launched impulsively to the donor-excited electronic state (shown as elongated ovals for clarity) overlaying contour plots for the harmonic surfaces of two site-excited states, each of which is displaced by $d$ from the ground state potential minimum along the appropriate nuclear coordinate. Bottom Panel: Population $P_1(t)$ of the donor-excited state for the wave packets whose initial trajectories are shown in the top panel; in this simulation any population lost by the donor is gained by the acceptor and vice versa. Shown are the cases in which the wave packet prior to electronic excitation is the vibrational ground state (solid line) or a coherent state executing harmonic motion along the *acceptor* vibrational coordinate (dashed line).

This picture of coherent wave-packet motion accompanying energy transfer suggests a method for controlling the short-time dynamics of energy transfer. If the nuclear wave packet is displaced from the minimum of the ground-state potential along the acceptor vibrational coordinate at the instant of resonant donor excitation (as shown by the dashed trace in the top panel of Fig. 1), then the trajectory of the photo-excited wave packet avoids the intersection between the two site-excited potential surfaces, which should significantly curtail short-time EET. We envisage the use of ISRS or coherent infrared absorption to generate this coherent nuclear displacement in the electronic ground state.

A numerical test calculation using the same model system as Ref. 30 shows how this kind of nuclear displacement prior to electronic excitation can alter the ensuing EET. In these simulations, the initial expectation values of the nuclear coordinates are set to $\langle q_a \rangle = 0$ and $\langle q_b \rangle = -d$ at the instant when the system is promoted to the donor-excited state (whose potential-energy minimum is at $(q_a, q_b) = (d, 0)$ ). The effect on the EET dynamics of this non-stationary initial state is visible in the evolving population of the donor-excited state (bottom panel of Fig. 1). Following a vertical transition from the equilibrium geometry at time zero (solid line), EET proceeds in a stepwise fashion, as the local transition energies for the donor-excited $\rightarrow$ ground and acceptor-excited $\leftarrow$ ground electronic transitions coincide whenever $t = 2\pi n / \omega$ (where ω is the vibrational frequency).[34] When the initial nuclear wave function is the displaced state (dashed line) on the other hand, the short-time population transfer occurs more slowly and at a steadier rate.

## B. nl-WPI and pump-probe observations of vibrational control over EET

The demonstration calculation of Fig. 1 shows that vibrational control can be exerted over EET and motivates our two main goals of outlining a framework for the spectroscopic study of this process (in this paper) and beginning its detailed numerical simulation (in the following paper).[35] In a WPI experiment, the system is subjected to one or more pairs of phase-locked (or phase-modulated) ultra-short laser pulses resonant with electronic transitions in the system, and the fluorescence (or some other observable indicative of excited-state population) is measured as a function of the various interpulse delays.[36,37,38,39,40,41] By cycling the intrapulse-pair phase differences or detecting the third-order signal field traveling in a particular phase-matched direction, it is sometimes possible to isolate the complex-valued overlap of a target wave packet prepared by one pulse in the sequence with a collection of reference wave packets prepared by the others. Theoretical[22,37,42,43,44,45,46,47,48] and experimental[49,50,51,52,53,54,55,56,57,58,59,60,61,62,63,64] studies have addressed the application of WPI and the broader methods of multidimensional phase-coherent electronic spectroscopy to a variety of molecular systems. By providing amplitude-level information on nuclear wave packets as they evolve following electronic excitation, wave-packet interferometry should, in particular, provide uniquely detailed records of the nuclear dynamics accompanying surface-crossing processes such as electronic excitation transfer.

Below, we derive the basic expressions for the nl-WPI signal from a sample of isotropically oriented EET dimers following prior vibrational excitation through ISRS by a pre-resonant control pulse. Ideally, the target wave packet—the alteration of whose dynamics by control-pulse-induced vibrational motion we wish to monitor—would be prepared by one of the pulses in the first pulse-pair of the four-pulse WPI sequence, while the remaining WPI pulses would generate a family of three-pulse reference wave packets, whose overlaps with the target constitute the signal (the relevant interference contribution to the one-exciton population). The control pulse and each of the phase-controlled pulse-pairs have independently selected linear polarizations, allowing a measure of external choice—in a statistical sense—of acceptor and donor transition-dipole lab-frame orientations. The control-influenced nl-WPI signal proves in general to be more complicated than the ideal situation described above; but the use of polarization selection, choice of pulse center-frequencies, and selection of interpulse delays and optical phase signature, all serve to simplify the observed signal.[42]

Although the full nl-WPI signal provides a more complete amplitude-level characterization of the control-influenced energy-transfer dynamics, a relatively simpler polarized fluorescence-detected pump-probe experiment may be sufficient to test the efficacy of vibrational control of EET. Such an experiment reveals the population transfer between monomers (nuclear probability density),[30] but provides less detailed information on the evolution of the accompanying amplitudes (nuclear wave packets). A nl-WPI experiment reduces to a pump-probe experiment when the pulses within each pair arrive simultaneously, and an expression for the pump-induced change in the probe-induced fluorescence is derived from the nl-WPI signal by taking this limit.

## 2. THEORY

We consider an energy-transfer dimer and treat each monomer as a system of two electronic levels. The Hamiltonian of the molecular complex is

$$H = |0\rangle H_0 \langle 0| + |1\rangle H_1 \langle 1| + |1'\rangle H_{1'} \langle 1'| + |2\rangle H_2 \langle 2| + J\left\{ |1'\rangle\langle 1| + |1\rangle\langle 1'| \right\} \ . \quad (1)$$

$H_n$ ( $n = 0, 1, 1', 2$ ) is the nuclear Hamiltonian in the electronic state $|n\rangle$, where

$$
\begin{aligned}
&|0\rangle = |gg\rangle && \text{(both chromophores unexcited)}\\
&|1\rangle = |eg\rangle && \text{("first" molecule electronically excited)}\\
&|1'\rangle = |ge\rangle && \text{("second" molecule excited)}\\
&|2\rangle = |ee\rangle && \text{(both molecules excited)} \quad .
\end{aligned}
\tag{2}
$$

Note that the $H_n$ govern the nuclear motion essential to our control strategy. The energy-transfer coupling $J$ may, in general, depend on nuclear coordinates.

The complex interacts with five independently polarized laser pulses. An electronically non-resonant control pulse, $P$, comes first; it is responsible for initiating vibrational motion in the electronic ground state by impulsive stimulated Raman scattering[65] (it could be replaced by a timed sequence of pulses in order to selectively excite a single mode of a particular frequency[12, 13, 66, 67]). The control pulse is followed by a nonlinear WPI sequence comprising a pair of phase-locked (or phase-modulated) electronically resonant pulse-pairs, $A$-$B$ and $C$-$D$ (see Fig. 2). The laser-molecule interactions are governed by

$$V(t) = -\sum_I \boldsymbol{\mu}\cdot\boldsymbol{E}_I(t)\,; \qquad I = P, A, B, C, D\,. \tag{3}$$

The dipole moment operator,

$$\boldsymbol{\mu} = \mathrm{m}\big(|0\rangle\langle 1| + |1'\rangle\langle 2|\big) + \mathrm{m}'\big(|0\rangle\langle 1'| + |1\rangle\langle 2|\big) + \mathrm{H.c.}\,, \tag{4}$$

facilitates electronic transitions for the individual chromophores. The dimer complex is assumed to be "rigid" in the sense that the relative angle between $\boldsymbol{m}$ and $\boldsymbol{m}'$ is taken to be a fixed parameter, rather than a dynamical variable. The electric field of the $I^{\text{th}}$ pulse is written as

$$\mathrm{E}_I(t) = \mathrm{e}_I E_I f_I(t)\cos\big(\Phi_I(t) + \varphi_I\big)\,, \tag{5}$$

with a well-defined polarization vector, field amplitude, envelope function (peaked at arrival time $t_I$ with duration $\sigma_I$), phase function, and constant phase.

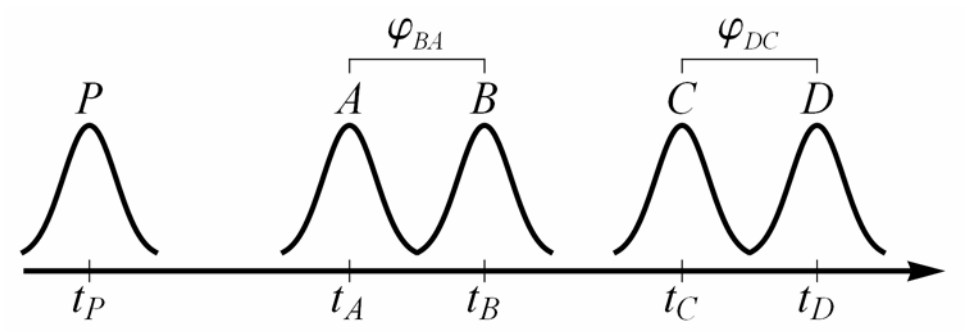

FIG. 2. Schematic representation of the pulse sequence for the experiments considered here. The pre-resonant control pulse (*P*) precedes the nl-WPI sequence (*A*-*D*). The absolute phase of any pulse is considered random, while the phase differences $\varphi_{BA}$ and $\varphi_{DC}$ are precisely controlled.

The two pulse-pairs are assumed to be temporally non-overlapping, and neither overlaps the non-resonant control pulse ($t_P \ll t_A, t_B \ll t_C, t_D$ with $t_{BA} \equiv t_B - t_A \geq 0$ and $t_{DC} \equiv t_D - t_C \geq 0$). We shall account for the possibility that pulses within a pair may overlap in time, so that pulse *B* (*D*) may act on the system before pulse *A* (*C*). $\varphi_{BA} \equiv \varphi_B - \varphi_A$ and $\varphi_{DC} \equiv \varphi_D - \varphi_C$ are assumed to be externally controlled or modulated, but the absolute phases $\varphi_I$ are unknown and subject to random jitter between successive laser shots. The relative phases between *P* and *A*, *B*, *C*, and *D* need not be specified, and those between *A* or *B* and *C* or *D* are similarly unknown.

The state ket of the system evolves according to

$$i\frac{\partial}{\partial t}\left|\Psi(t)\right\rangle = \left(H + V(t)\right)\left|\Psi(t)\right\rangle \ , \tag{6}$$

(we set $\hbar$ equal to 1 throughout) with an initial condition

$$\left|\Psi(t \ll t_P)\right\rangle = e^{-iH(t-t_A)}\left|0\right\rangle\left|\psi_0\right\rangle \ , \tag{7}$$

prior to any of the laser pulses, where $\left|\psi_0\right\rangle$ is an eigenstate of $H_0$.[68,69,70] Switching to the interaction picture, we evaluate

$$i\frac{\partial}{\partial t}e^{iH(t-t_A)}\left|\Psi(t)\right\rangle = e^{iH(t-t_A)}\left\{-H+\left(H+V(t)\right)\left|\Psi(t)\right\rangle\right\}$$
$$= e^{iH(t-t_A)}V(t)\left|\Psi(t)\right\rangle \ . \tag{8}$$

The formal solution is

$$e^{iH(t-t_A)}\left|\Psi(t)\right\rangle = \left|0\right\rangle\left|\psi_0\right\rangle + \frac{1}{i}\int_{-\infty}^{t} dt'\, e^{iH(t'-t_A)}V(t')\left|\Psi(t')\right\rangle \ , \tag{9}$$

or, reverting to the Schrödinger picture,

$$\left|\Psi(t)\right\rangle = e^{-iH(t-t_A)}\left|0\right\rangle\left|\psi_0\right\rangle + \frac{1}{i}\int_{-\infty}^{t} dt' e^{-iH(t-t')}V(t')\left|\Psi(t')\right\rangle \ . \tag{10}$$

This standard integral equation is mathematically equivalent to Eq. (6). It can be iterated—with the expression for $\left|\Psi(t)\right\rangle$ on the RHS being repeatedly substituted into the integral—in order to obtain solutions perturbative to any desired combination of orders of the incident fields.

As the observables of interest we seek all contributions to the one-exciton population $\left\langle\Psi(t)\right|\left(\left|1\right\rangle\left\langle 1\right| + \left|1'\right\rangle\left\langle 1'\right|\right)\left|\Psi(t)\right\rangle$ that are quadratic in $E_P$; quadrilinear in $E_A, E_B, E_C$ and $E_D$; and carry a specific dependence on the optical phase shifts $\varphi_{BA}$ and $\varphi_{DC}$. These observables represent the control-pulse-induced change in the nl-WPI signals of various phase signatures. Contributions of this kind can be isolated experimentally by optical phase cycling or other fundamentally equivalent procedures.[40, 37, 49, 50, 57, 58, 59, 60, 61, 62, 63, 64]

In order to keep things relatively simple, we specialize to a complex that does not rotate in the interval $t_{DP}$ between the first and last pulses. We further assume that the time-and-frequency-integrated fluorescence from the one-exciton manifold can be selectively detected without contamination from the two-exciton emission.[71] As an additional experimentally realizable simplification, we assume that the pulse durations $\sigma_I$ are significantly shorter than the characteristic timescale, $2\pi/J$, for electronic excitation transfer; this assumption allows us to neglect energy transfer in calculating the action of the individual pulses.[72]

Because contributions to $\langle 1|\Psi(t)\rangle$ and $\langle 1'|\Psi(t)\rangle$ come at odd orders in the external fields, we must determine the first-, third-, and fifth-order terms in perturbation theory. Using the notation $[t] = \exp(-i\,H\,t)$ for the free-evolution operator, we can write an expansion of Eq. (10) through fifth order,

$$\begin{aligned} |\Psi(t)\rangle = \{ & [t - t_A] + \frac{1}{i}\int_{-\infty}^{t} dt'[t-t']V(t')[t'-t_A] \\ & + \frac{1}{i^2}\int_{-\infty}^{t} dt'' \int_{-\infty}^{t''} dt'[t-t'']V(t'')[t''-t']V(t')[t'-t_A] \\ & + \cdots \text{ through fifth order}\}|0\rangle|\psi_0\rangle \quad . \end{aligned} \tag{11}$$

We define the pulse propagators

$$P_I(t;t') = -\frac{1}{i}\int_{-\infty}^{t} dt'[t_I - t']\boldsymbol{\mu}\cdot\boldsymbol{E}_I(t')[t'-t_I] \ , \tag{12}$$

where the first argument is the upper limit of integration and the second is the integration variable. With this notation, the perturbed state (11) can be re-expressed as

$$\begin{aligned} |\Psi(t)\rangle = \{ & [t - t_A] + \sum_I [t - t_I]P_I(t;t')[t_{IA}] \\ & + \sum_J\sum_I [t-t_J]P_J(t;t'')[t_{JI}]P_I(t'';t')[t_{IA}] \\ & + \cdots \text{ through fifth order}\}|0\rangle|\psi_0\rangle \quad . \end{aligned} \tag{13}$$

We evaluate the necessary electronic matrix elements of the pulse propagators with the help of the rotating wave approximation and neglect energy transfer (but not vibrational motion) during the interaction due to the impulsive nature of the excitation.

For an upward transition such as $1 \leftarrow 0$, we have

$$\begin{aligned}
\langle 1|P_I(t;t')|0\rangle &\cong \frac{iE_I}{2}\int_{-\infty}^{t} dt'[t_I - t']_{11}\langle 1|\boldsymbol{\mu}|0\rangle\cdot \mathrm{e}_I f_I(t')\, e^{-i\Phi_I(t')-i\varphi_I}[t'-t_I]_{00} \\
&= e^{-i\varphi_I} e_I \frac{iE_I m}{2}\int_{-\infty}^{t} dt'\, f_I(t')\, e^{-i\Phi_I(t')}[t_I - t']_{11}[t'-t_I]_{00} \\
&\equiv e^{-i\varphi_I} e_I\, p_I^{(10)}(t;t') \quad ,
\end{aligned} \tag{14}$$

where $\mathrm{e}_I \cdot \mathrm{m} = e_I m$ (i.e. $e_I$ is the component of $\mathrm{e}_I$ along $\mathrm{m}$).[73] With the neglect of intrapulse energy transfer, *J* is to be set to zero when evaluating the free-evolution operators in the integrand, so $[t]_{11} \cong \exp(-iH_1 t)$, $[t]_{1'1'} \cong \exp(-iH_{1'}t)$, and $[t]_{1'1} = [t]_{11'} \cong 0$.

For a downward transition such as $1 \rightarrow 0$ we have, similarly,

$$\begin{aligned}
\langle 0|P_I(t;t')|1\rangle &\cong e^{i\varphi_I} e_I \frac{iE_I m}{2}\int_{-\infty}^{t} dt'[t_I - t']_{00} f_I(t')\, e^{i\Phi_I(t')}[t'-t_I]_{11} \\
&\equiv e^{i\varphi_I} e_I\, p_I^{(01)}(t;t') \quad .
\end{aligned} \tag{15}$$

Notice the anti-Hermitian property of the *reduced pulse propagators*

$$p_I^{(01)}(t;t') = -\left(p_I^{(10)}(t;t')\right)^{\dagger} . \tag{16}$$

The reduced pulse propagators introduced in Eq. (15) are purely nuclear operators describing the pulse-dependent distortion of vibrational wave packets accompanying electronic transitions. Extending this notation to the other electronic matrix elements of $P_I$ allows us to write

$$\begin{aligned}
P_I(t;t') = {} & e^{-i\varphi_I} e_I \left(|1\rangle\langle 0|\, p_I^{(10)}(t;t') + |2\rangle\langle 1'|\, p_I^{(21')}(t;t')\right) \\
& + e^{-i\varphi_I} e_I' \left(|1'\rangle\langle 0|\, p_I^{(1'0)}(t;t') + |2\rangle\langle 1|\, p_I^{(21)}(t;t')\right) \\
& + e^{i\varphi_I} e_I \left(|0\rangle\langle 1|\, p_I^{(01)}(t;t') + |1'\rangle\langle 2|\, p_I^{(1'2)}(t;t')\right) \\
& + e^{i\varphi_I} e_I' \left(|0\rangle\langle 1'|\, p_I^{(01')}(t;t') + |1\rangle\langle 2|\, p_I^{(12)}(t;t')\right) \quad .
\end{aligned} \tag{17}$$

For the purpose of calculating the non-resonantly pumped nl-WPI signal, we need not determine each and every nuclear wave packet linearly superposed to form $\langle 1|\Psi(t)\rangle$ and $\langle 1'|\Psi(t)\rangle$. Only those packets linear or trilinear in the WPI-pulse field amplitudes and zeroth- or second-order in the non-resonant control field contribute to the signal (because of its non-resonance, the control pulse cannot make a linear contribution to these nuclear amplitudes alone or in combination with the temporally separated WPI pulses). The relevant first-order contributions to $\langle \varepsilon|\Psi(t)\rangle$ $(\varepsilon = 1,1')$ are

$$\left|(A)_{\varepsilon}\right\rangle,\ \left|(B)_{\varepsilon}\right\rangle,\ \left|(C)_{\varepsilon}\right\rangle,\ \left|(D)_{\varepsilon}\right\rangle \ . \tag{18}$$

The necessary third-order terms are

$$\begin{aligned}&\left|(APP)_{\varepsilon}\right\rangle,\ \left|(BPP)_{\varepsilon}\right\rangle,\ \left|(CPP)_{\varepsilon}\right\rangle,\ \left|(DPP)_{\varepsilon}\right\rangle\\ &\left|(CBA)_{\varepsilon}\right\rangle,\ \left|(CAB)_{\varepsilon}\right\rangle,\ \left|(DBA)_{\varepsilon}\right\rangle,\ \left|(DAB)_{\varepsilon}\right\rangle\\ &\left|(DCA)_{\varepsilon}\right\rangle,\ \left|(CDA)_{\varepsilon}\right\rangle,\ \left|(DCB)_{\varepsilon}\right\rangle,\ \left|(CDB)_{\varepsilon}\right\rangle,\end{aligned} \tag{19}$$

and contributing fifth-order wave packets consist of

$$\begin{aligned}&\left|(CBAPP)_{\varepsilon}\right\rangle,\ \left|(CABPP)_{\varepsilon}\right\rangle,\ \left|(DBAPP)_{\varepsilon}\right\rangle,\ \left|(DABPP)_{\varepsilon}\right\rangle\\ &\left|(DCAPP)_{\varepsilon}\right\rangle,\ \left|(CDAPP)_{\varepsilon}\right\rangle,\ \left|(DCBPP)_{\varepsilon}\right\rangle,\ \left|(CDBPP)_{\varepsilon}\right\rangle \ .\end{aligned}$$

(20)

The meaning of our notation should already be clear. For example,

$$\left|(APP)_{\varepsilon}\right\rangle = \langle \varepsilon|[t-t_A]P_A(t;t''')[t_{AP}]P_P(t''';t'')P_P(t'';t')[t_{PA}]|0\rangle|\psi_0\rangle \ . \tag{21}$$

The nuclear state-kets listed in Eqs. (18) through (20) and discussed extensively in the rest of this paper are the vibrational wave packets referred to in the term "wave-packet interferometry." It is a central feature of our analysis that the sought-after multi-dimensional electronic spectroscopy signals are explicitly expressed and interpreted in terms of the quantum mechanical overlaps between various pairs of these vibrational wave packets.

A special feature of the experimental observable further simplifies the calculation. We are interested in various contributions to the population of the one-exciton manifold at a time $t > t_D$ (i.e. well after the last pulse has acted). Any contribution to the $\varepsilon$-state amplitude continues to evolve during $t$-$t_D$ and, in particular, amplitude may be gained or lost through excitation transfer. But the Hermitian operator for the one-exciton population $|1\rangle\langle 1| + |1'\rangle\langle 1'|$ commutes with $H$, so $[-t+t_D]\big(|1\rangle\langle 1| + |1'\rangle\langle 1'|\big)[t-t_D] = |1\rangle\langle 1| + |1'\rangle\langle 1'|$. Without approximation, we may therefore formally set $t = t_D$ in the final free-evolution operator of the contributing first-, third-, and fifth-order nuclear amplitudes used in calculating the one-exciton population.

Each of the first-, third-, and fifth-order nuclear wave packets listed above involves multiple interfering pathways through electronic state space.[74] For instance, the $\varepsilon$-state wave packet linear in $E_A$ alone has contributions from nuclear amplitudes accompanying the direct excitation of both $|1\rangle$ and $|1'\rangle$ states:

$$\begin{aligned} \big|(A)_\varepsilon\big\rangle &= \langle\varepsilon|[t_{DA}]P_A(\infty;t')|0\rangle|\psi_0\rangle \\ &= e^{-i\varphi_A} e_A [t_{DA}]_{\varepsilon 1}\, p_A^{(10)}(\infty;t')|\psi_0\rangle \\ &\quad + e^{-i\varphi_A} e'_A [t_{DA}]_{\varepsilon 1'}\, p_A^{(1'0)}(\infty;t')|\psi_0\rangle \ . \end{aligned} \tag{22}$$

The quantities $[t_{DA}]_{\varepsilon 1}$ and $[t_{DA}]_{\varepsilon 1'}$ are nuclear operators that incorporate the energy-transfer events occurring after electronic excitation $(J \neq 0)$. The upper integration limit of the reduced pulse propagators in Eq. (22) has been set to infinity, since the pulse is assumed to have passed entirely before the excited-state amplitude is reckoned. This assumption is appropriate, because the one-exciton population is to be determined by fluorescence, whose lifetime greatly exceeds the pulse durations.

We can represent this situation with a sketch illustrating the branching pathways through electronic state space that contribute to $\left|(A)_\varepsilon\right\rangle$ :

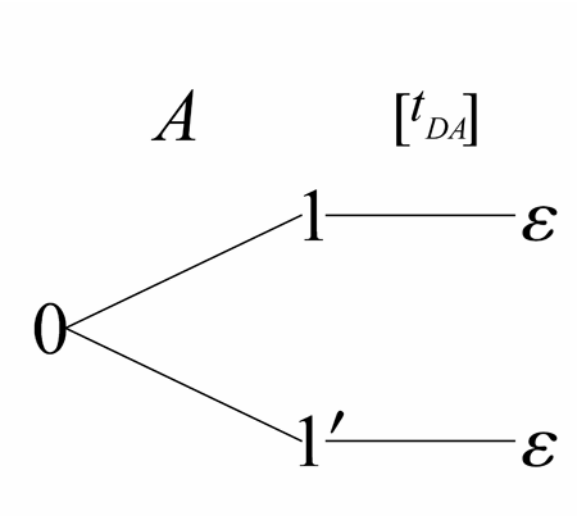

(23)

We now introduce a streamlined notation for $\left|(A)_\varepsilon\right\rangle$ that makes explicit the optical phase-dependence, the polarization-moment direction cosines, and the electronic state-space pathway of each contributing term, rewriting Eq. (22) as

$$\left|(A)_\varepsilon\right\rangle = e^{-i\varphi_A}\left[e_A\left|\{a(10)\}_\varepsilon\right\rangle + e'_A\left|\{a(1'0)\}_\varepsilon\right\rangle\right] \quad . \qquad (24)$$

Both terms carry the same optical phase factor, as both result from an upward electronic transition driven by pulse *A*. In the case $\varepsilon = 1(1')$ the first (second) term is the probability amplitude for the excitation to remain in the $1(1')$ state after $t_{DA}$ of evolution in the one-exciton manifold (back-transfer included); the second (first) term is the probability amplitude for energy transfer to have taken place to the $1'(1)$ state. This example should illustrate sufficiently the notation we will use to express the other first, third, and fifth-order $\varepsilon$-state wave packets resulting from multiple interfering pathways through electronic state-space. The other first-order wave packets follow by direct analogy.

The state-space pathways for $\left|(APP)_\varepsilon\right\rangle$ can be sketched as

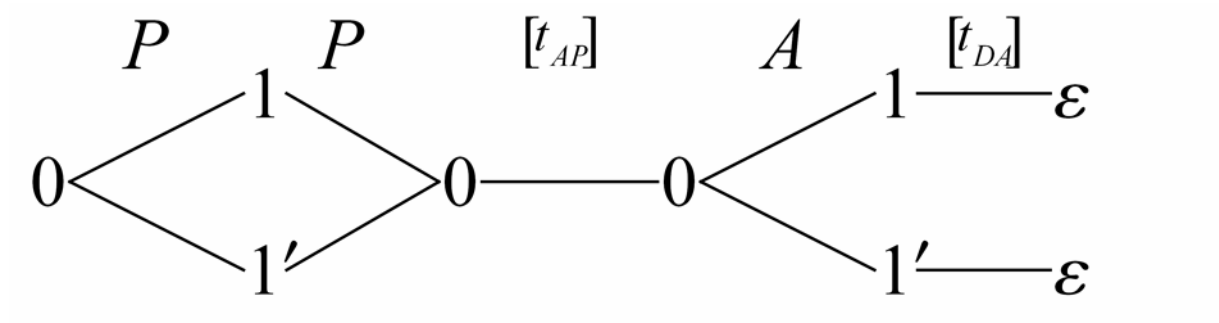

. (25)

The second-order action of the non-resonant control pulse can return amplitude only to the electronic ground state; transient occupation of the 1 or $1'$ state imparts momentum to Franck-Condon active vibrational modes. We wish to investigate the opportunity this impulsive Raman process offers for vibrational control over the energy transfer that takes place after subsequent resonant excitation to a site-excited electronic state. An expression for $\left|(APP)_\varepsilon\right\rangle$ can be obtained by applying the second-order action of $P$ along with $t_{AP}$ of ground-state evolution as a "prefix" to $\left|(A)_\varepsilon\right\rangle$:

$$\begin{aligned}\left|(APP)_\varepsilon\right\rangle = e^{-i\varphi_A}\big[e_A e_P^2\left|\{a(10)p(01)p(10)\}_\varepsilon\right\rangle + e_A e_P'^2\left|\{a(10)p(01')p(1'0)\}_\varepsilon\right\rangle \\ + e_A' e_P^2\left|\{a(1'0)p(01)p(10)\}_\varepsilon\right\rangle + e_A' e_P'^2\left|\{a(1'0)p(01')p(1'0)\}_\varepsilon\right\rangle\big] \quad . \end{aligned} \tag{26}$$

Due to the up-and-down action of electronic excitation and de-excitation by the non-resonant control pulse, its optical phase does not appear. The remaining third-order terms which are quadratic in the non-resonant control pulse, $\left|(BPP)_\varepsilon\right\rangle$, $\left|(CPP)_\varepsilon\right\rangle$, and $\left|(DPP)_\varepsilon\right\rangle$ can be obtained by simply changing $A$ and $a$ to $B$ and $b$, $C$ and $c$, or $D$ and $d$, respectively, in Eq. (26) (but $\left|(DPP)_\varepsilon\right\rangle$ reduces to only two nonzero terms for each choice of $\varepsilon$).

The electronic state-space pathways for $\left|(CBA)_\varepsilon\right\rangle$ are shown below:

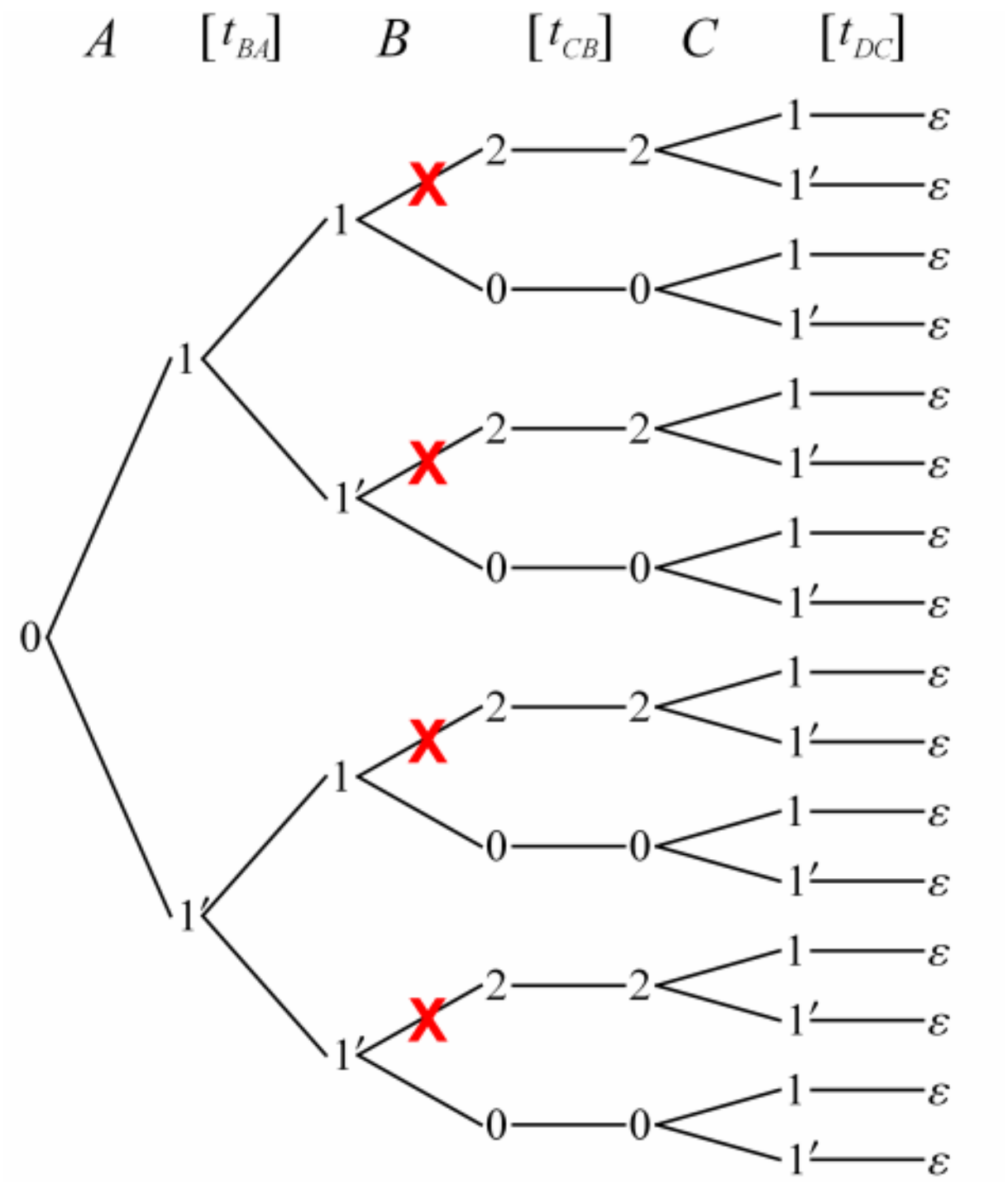

. (27)

Eight of the sixteen superposed amplitudes in $\left|(CBA)_{\varepsilon}\right\rangle$ cannot contribute in practice to the nl-WPI signal. In those marked with an **X**, both *A* and *B* pulses effect upward electronic transitions, $2 \leftarrow 1 \leftarrow 0$ or $2 \leftarrow 1' \leftarrow 0$. In either instance, the overall optical phase factor (including that from a subsequent downward transition under pulse $C$) is $\exp(-i\varphi_A - i\varphi_B + i\varphi_C)$. Any overlap involving a wave packet of this phase dependence will average to zero over many laser shots, as the uncontrolled phase factor $\exp(-i\varphi_A - i\varphi_B)$ samples points distributed randomly on a unit circle in the complex plane.

The eight remaining amplitudes in (27) carry a phase factor $\exp(-i\varphi_A + i\varphi_B - i\varphi_C)$ resulting from the upward, downward, and upward action of pulses *A*, *B*, and *C*, respectively. Overlaps between these wave packets and those involving an upward transition under pulse $D$, which carry a phase factor $\exp(-i\varphi_D)$, therefore have a phase factor $\exp(i\varphi_{BA}+i\varphi_{DC})$; since this factor involves only the controlled phase *differences,* these overlaps do not automatically average to zero and can contribute to the nl-WPI signal. The sum of amplitudes forming $\left|(CBA)_\varepsilon\right\rangle$ can now be written out from the state-space sketch:

$$
\begin{aligned}
\left|(CBA)_\varepsilon\right\rangle = e^{i\varphi_{BA}-i\varphi_C}\big[ & e_A e_B e_C \left|\{c(10)b(01)a(10)\}_\varepsilon\right\rangle \\
& + e_A e_B e'_C \left|\{c(1'0)b(01)a(10)\}_\varepsilon\right\rangle + e_A e'_B e_C \left|\{c(10)b(01')a(10)\}_\varepsilon\right\rangle \\
& + e_A e'_B e'_C \left|\{c(1'0)b(01')a(10)\}_\varepsilon\right\rangle + e'_A e_B e_C \left|\{c(10)b(01)a(1'0)\}_\varepsilon\right\rangle \\
& + e'_A e_B e'_C \left|\{c(1'0)b(01)a(1'0)\}_\varepsilon\right\rangle + e'_A e'_B e_C \left|\{c(10)b(01')a(1'0)\}_\varepsilon\right\rangle \\
& + e'_A e'_B e'_C \left|\{c(1'0)b(01')a(1'0)\}_\varepsilon\right\rangle\big] \quad .
\end{aligned}
\tag{28}
$$

A similar expression for $\left|(DBA)_\varepsilon\right\rangle$ follows by replacing *C* and *c* with *D* and *d*, respectively.

In $\left|(CAB)_\varepsilon\right\rangle$ the roles of *A*(*a*) and *B*(*b*) are reversed, despite the fact that the arrival time for the former precedes that for the latter; the *B* pulse acts first to electronically excite the system before it is de-excited by the *A* pulse. Because the interval $t_{BA}$ is nonnegative, the expression for $\left|(CAB)_\varepsilon\right\rangle$ resulting from a simple substitution in Eq. (28) would introduce a slight inconsistency in our treatment. We neglect EET during the short interaction time, and must also do so during backward evolution from $t_B$ to $t_A$ in $\left|(CAB)_\varepsilon\right\rangle$, as $t_{BA}$ must be shorter than the pulse duration for this amplitude to be nonzero. This neglect amounts to omitting the four kets whose labels would include either $a(01')b(10)$ or $a(01)b(1'0)$. $\left|(DAB)_\varepsilon\right\rangle$ can be obtained from $\left|(CAB)_\varepsilon\right\rangle$ by again replacing *C* with *D* and *c* with *d*.

The next step is to draw the state-space pathways contributing to $\left|(DCA)_\varepsilon\right\rangle$:

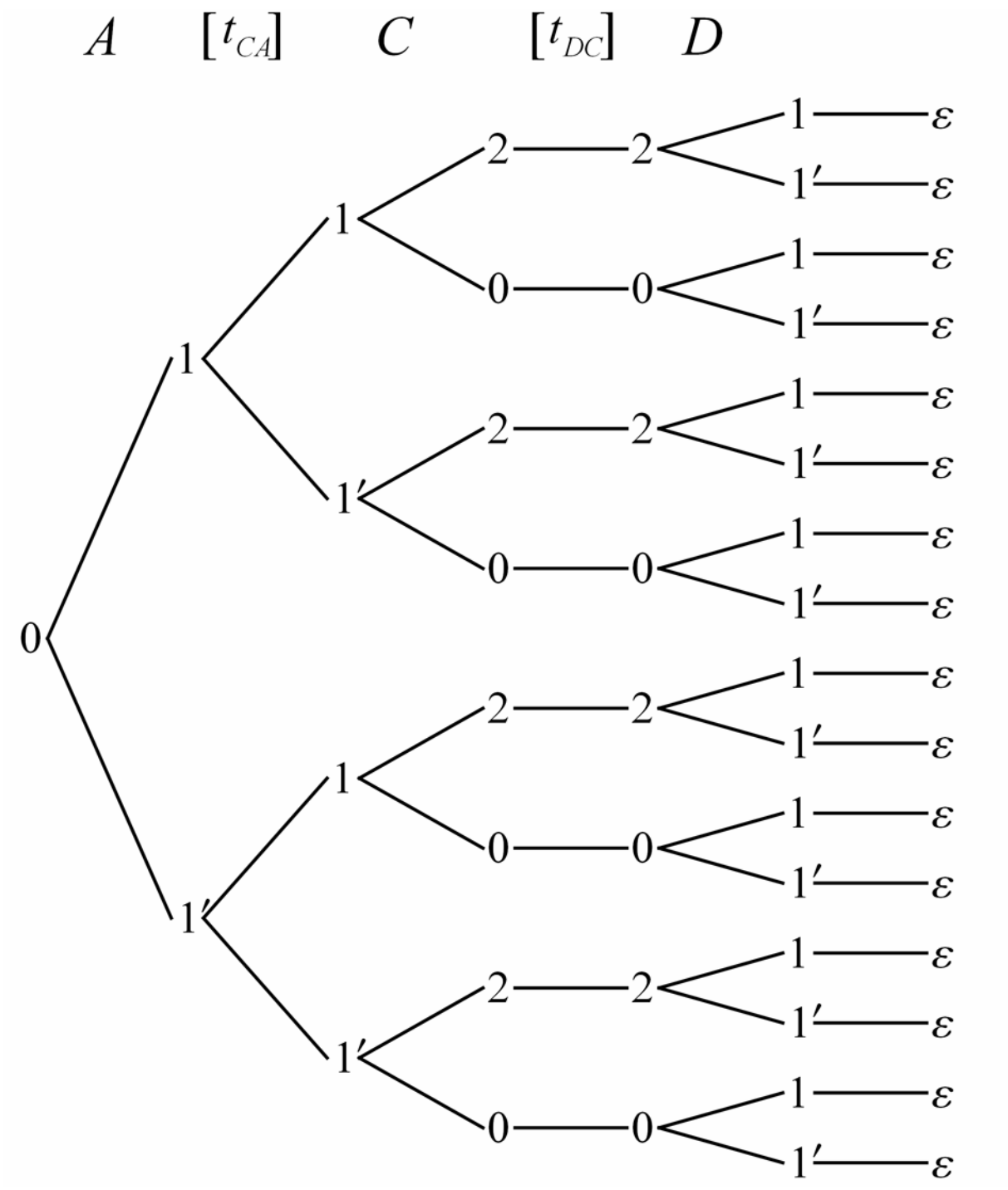

. (29)

The $C$ and $D$ pulses act in opposite directions in every instance (upwards and downwards, respectively, or vice versa). The optical phase factors accompanying all these amplitudes involve the controlled difference $\varphi_{DC}$ rather than the uncontrolled sum $\varphi_D + \varphi_C$, and none of the signal contributions to which these amplitudes contribute averages to zero due to shot-to-shot fluctuations in the absolute phase. The corresponding amplitude is

$$
\begin{aligned}
\left|(DCA)_\varepsilon\right\rangle = e^{-i\varphi_A+i\varphi_{DC}}\big[ & e_A e'_C e'_D \left|\{d(12)c(21)a(10)\}_\varepsilon\right\rangle \\
& + e_A e'_C e_D \left|\{d(1'2)c(21)a(10)\}_\varepsilon\right\rangle + e_A e_C e'_D \left|\{d(12)c(21')a(10)\}_\varepsilon\right\rangle \\
& + e_A e_C e_D \left|\{d(1'2)c(21')a(10)\}_\varepsilon\right\rangle + e'_A e'_C e'_D \left|\{d(12)c(21)a(1'0)\}_\varepsilon\right\rangle \\
& + e'_A e'_C e_D \left|\{d(1'2)c(21)a(1'0)\}_\varepsilon\right\rangle + e'_A e_C e'_D \left|\{d(12)c(21')a(1'0)\}_\varepsilon\right\rangle \\
& + e'_A e_C e_D \left|\{d(1'2)c(21')a(1'0)\}_\varepsilon\right\rangle\big] \\
& + e^{-i\varphi_A-i\varphi_{DC}}\big[e_A e_C e_D \left|\{d(10)c(01)a(10)\}_\varepsilon\right\rangle \\
& + e_A e_C e'_D \left|\{d(1'0)c(01)a(10)\}_\varepsilon\right\rangle + e_A e'_C e_D \left|\{d(10)c(01')a(10)\}_\varepsilon\right\rangle \\
& + e_A e'_C e'_D \left|\{d(1'0)c(01')a(10)\}_\varepsilon\right\rangle + e'_A e_C e_D \left|\{d(10)c(01)a(1'0)\}_\varepsilon\right\rangle \\
& + e'_A e_C e'_D \left|\{d(1'0)c(01)a(1'0)\}_\varepsilon\right\rangle + e'_A e'_C e_D \left|\{d(10)c(01')a(1'0)\}_\varepsilon\right\rangle \\
& + e'_A e'_C e'_D \left|\{d(1'0)c(01')a(1'0)\}_\varepsilon\right\rangle\big] \quad .
\end{aligned}
\tag{30}
$$

$\left|(DCB)_\varepsilon\right\rangle$ can be obtained by replacing $A$ with $B$ and $a$ with $b$ in Eq. (30). There is no interval of free evolution after the action of pulse $D$ in either $\left|(DCA)_\varepsilon\right\rangle$ or $\left|(DCB)_\varepsilon\right\rangle$. Hence eight of the contributing sixteen wave packets—those for which the left index of $d$ differs from $\varepsilon$—vanish identically.

Expressions for $\left|(CDA)_\varepsilon\right\rangle$ and $\left|(CDB)_\varepsilon\right\rangle$ can also be obtained by suitable substitutions in the full expression (30). Once again, eight of the sixteen terms vanish, in this instance approximately (those for which the left index of $c$ differs from $\varepsilon$). Because we have ignored energy transfer inside the pulse propagators, we must for consistency neglect it as well during forward propagation under $[t_{DC}]$ following the action of the $C$-pulse propagator and during backward propagation under $[-t_{DC}]$ between $D$ and $C$, as this interval must be shorter than the pulse duration for $C$ and $D$ to overlap in time.

The fifth-order wave packets listed in Eq. (20) incorporate second-order action of the control pulse and can be obtained by inserting the appropriate additional factors in the third-order amplitudes (28), (30), and their analogs. The pattern is the same as that used to obtain $\left|(APP)_{\varepsilon}\right\rangle$ from $\left|(A)_{\varepsilon}\right\rangle$ (see Eqs. (24) and (26)).

## 3. nl-WPI SIGNAL

The demonstration calculations of Section 1 illustrate one possible effect of coherent nuclear displacement on subsequent EET population dynamics (see Figure 1). Experimental verification of this control mechanism and the investigation of others by nl-WPI would perhaps be clearest on a molecular complex with a specific orientation in the lab frame.[42] A definite fixed orientation would allow polarization selection of specific “donor” and “acceptor” chromophores. In a liquid-solution or gas-phase sample of many randomly oriented complexes of fixed internal structure, a given laser pulse may instead interact with either or both of the constituent monomers. Vibrational control over EET should still be detectable, however. The polarization of the non-resonant control pulse (*P*) could, for instance, be set perpendicular to that of the first electronically resonant pulse-pair (*A* & *B*). The latter selects the donor in a statistical sense, and the former would therefore preferentially vibrationally excite the acceptor chromophore.[75] The effect of the control pulse could be sought in the signal anisotropy with respect to the polarization of the second pulse-pair (*C* & *D*), chosen parallel to the first (preferentially monitoring the donor-excited $\leftrightarrow$ ground state and acceptor-excited $\leftrightarrow$ two-exciton state transitions) or perpendicular to it (preferentially monitoring the acceptor-excited $\leftrightarrow$ ground state and donor-excited $\leftrightarrow$ two-exciton state transitions).[30]

The signal components of various phase signatures together comprise all control-induced changes in the portion of the one-exciton population that are quadrilinear in the WPI-pulse field strengths (i.e. the quadrilinear one-exciton population with the control pulse minus the quadrilinear one-exciton population without the control pulse). We therefore write the signal as

$$S = S_1 + S_{1'} \quad , \tag{31}$$

and denote that portion of $\varepsilon$-state population that is quadratic in $E_P$ and quadrilinear the WPI fields as

$$\begin{aligned} S_\varepsilon = 2\,\mathrm{Re}\Big[ & \langle (A)_\varepsilon | (DCBPP)_\varepsilon \rangle + \langle (A)_\varepsilon | (CDBPP)_\varepsilon \rangle + \langle (B)_\varepsilon | (DCAPP)_\varepsilon \rangle \\ & + \langle (B)_\varepsilon | (CDAPP)_\varepsilon \rangle + \langle (C)_\varepsilon | (DABPP)_\varepsilon \rangle + \langle (C)_\varepsilon | (DBAPP)_\varepsilon \rangle \\ & + \langle (D)_\varepsilon | (CABPP)_\varepsilon \rangle + \langle (D)_\varepsilon | (CBAPP)_\varepsilon \rangle + \langle (APP)_\varepsilon | (DCB)_\varepsilon \rangle \\ & + \langle (APP)_\varepsilon | (CDB)_\varepsilon \rangle + \langle (BPP)_\varepsilon | (DCA)_\varepsilon \rangle + \langle (BPP)_\varepsilon | (CDA)_\varepsilon \rangle \\ & + \langle (CPP)_\varepsilon | (DAB)_\varepsilon \rangle + \langle (CPP)_\varepsilon | (DBA)_\varepsilon \rangle + \langle (DPP)_\varepsilon | (CAB)_\varepsilon \rangle \\ & + \langle (DPP)_\varepsilon | (CBA)_\varepsilon \rangle \Big] \quad . \end{aligned} \tag{32}$$

Among the contributions to the signal (32) are the overlaps of each of the four single-pulse wave packets (18) with its two complementary four-pulse (fifth-order) packets from Eq. (20), with the latter differing only in the order of action of the pulses making up a pulse-pair. In addition, there are the overlaps of the four two-pulse (third-order) wave packets from Eq. (19), each of which involves a single WPI-pulse, with the two complementary three-pulse (third-order) packets from the same equation, with the latter again differing in the order of action of the pulses making up a pulse-pair.

Phase-sensitive detection methods (see especially the fluorescence-detected phase-modulation technique recently demonstrated by Marcus and co-workers in Refs. 49 and 50) can isolate the (complex-valued) portion of the signal bearing any of the four possible dependencies on the externally controlled phase differences $\varphi_{BA}$ and $\varphi_{DC}$. We have

$$S_\varepsilon = e^{i(\varphi_{BA}+\varphi_{DC})} S_\varepsilon^{++} + e^{i(\varphi_{BA}-\varphi_{DC})} S_\varepsilon^{+-} + e^{i(-\varphi_{BA}+\varphi_{DC})} S_e^{-+} + e^{i(-\varphi_{BA}-\varphi_{DC})} S_\varepsilon^{--}, \tag{33}$$

in which, for instance,

$$S_\varepsilon^{+-} = \frac{1}{4\pi^2} \int_0^{2\pi} d\varphi_{BA} \int_0^{2\pi} d\varphi_{DC}\, e^{-i(\varphi_{BA}-\varphi_{DC})}\, S_\varepsilon \tag{34}$$

is the component of $S_\varepsilon$ with phase $\varphi_{BA} - \varphi_{DC}$. We can extend this notation to identify the portion of each wave-packet overlap contributing to the signal component of a given phase signature:

$$\begin{aligned} S_\varepsilon^{++} = &\left\langle (B)_\varepsilon \middle| (DCAPP)_\varepsilon \right\rangle^{++} + \left\langle (B)_\varepsilon \middle| (CDAPP)_\varepsilon \right\rangle^{++} + \left\langle (BPP)_\varepsilon \middle| (DCA)_\varepsilon \right\rangle^{++} \\ &+ \left\langle (BPP)_\varepsilon \middle| (CDA)_\varepsilon \right\rangle^{++} + \left\langle (DCBPP)_\varepsilon \middle| (A)_\varepsilon \right\rangle^{++} + \left\langle (CDBPP)_\varepsilon \middle| (A)_\varepsilon \right\rangle^{++} \\ &+ \left\langle (DCB)_\varepsilon \middle| (APP)_\varepsilon \right\rangle^{++} + \left\langle (CDB)_\varepsilon \middle| (APP)_\varepsilon \right\rangle^{++} + \left\langle (D)_\varepsilon \middle| (CBAPP)_\varepsilon \right\rangle^{++} \\ &+ \left\langle (DPP)_\varepsilon \middle| (CBA)_\varepsilon \right\rangle^{++} + \left\langle (DABPP)_\varepsilon \middle| (C)_\varepsilon \right\rangle^{++} + \left\langle (DAB)_\varepsilon \middle| (CPP)_\varepsilon \right\rangle^{++} \end{aligned} \tag{35}$$

$$\begin{aligned} S_\varepsilon^{+-} = &\left\langle (B)_\varepsilon \middle| (CDAPP)_\varepsilon \right\rangle^{+-} + \left\langle (B)_\varepsilon \middle| (DCAPP)_\varepsilon \right\rangle^{+-} + \left\langle (BPP)_\varepsilon \middle| (CDA)_\varepsilon \right\rangle^{+-} \\ &+ \left\langle (BPP)_\varepsilon \middle| (DCA)_\varepsilon \right\rangle^{+-} + \left\langle (CDBPP)_\varepsilon \middle| (A)_\varepsilon \right\rangle^{+-} + \left\langle (DCBPP)_\varepsilon \middle| (A)_\varepsilon \right\rangle^{+-} \\ &+ \left\langle (CDB)_\varepsilon \middle| (APP)_\varepsilon \right\rangle^{+-} + \left\langle (DCB)_\varepsilon \middle| (APP)_\varepsilon \right\rangle^{+-} + \left\langle (C)_\varepsilon \middle| (DBAPP)_\varepsilon \right\rangle^{+-} \\ &+ \left\langle (CPP)_\varepsilon \middle| (DBA)_\varepsilon \right\rangle^{+-} + \left\langle (CABPP)_\varepsilon \middle| (D)_\varepsilon \right\rangle^{+-} + \left\langle (CAB)_\varepsilon \middle| (DPP)_\varepsilon \right\rangle^{+-} \end{aligned} \tag{36}$$

$$S_\varepsilon^{-+} = (S_\varepsilon^{+-})^* \tag{37}$$

$$S_\varepsilon^{--} = (S_\varepsilon^{++})^* \quad . \tag{38}$$

Full expressions for the components (35) – (38) in terms of overlaps between wave packets arising from specified state-space pathways, for an arbitrarily oriented EET complex, are given in Appendix B. Of particular interest is $S_\varepsilon^{+-}$ (or $S_\varepsilon^{-+}$), which may be referred to as echo-like, as it should be the least affected by inhomogeneous dephasing due to differences in the local molecular environment, provided the two site energies of the dimer are positively correlated (see for example Section V.C. of Ref. 42). This distinction may become less significant if nl-WPI experiments can eventually be performed on a single-molecule basis[76], as is possible in principle with fluorescence-detected phase-cycling methods.[49,50, 60, 61, 62]

## 4. PUMP-PROBE LIMIT

Although nl-WPI more fully exposes the amplitude-level nuclear dynamics accompanying electronic excitation transfer, the control scheme proposed here could also be tested with a simpler three-pulse experiment. In such a measurement, the population dynamics following vibrational excitation by the control pulse would be monitored by polarized fluorescence-detected pump-probe spectroscopy. The nl-WPI measurement becomes equivalent to a pump-probe experiment in the limit that pulses *A* and *B* share the same pulse shape and arrival time (the pump), and pulses *C* and *D* have a common pulse shape and arrival time (the probe).

We define $\Theta = \Theta_1 + \Theta_{1'}$ as the change in the pump-probe fluorescence signal caused by the non-resonant control pulse. All four components of the nl-WPI signal (35) – (38) contain identical information when the intrapulse-pair delays are both set to zero. Notice that the Eq. (36) can be obtained from Eq. (35) simply by interchanging the actions of pulses *C* and *D*. Hence $S_\varepsilon^{++} = S_\varepsilon^{+-}$ (and $S_\varepsilon^{--} = S_\varepsilon^{-+}$) in the pump-probe limit; and the further equivalence of pulses *A* and *B* then leads to

$$S_\varepsilon^{++} = S_\varepsilon^{+-} = S_\varepsilon^{--} = S_\varepsilon^{-+} \ , \tag{39}$$

whence

$$\Theta_\varepsilon = 4S_\varepsilon^{+-} \quad (t_{BA} = t_{DC} = 0) \ . \tag{40}$$

An additional simplification is possible, due to the fact that relationships of the form

$$\left\langle (CABPP)_{\varepsilon} \middle| (D)_{\varepsilon} \right\rangle^{+-} = \left( \left\langle (C)_{\varepsilon} \middle| (DBAPP)_{\varepsilon} \right\rangle^{+-} \right)^{*}$$

(41)

obtain in the pump-probe limit, and we arrive at

$$\begin{aligned} \Theta_{\varepsilon} = & 8\,\mathrm{Re}\left[ \left\langle (C)_{\varepsilon} \middle| (DBAPP)_{\varepsilon} \right\rangle^{+-} + \left\langle (CPP)_{\varepsilon} \middle| (DBA)_{\varepsilon} \right\rangle^{+-} \right] \\ & + 8\,\mathrm{Re}\left[ \left\langle (B)_{\varepsilon} \middle| (CDAPP)_{\varepsilon} \right\rangle^{+-} + \left\langle (BPP)_{\varepsilon} \middle| (CDA)_{\varepsilon} \right\rangle^{+-} \right] \\ & + 8\,\mathrm{Re}\left[ \left\langle (B)_{\varepsilon} \middle| (DCAPP)_{\varepsilon} \right\rangle^{+-} + \left\langle (BPP)_{\varepsilon} \middle| (DCA)_{\varepsilon} \right\rangle^{+-} \right] \quad . \end{aligned} \tag{42}$$

Grouping wave-packet overlaps based on the nature of their interactions with the resonant pump (*A*-*B*) and probe (*C*-*D*) pulses leads to the identification of the three terms in Eq. (42) as the ground-state bleach, excited-state absorption, and stimulated emission components of the control-induced change in the pump-probe signal, respectively ($\Theta_{\varepsilon} = \Theta_{\varepsilon}^{GSB} + \Theta_{\varepsilon}^{ESA} + \Theta_{\varepsilon}^{SE}$). In the ground-state bleach terms, the *A*-*B* pulse acts twice ($0 \leftarrow \varepsilon \leftarrow 0$) to return amplitude to the ground electronic state (and, loosely speaking, thereby cancel ground-state amplitude through destructive interference).[77] In the remaining terms, the *C*-*D* pulse acts twice on a wave packet already prepared in an excited electronic state by the *A*-*B* pulse ($\varepsilon \leftarrow 2 \leftarrow \varepsilon$ for excited-state absorption and $\varepsilon \leftarrow 0 \leftarrow \varepsilon$ for stimulated emission). Tuning the pulse parameters, such as their duration and center frequency, should enable the isolation of one or two of these contributions in many cases.[30] Explicit expressions for these contributions to the pump-probe difference signal, in terms of overlaps between nuclear wave packets generated by specific pathways through electronic state-space, can be found in Appendix C. Numerical calculations based on the formulas given in Appendix C are presented in the following article.[35]

## 5. CONCLUDING DISCUSSION

We have proposed a strategy both for influencing the short-time dynamics of electronic excitation transfer by imparting coherent nuclear motion through impulsive vibrational excitation prior to electronic absorption and for monitoring the operation of this control process by nonlinear wave-packet interferometry. The theoretical framework necessary to calculate the polarized nonlinear wave-packet interferometry signal from a randomly oriented EET complex of well-defined internal structure after vibrational excitation by impulsive stimulated Raman scattering is presented in full detail, along with the corresponding expressions in the pump-probe limit of nl-WPI. The amplitude-level picture of system dynamics provided by nl-WPI will give useful information on surface-crossing dynamics in multi-chromophoric systems. The effect of coherent nuclear motion on molecular processes in excited electronic states could be studied by similar means in the context of internal conversion in large molecules, systems exhibiting Jahn-Teller dynamics, or those with conically intersecting adiabatic electronic potential surfaces generally.[78,79]

Good candidates for experimental and further theoretical study of the control processes envisaged here would be dimer complexes with EET coupling sufficiently strong to compete with other electronic decay mechanisms, yet weak enough to allow the neglect of excitation transfer on the timescale of ultra-short pulse lengths. The linear absorption and fluorescence spectra of such species would be expected to resemble qualitatively those of the constituent monomers. In order to separately address the internal vibrations (by ISRS) and electronic transitions of the monomers, it will be important that they possess non-parallel electronic transition dipole moments. In the simplest case, allowing single-pulse pre-resonant impulsive excitation of a controlling vibrational mode, the monomer should have a Raman-active low-frequency mode that can be selectively excited by a pulse whose duration is somewhat shorter than the period of that mode, but longer than that of other active modes.

Various anthracene dimers investigated by Yamazaki and co-workers might be appropriate systems in which to study the vibrational control of EET.[80,81] The low-frequency vibrational progression in the $S_1 \leftarrow S_0$ absorption (and $S_1 \rightarrow S_0$ fluorescence) spectrum of anthracene is dominated by the $\nu_{12}$ fundamental and its first overtone[82] ($\omega \cong 390\,\mathrm{cm}^{-1}$), which suggests that this mode could be selectively driven to coherent motion of non-negligible amplitude by pre-resonant ISRS. Dithia-anthracenophane (DTA)[80] is one promising system, with two anthracene monomers held rigidly at an angle of 88.5°. Other anthracene dimers include *ortho-* and *meta-*dianthrylbenzene.[81] All three of these dimer complexes exhibit quantum beats in their time-resolved fluorescence anisotropy indicative of EET. Several naphthalene dimers have recently been synthesized by Johnson and co-workers.[83] DTA is among the model complexes for which calculated pump-probe difference signals are reported in the paper following.[35]

## ACKNOWLEDGEMENTS

We thank Ms. Yanling Zhao and Mr. Chao Li for performing preliminary calculations on the vibrational control of EET. This research was supported by grants from the American Chemical Society Petroleum Research Fund and the US National Science Foundation.

## APPENDIX A: ORIENTATIONAL AVERAGING

If the experiment is performed on a sample in liquid or solid solution, or in the gas-phase, then we must average the direction cosine terms such as $e_A e_B e_C e_D e_P^2$ to account for the isotropic distribution of dimer orientations (see Appendices B and C).[84] Although the ideal system to demonstrate our control mechanism would have perpendicular transition dipole moments, any system whose monomer transition moments are non-parallel should be susceptible to coherent vibrational control. The nl-WPI signal is a portion of the total fluorescence from the one-exciton manifold, and is therefore insensitive to molecular rotations following the last-acting pulse.

We model the system as possessing two transition dipoles whose relative orientation remains fixed while the lab-frame orientation of the system is randomly distributed. With $\alpha$ as the angle between the monomer transition moments, we can specify the orientation the molecular dipoles in the lab frame ($X,Y,Z$) in terms of Euler angles through

$$\begin{aligned} \mathrm{m} &= m\,\mathrm{R}_Z(\phi)\mathrm{R}_Y(\theta)\mathrm{R}_Z(\psi)\hat{\mathrm{Z}} \\ \mathrm{m}' &= m'\,\mathrm{R}_Z(\phi)\mathrm{R}_Y(\theta)\mathrm{R}_Z(\psi)\mathrm{R}_Y(\alpha)\hat{\mathrm{Z}}. \end{aligned} \tag{43}$$

Although the number of possible polarization combinations is large, we restrict ourselves here to two arrangements. In order for the non-resonant control pulse to influence EET in a manner similar to that illustrated in the demonstration calculation of Sec. 1, it should excite coherent motion on the chromophore that is *not* electronically excited by the first pulse-pair ($A$ & $B$). Thus we choose $\boldsymbol{e}_P$ perpendicular to $\boldsymbol{e}_A$ (we also choose $\boldsymbol{e}_A$ parallel to $\boldsymbol{e}_B$ and $\boldsymbol{e}_C$ parallel to $\boldsymbol{e}_D$). We consider the cases in which the second pulse-pair ($C$ & $D$) is polarized either parallel to the first pulse-pair or parallel to the control pulse. Defining the orientational average as

$$\langle F \rangle = \frac{1}{8\pi^2} \int_0^{2\pi} \mathrm{d}\phi \int_0^{\pi} \sin\theta \; \mathrm{d}\theta \int_0^{2\pi} \mathrm{d}\psi \, F \quad , \tag{44}$$

we find values for the direction-cosine prefactors parametrized by the inter-chromophore angle $\alpha$. These are listed in Tables 1 and 2.

TABLE 1. Rotationally averaged values for the direction-cosine terms multiplying the wave-packet overlaps that compose the signals considered here. For the case of "parallel probing," we take the four WPI pulses (*A*,*B*,*C*, and *D*) polarized parallel to each other and perpendicular to the control pulse (*P*). $\alpha$ is the angle between the monomer transition-dipole moments.

| **Term** | **Parallel Probing** |
|---|---|
| $\langle e_A e_B e_C e_D e_P^2 \rangle, \langle e'_A e'_B e'_C e'_D e'^2_P \rangle$ | $1/35$ |
| $\langle e'_A e_B e_C e_D e_P^2 \rangle, \langle e_A e'_B e_C e_D e_P^2 \rangle$ | |
| $\langle e_A e_B e'_C e_D e_P^2 \rangle, \langle e_A e_B e_C e'_D e_P^2 \rangle$ | $(\cos\alpha)/35$ |
| $\langle e_A e'_B e'_C e'_D e'^2_P \rangle, \langle e'_A e_B e'_C e'_D e'^2_P \rangle$ | |
| $\langle e'_A e'_B e_C e'_D e'^2_P \rangle, \langle e'_A e'_B e'_C e_D e'^2_P \rangle$ | |
| $\langle e'_A e'_B e_C e_D e_P^2 \rangle, \langle e'_A e_B e'_C e_D e_P^2 \rangle$ | |
| $\langle e'_A e_B e_C e'_D e_P^2 \rangle, \langle e_A e'_B e'_C e_D e_P^2 \rangle$ | |
| $\langle e_A e_B e'_C e'_D e_P^2 \rangle, \langle e_A e'_B e_C e'_D e_P^2 \rangle$ | |
| $\langle e'_A e'_B e_C e_D e'^2_P \rangle, \langle e'_A e_B e'_C e_D e'^2_P \rangle$ | $(5+\cos 2\alpha)/210$ |
| $\langle e'_A e_B e_C e'_D e'^2_P \rangle, \langle e_A e'_B e'_C e_D e'^2_P \rangle$ | |
| $\langle e_A e_B e'_C e'_D e'^2_P \rangle, \langle e_A e'_B e_C e'_D e'^2_P \rangle$ | |
| $\langle e'_A e'_B e'_C e_D e_P^2 \rangle, \langle e'_A e'_B e_C e'_D e_P^2 \rangle$ | |
| $\langle e'_A e_B e'_C e'_D e_P^2 \rangle, \langle e_A e'_B e'_C e'_D e_P^2 \rangle$ | |
| $\langle e'_A e_B e_C e_D e'^2_P \rangle, \langle e_A e'_B e_C e_D e'^2_P \rangle$ | $(5\cos\alpha - \cos 3\alpha)/140$ |
| $\langle e_A e_B e'_C e_D e'^2_P \rangle, \langle e_A e_B e_C e'_D e'^2_P \rangle$ | |
| $\langle e_A e_B e_C e_D e'^2_P \rangle, \langle e'_A e'_B e'_C e'_D e_P^2 \rangle$ | $(2-\cos 2\alpha)/35$ |

TABLE 2. The same as Table 1, but in this case the "probe pulse-pair" ($C$ & $D$) is polarized parallel to the control pulse ($P$) and perpendicular to the "pump pulse-pair" ($A$ & $B$).

| **Term** | **Perpendicular Probing** |
|---|---|
| $\langle e_A e_B e_C e_D e_P^2 \rangle, \langle e'_A e'_B e'_C e'_D e'^2_P \rangle$ | $1/35$ |
| $\langle e'_A e_B e_C e_D e_P^2 \rangle, \langle e_A e'_B e_C e_D e_P^2 \rangle$<br>$\langle e_A e_B e'_C e_D e_P^2 \rangle, \langle e_A e_B e_C e'_D e_P^2 \rangle$<br>$\langle e_A e'_B e'_C e'_D e'^2_P \rangle, \langle e'_A e_B e'_C e'_D e'^2_P \rangle$<br>$\langle e'_A e'_B e_C e'_D e'^2_P \rangle, \langle e'_A e'_B e'_C e_D e'^2_P \rangle$ | $(\cos\alpha)/35$ |
| $\langle e'_A e'_B e_C e_D e_P^2 \rangle, \langle e_A e_B e'_C e'_D e'^2_P \rangle$ | $(2-\cos 2\alpha)/35$ |
| $\langle e'_A e_B e'_C e_D e_P^2 \rangle, \langle e'_A e_B e_C e'_D e_P^2 \rangle$<br>$\langle e_A e'_B e'_C e_D e_P^2 \rangle, \langle e_A e'_B e_C e'_D e_P^2 \rangle$<br>$\langle e_A e'_B e_C e'_D e'^2_P \rangle, \langle e'_A e_B e'_C e_D e'^2_P \rangle$<br>$\langle e'_A e_B e_C e'_D e'^2_P \rangle, \langle e_A e'_B e'_C e_D e'^2_P \rangle$ | $(1+3\cos 2\alpha)/140$ |
| $\langle e'_A e'_B e'_C e_D e_P^2 \rangle, \langle e'_A e'_B e_C e'_D e_P^2 \rangle$<br>$\langle e_A e_B e'_C e_D e'^2_P \rangle, \langle e_A e_B e_C e'_D e'^2_P \rangle$ | $(5\cos\alpha-\cos 3\alpha)/140$ |
| $\langle e'_A e_B e'_C e'_D e_P^2 \rangle, \langle e_A e'_B e'_C e'_D e_P^2 \rangle$<br>$\langle e'_A e_B e_C e_D e'^2_P \rangle, \langle e_A e'_B e_C e_D e'^2_P \rangle$ | $(2\cos\alpha+\cos 3\alpha)/105$ |
| $\langle e'_A e'_B e'_C e'_D e_P^2 \rangle, \langle e_A e_B e'_C e'_D e_P^2 \rangle$<br>$\langle e_A e_B e_C e_D e'^2_P \rangle, \langle e'_A e'_B e_C e_D e'^2_P \rangle$ | $(5+\cos 2\alpha)/210$ |

## References

[1] P. S. Christopher, M. Shapiro, and P. Brumer, "Quantum control of internal conversion in 24-vibrational-mode pyrazine," J. Chem. Phys. **125**, 124310 (2006).

[2] M. Abe, Y. Ohtsuki, Y. Fujimura, and W. Domcke, "Optimal control of ultrafast cis-trans photoisomerization of retinal in rhodopsin via a conical intersection," J. Chem. Phys. **123**, 144508 (2005).

[3] M. Sugawara, S. Yoshizawa, and S. Yabushita, "Coherent control of wavepacket dynamics by locally designed external field," Chem. Phys. Lett. **350**, 253 (2001).

[4] T. Buckup, J. Hauer, J. Möhring, and M. Motzkus, "Multidimensional spectroscopy of β-carotene: Vibrational cooling in the excited state," Arch. Biochem. Biophys. **483**, 219 (2009).

[5] J. Hauer, H. Skenderovic, K.-L. Kompa, and M. Motzkus, "Enhancement of Raman modes by coherent control in β-carotene," Chem. Phys. Lett. **421**, 523 (2006).

[6] J. B. Ballard, H. U. Stauffer, Z. Amitay, and S. R. Leone, "Optimization of wave packet coefficients in $Li_2$ using an evolutionary algorithm: The role of resonant and nonresonant wavelengths," J. Chem. Phys. **116**, 1350 (2002).

[7] B. Amstrup and N. E. Henriksen, "Control of HOD photodissociation dynamics via bond-selective infrared multiphoton excitation and a femtosecond ultraviolet laser pulse," J. Chem. Phys. **97**, 8285 (1992).

[8] N. Elghobashi, P. Krause, J. Manz, and M. Oppel, "IR + UV laser pulse control of momenta directed to specific products: Quantum model simulations for HOD* $\rightarrow$ H + OD *versus* HO + D," Phys. Chem. Chem. Phys. **5**, 4806 (2003).

[9] L. Dhar, J. A. Rogers, and K. A. Nelson, “Time-resolved vibrational spectroscopy in the impulsive limit,” Chem. Rev. **94**, 157 (1994).

[10] A. M. Walsh and R. A. Loring, “Quantum mechanical analysis of impulsive stimulated light scattering from phonons,” J. Chem. Phys. **93**, 7566 (1990).

[11] U. Banin, A. Bartana, S. Ruhman, and R. Kosloff, “Impulsive Excitation of Coherent Vibrational Motion Ground Surface Dynamics Induced by Intense Short Pulses,” J. Chem. Phys. **101,** 8461 (1994).

[12] T. J. Smith and J. A. Cina, “Toward preresonant impulsive Raman preparation of large amplitude vibrational motion,” J. Chem. Phys. **104,** 1272 (1996).

[13] E. M. Hiller and J. A. Cina, “Can chirp enhance cumulative pre-resonant impulsive stimulated Raman excitation?” J. Chem. Phys. **105,** 3419 (1996).

[14] C. J. Bardeen, Q. Wang, and C. V. Shank, “Femtosecond chirped pulse excitation of vibrational wave packets in LD690 and bacteriorhodopsin,” J. Phys. Chem. A **102,** 2759 (1998).

[15] D. Gelman and R. Kosloff, “Minimizing broadband excitation under dissipative conditions,” J. Chem. Phys. **123**, 234506 (2005).

[16] S. Fujiyoshi, S. Takeuchi, and T. Tahara, “Time-resolved impulsive stimulated Raman scattering from excited-state polyatomic molecules in solution,” J. Phys. Chem. A **107**, 494 (2003).

[17] S. Fujiyoshi, S. Takeuchi, and T. Tahara, “Time-resolved impulsive stimulated Raman studies of 1,1´-binaphthyl in the excited state: Low-frequency vibrations and conformational relaxation,” J. Phys. Chem. A **108**, 5938 (2004).

[18] A. Kahan, O. Nahmias, N. Friedman, M. Sheves, and S. Ruhman, “Following photoinduced dynamics in bacteriorhodopsin with 7-fs impulsive vibrational spectroscopy,” J. Am. Chem. Soc. **129**, 537 (2007).

[19] P. N. Butcher and D. Cotter, *The Elements of Nonlinear Optics* (Cambridge University Press, Cambridge, 1991).

[20] S. Mukamel, *Principles of Nonlinear Optical Spectroscopy* (Oxford University Press, New York, 1995).

[21] M. Cho, *Two-Dimensional Optical Spectroscopy* (CRC press, Boca Raton, 2009).

[22] An approach somewhat similar to ours is taken in a recent paper on the two-dimensional optical spectroscopy of molecular dimers: J. Seibt, K. Renziehausen, D. V. Voronine, and V. Engel, “Probing the geometry dependence of molecular dimers with two-dimensional-vibronic spectroscopy,” J. Chem. Phys. **130**, 134318 (2009).

[23] J. L. Herek, W. Wohlleben, R. J. Cogdell, D. Zeidler, and M. Motzkus, “Quantum control of energy flow in light harvesting,” Nature **417**, 533 (2002).

[24] S. E. Bradforth, R. Jimenez, F. van Mourik, R. van Grondelle, and G. R. Fleming, “Excitation transfer in the core light-harvesting complex (LH-1) of *Rhodobacter sphaeroides*: An ultrafast fluorescence depolarization and annihilation study,” J. Phys. Chem. **99**, 16179 (1995).

[25] V. Subramanian and D. G. Evans, “Excitation energy transfer in model light-harvesting antennae,” J. Phys. Chem. B **108**, 1085 (2004).

[26] T. Renger, V. May, and O. Kühn, “Ultrafast excitation energy transfer dynamics in photosynthetic pigment-protein complexes,” Phys. Rep. **343**, 137 (2001).

[27] G. S. Engel, T. R. Calhoun, E. L. Read, T. K. Ahn, T. Mancal, Y. C. Cheng, R. E. Blankenship, and G. R. Fleming, "Evidence for wavelike energy transfer through quantum coherence in photosynthetic systems," Nature **446**, 782 (2007).

[28] T. Brixner, J. Stenger, H. M. Vaswani, M. Cho, R. E. Blankenship, and G. R. Fleming, "Two-dimensional spectroscopy of electronic couplings in photosynthesis," Nature **434**, 625 (2005).

[29] M. Cho, "Coherent two-dimensional optical spectroscopy," Bull. Korean Chem. Soc. **27**, 1940 (2006).

[30] J. A. Cina and G. R. Fleming, "Vibrational coherence transfer and trapping as sources for long-lived quantum beats in polarized emission from energy transfer complexes," J. Phys. Chem. A **108**, 11196 (2004).

[31] D. S. Kilin, O. V. Prezhdo, and M. Schreiber, "Photoinduced vibrational coherence transfer in molecular dimers," J. Phys. Chem. A **111**, 10212 (2007).

[32] R. Monshouwer, A. Baltuska, F. van Mourik, and R. van Grondelle, "Time-resolved absorption difference spectroscopy of the LH-1 antenna of *Rhodopseudomonas viridis*," J. Phys. Chem. A **102**, 4360 (1998).

[33] V. Novoderezhkin, R. Monshouwer, and R. van Grondelle, "Electronic and vibrational coherence in the core light-harvesting antenna of *Rhodopseudomonas viridis*," J. Phys. Chem. B **104**, 12056 (2000).

[34] S. Jang, Y. J. Jung, and R. J. Silbey, "Nonequilibrium generalization of Förster–Dexter theory for excitation energy transfer," Chem. Phys. **275**, 319 (2002).

[35] J. D. Biggs and J. A. Cina, "Calculations of nonlinear wave-packet interferometry signals in the pump-probe limit as tests for vibrational control over electronic excitation transfer," J. Chem. Phys. xxx.

[36] J. A. Cina, "Wave-packet interferometry and molecular state reconstruction: Spectroscopic adventures on the left-hand side of the Schrödinger equation," Annu. Rev. Phys. Chem. **59**, 319 (2008).

[37] T. S. Humble and J. A. Cina, "Nonlinear wave-packet interferometry and molecular state reconstruction in a vibrating and rotating diatomic molecule," J. Phys. Chem. B **110**, 18879 (2006).

[38] T. S. Humble and J. A. Cina, "Molecular state reconstruction by nonlinear wave packet interferometry," Phys. Rev. Lett. **93**, 060402 (2004).

[39] J. A. Cina, "Nonlinear wavepacket interferometry for polyatomic molecules," J. Chem. Phys. **113**, 9488 (2000).

[40] N. F. Scherer, A. Matro, L. D. Ziegler, M. Du, R. J. Carlson, J. A. Cina, and G. R. Fleming, "Fluorescence-detected wave packet interferometry. II. Role of rotations and determination of the susceptibility," J. Chem. Phys. **96**, 4180 (1992).

[41] N. F. Scherer, R. J. Carlson, A. Matro, M. Du, A. J. Ruggiero, V. Romero-Rochín, J. A. Cina, G. R. Fleming, and S. A. Rice, "Fluorescence-detected wave packet interferometry: Time resolved molecular spectroscopy with sequences of femtosecond phase-locked pulses," J. Chem. Phys. **95**, 1487 (1991).

[42] J. A. Cina, D. S. Kilin, and T. S. Humble, "Wave packet interferometry for short-time electronic energy transfer: Multidimensional optical spectroscopy in the time domain," J. Chem. Phys. **118**, 46 (2003).

[43] R. Martínez-Galicia and V. Romero-Rochín, "Molecular wave packet interferometry and quantum entanglement," J. Chem. Phys. **122**, 094101 (2005).

[44] S. Ramos-Sánchez and V. Romero-Rochín, "Preparation and resolution of molecular states by coherent sequences of phase-locked ultrashort laser pulses," J. Chem. Phys. **121**, 2117 (2004).

[45] P. Kjellberg, B. Brüggemann, and T. Pullerits, "Two-dimensional electronic spectroscopy of an excitonically coupled dimer," Phys. Rev. B **74**, 024303 (2006).

[46] A. Donoso, D. Kohen, and C. C. Martens, "Simulation of nonadiabatic wave packet interferometry using classical trajectories" J. Chem. Phys. **112**, 7345 (2000).

[47] S. Mukamel, "Multidimensional femtosecond correlation spectroscopies of electronic and vibrational excitations," Annu. Rev. Phys. Chem. **51**, 691 (2000).

[48] D. Egorova, M. F. Gelin, and W. Domcke, "Analysis of cross peaks in two-dimensional electronic photon-echo spectroscopy for simple models with vibrations and dissipation," J. Chem. Phys. **126**, 074314 (2007).

[49] P. F. Tekavec, T. R. Dyke, and A. H. Marcus, "Wave packet interferometry and quantum state reconstruction by acousto-optic phase modulation," J. Chem. Phys. **125**, 194303 (2006).

[50] P. F. Tekavec, G. A. Lott, and A. H. Marcus, "Fluorescence-detected two-dimensional electronic coherence spectroscopy by acousto-optic phase modulation," J. Chem. Phys. **127**, 214307 (2007).

[51] H. Katsuki, H. Chiba, C. Meier, B. Girard, and K. Ohmori, "Actively tailored spatiotemporal images of quantum interference on the picometer and femtosecond scales," Phys. Rev. Lett. **102**, 103602 (2009).

[52] K. Ohmori, "Wave-packet and coherent control dynamics," Annu. Rev. Phys. Chem. **60**, 487 (2009).

[53] K. Ohmori, H. Katsuki, H. Chiba, M. Honda, Y. Hagihara, K. Fujiwara, Y. Sato, and K. Ueda, "Real-time observation of phase-controlled molecular wave-packet interference," Phys. Rev. Lett. **96**, 093002 (2006).

[54] K. Ohmori, Y. Sato, E. E. Nikitin, and S. A. Rice, "High-precision molecular wave-packet interferometry with HgAr dimers," Phys. Rev. Lett. **91**, 243003 (2003).

[55] M. A. Bouchene, C. Nicole, and B. Girard, "Wavepacket interferometry with chirped pulses," J. Phys. B-At. Mol. Opt. **32**, 5167 (1999).

[56] Y. Cao, L. Zhang, Y. Yang, Z. Sun, and Z. Wang, "Molecular rovibrational dynamics investigated by two-photon wavepacket interferometry with phase-locked pulse pairs," Chem. Phys. Lett. **442**, 53 (2007).

[57] J. C. Vaughan, T. Hornung, K. W. Stone, and K. A. Nelson, "Coherently controlled ultrafast four-wave mixing spectroscopy," J. Phys. Chem. A **111**, 4873 (2007).

[58] M. Fushitani, M. Bargheer, M. Gühr, H. Ibrahim, and N. Schwentner, "Control of chromophore-to-bath coupling by interferometry: $Cl_2$ vibrational wave packets in solid

Ar," J. Phys. B: At. Mol. Opt. Phys. **41**, 074013 (2008).

[59] D. M. Jonas, "Two-dimensional femtosecond spectroscopy," Annu. Rev. Phys Chem. **54**, 425 (2003).

[60] D. Keusters, H. S. Tan, and W. S. Warren, "Role of pulse phase and direction in two-dimensional optical spectroscopy," J. Phys. Chem. A **103**, 10369 (1999).

[61] P. Tian, D. Keusters, Y. Suzaki, and W. S. Warren, "Femtosecond phase-coherent two-dimensional spectroscopy," Science **300**, 1553 (2003).

[62] C. Li, W. Wagner, M. Ciocca, and W. S. Warren, "Multiphoton femtosecond phase-coherent two-dimensional electronic spectroscopy," J. Chem. Phys. **126**, 164307 (2007).

[63] M. L. Cowan, J. P. Ogilvie, and R. J. D. Miller, "Two-dimensional spectroscopy using diffractive optics based phased-locked photon echoes," Chem. Phys. Lett. **386**, 184 (2004).

[64] T. Brixner, T. Mancal, I. V. Stiopkin, and G. R. Fleming, "Phase-stabilized two-dimensional electronic spectroscopy," J. Chem. Phys. **121**, 4221 (2004).

[65] For present purposes this pulse should perhaps be regarded as *pre*-resonant with the $1 \leftrightarrow 0$ and $1' \leftrightarrow 0$ transitions, so that effects of interactions with higher-lying electronic states that are absent from our model Hamiltonian can safely be neglected.

[66] A. M. Weiner, D. E. Leaird, G. P. Wiederrecht, and K. A. Nelson, "Femtosecond pulse sequences used for optical manipulation of molecular motion," Science **247**, 1317 (1990).

[67] M. M. Wefers, H. Kawashima, and K. A. Nelson, "Optical control over two-

dimensional lattice vibrational trajectories in crystalline quartz," J. Chem. Phys. **108**, 10248 (1998).

[68] The nuclear degrees of freedom governed by the $H_n$ may comprise any number of molecular and environmental modes. The pure state $|g\rangle|\psi_0\rangle$ may, in particular, describe the energy-transfer complex in thermal equilibrium with a surrounding medium at any temperature such that $k_BT$ is much smaller than the monomer electronic excitation energies. For an interesting discussion on this point in a larger context, see Refs. 69 and 70.

[69] S. Popescu, A. J. Short, and A. Winter, "Entanglement and the foundations of statistical mechanics," Nature Physics **2**, 754 (2006).

[70] S. Popescu, A. J. Short, and A. Winter, "The foundations of statistical mechanics from entanglement: Individual states vs. averages." Preprint at http://arxiv.org/abs/quant-ph/0511225 (2006).

[71] This choice of observable differs from that of Ref. 42, where the population of the "acceptor" state $|1'\rangle$ was monitored for an oriented complex. If the similarity in emission frequency from the two- and one-exciton states were to make discrimination between them difficult, the theoretical treatment would have to be modified to calculate the expected number of "stored photons," $\langle\Psi(t)|\big(2|2\rangle\langle 2| + |1\rangle\langle 1| + |1'\rangle\langle 1'|\big)|\Psi(t)\rangle$.

[72] Under stronger EET coupling, when the timescale for energy transfer may be less than or on the order of pulse durations, vibrational periods, or both, it can be advantageous to frame the treatment in terms of delocalized exciton states, rather than site-states. See Ref. 29 and Sec. III.C of Ref. 30.

[73] In the analogous matrix elements for a $1' \leftarrow 0$ transition, $\mathrm{e}_I \cdot \mathrm{m}' \equiv e_I' m'$, and $e_I'$ is the component of $\mathrm{e}_I$ along $\mathrm{m}'$.

[74] $\left|(D)_\varepsilon\right\rangle$ is an exception, however. Because energy transfer during the pulse is presumed negligible, this amplitude involves a single $\varepsilon \leftarrow 0$ excitation pathway.

[75] Another possibility would be to set $\boldsymbol{e}_{\boldsymbol{P}}$, $\boldsymbol{e}_{\boldsymbol{A}}$ and $\boldsymbol{e}_{\boldsymbol{B}}$ all parallel to each other and the delay time $t_{AP}$ equal to one quarter of the vibrational period for the mode in question. In this case the ground nuclear state of the donor-excited electronic state is preferentially populated and excited state nuclear motion is essentially quenched.

[76] J. Hernando, E. M. H. P. van Dijk, J. P. Hoogenboom, J. García-López, D. N. Reinhoudt, M. Crego-Calama, M. García-Parajó, and N. F. van Hulst, "Effect of Disorder on Ultrafast Exciton Dynamics Probed by Single Molecule Spectroscopy," Phys. Rev. Lett. **97**, 216403 (2006).

[77] T. J. Smith, L. W. Ungar, and J. A. Cina, "Resonant short-pulse effects on nuclear motion in the electronic ground state," J. Lumin. **58**, 66 (1994).

[78] A. Migani and M. Olivucci, "Conical intersections and organic reaction mechanisms," Chpt. 6 in *Conical Intersections. Electronic Structure, Dynamics, and Spectroscopy*, W. Domcke, D. R. Yarkony, and H. Köppel, Eds. (World Scientific, Singapore, 2004).

[79] D. A. Farrow, W. Qian, E. R. Smith, A. A. Ferro, and D. M. Jonas, "Polarized pump-probe measurements of electronic motion via a conical intersection," J. Chem. Phys. **128**, 144510 (2008).

[80] I. Yamazaki, S. Akimoto, T. Yamazaki, S. Sato, and Y. Sakata, "Oscillatory excitation transfer in dithiaanthracenophane: Quantum beat in a coherent photochemical process in solution," J. Phys. Chem. A **106**, 2122 (2002).

[81] I. Yamazaki, S. Akimoto, N. Aratani, and A. Osuka, "Observation of coherent recurrence motion of excitons in anthracene dimers," Bull. Chem. Soc. Jpn. **77**, 1959 (2004).

[82] W. R. Lambert, P. M. Felker, J. A. Syage, and A. H. Zewail, "Jet spectroscopy of anthracene and deuterated anthracenes," J. Chem. Phys. **81**, 2195 (1984).

[83] V. M. Cangelosi, A. C. Sather, L. N. Zakharov, O. B. Berryman, and D. W. Johnson, "Diastereoselectivity in the self-assembly of $As_2L_2Cl_2$ macrocycles is directed by the As-π interaction," Inorg. Chem. **46**, 9278 (2007).

[84] For another example of calculations involving the orientational average of six-fold products of transition-dipole directions, see A. Tokmakoff, "Orientational correlation functions and polarization selectivity for nonlinear spectroscopy of isotropic media. II. Fifth order," J. Chem. Phys. **105** 13 (1996).

# APPENDIX B: NL-WPI SIGNAL FOLLOWING CONTROL PULSE

Here we provide detailed expressions for the non-linear wave-packet interferometry signal from an energy transfer complex subjected to a pre-resonant control pulse (derived in *Using Wave-Packet Interferometry to Monitor the External Vibrational Control of Electronic Excitation Transfer*, Section 3). The signal is decomposed into four components of different phase signature,

$$\begin{aligned} S_\varepsilon &= e^{i(\phi_{BA}+\phi_{DC})} S_\varepsilon^{++} + e^{i(\phi_{BA}-\phi_{DC})} S_\varepsilon^{+-} \\ &+ e^{i(-\phi_{BA}+\phi_{DC})} S_\varepsilon^{-+} + e^{i(-\phi_{BA}-\phi_{DC})} S_\varepsilon^{--} \ , \end{aligned} \tag{B.1}$$

the first two of which are written out in detail below. The other components are easily obtained from the relations $S_\varepsilon^{--} = (S_\varepsilon^{++})^*$ and $S_\varepsilon^{-+} = (S_\varepsilon^{+-})^*$.

$$\begin{aligned} S_\varepsilon^{++} &= \langle (B)_\varepsilon | (DCAPP)_\varepsilon \rangle^{++} + \langle (B)_\varepsilon | (CDAPP)_\varepsilon \rangle^{++} + \langle (BPP)_\varepsilon | (DCA)_\varepsilon \rangle^{++} \\ &+ \langle (BPP)_\varepsilon | (CDA)_\varepsilon \rangle^{++} + \langle (DCBPP)_\varepsilon | (A)_\varepsilon \rangle^{++} + \langle (CDBPP)_\varepsilon | (A)_\varepsilon \rangle^{++} \\ &+ \langle (DCB)_\varepsilon | (APP)_\varepsilon \rangle^{++} + \langle (CDB)_\varepsilon | (APP)_\varepsilon \rangle^{++} + \langle (D)_\varepsilon | (CBAPP)_\varepsilon \rangle^{++} \\ &+ \langle (DPP)_\varepsilon | (CBA)_\varepsilon \rangle^{++} + \langle (DABPP)_\varepsilon | (C)_\varepsilon \rangle^{++} + \langle (DAB)_\varepsilon | (CPP)_\varepsilon \rangle^{++} \end{aligned} \tag{B.2}$$

$$\begin{aligned} S_\varepsilon^{+-} &= \langle (B)_\varepsilon | (DCAPP)_\varepsilon \rangle^{+-} + \langle (B)_\varepsilon | (CDAPP)_\varepsilon \rangle^{+-} + \langle (BPP)_\varepsilon | (DCA)_\varepsilon \rangle^{+-} \\ &+ \langle (BPP)_\varepsilon | (CDA)_\varepsilon \rangle^{+-} + \langle (DCBPP)_\varepsilon | (A)_\varepsilon \rangle^{+-} + \langle (CDBPP)_\varepsilon | (A)_\varepsilon \rangle^{+-} \\ &+ \langle (DCB)_\varepsilon | (APP)_\varepsilon \rangle^{+-} + \langle (CDB)_\varepsilon | (APP)_\varepsilon \rangle^{+-} + \langle (C)_\varepsilon | (DBAPP)_\varepsilon \rangle^{+-} \\ &+ \langle (CPP)_\varepsilon | (DBA)_\varepsilon \rangle^{+-} + \langle (CABPP)_\varepsilon | (D)_\varepsilon \rangle^{+-} + \langle (CAB)_\varepsilon | (DPP)_\varepsilon \rangle^{+-} \end{aligned} \tag{B.3}$$

$$
\begin{aligned}
\langle (B)_\varepsilon | (DCAPP)_\varepsilon \rangle^{++} = {} & e_A e_B e_C e_D {e_P}^2 \langle \{b(10)\}_\varepsilon | \{d(1'2)c(21')a(10)p(01)p(10)\}_\varepsilon \rangle \\
& + e'_A e_B e_C e_D {e_P}^2 \langle \{b(10)\}_\varepsilon | \{d(1'2)c(21')a(1'0)p(01)p(10)\}_\varepsilon \rangle \\
& + e'_A e'_B e_C e_D {e_P}^2 \langle \{b(1'0)\}_\varepsilon | \{d(1'2)c(21')a(1'0)p(01)p(10)\}_\varepsilon \rangle \\
& + e'_A e_B e'_C e_D {e_P}^2 \langle \{b(10)\}_\varepsilon | \{d(1'2)c(21)a(1'0)p(01)p(10)\}_\varepsilon \rangle \\
& + e'_A e_B e_C e'_D {e_P}^2 \langle \{b(10)\}_\varepsilon | \{d(12)c(21')a(1'0)p(01)p(10)\}_\varepsilon \rangle \\
& + e'_A e'_B e'_C e_D {e_P}^2 \langle \{b(1'0)\}_\varepsilon | \{d(1'2)c(21)a(1'0)p(01)p(10)\}_\varepsilon \rangle \\
& + e'_A e_B e'_C e'_D {e_P}^2 \langle \{b(10)\}_\varepsilon | \{d(12)c(21)a(1'0)p(01)p(10)\}_\varepsilon \rangle \\
& + e'_A e'_B e_C e'_D {e_P}^2 \langle \{b(1'0)\}_\varepsilon | \{d(12)c(21')a(1'0)p(01)p(10)\}_\varepsilon \rangle \\
& + e'_A e'_B e'_C e'_D {e_P}^2 \langle \{b(1'0)\}_\varepsilon | \{d(12)c(21)a(1'0)p(01)p(10)\}_\varepsilon \rangle \\
& + e_A e'_B e_C e_D {e_P}^2 \langle \{b(1'0)\}_\varepsilon | \{d(1'2)c(21')a(10)p(01)p(10)\}_\varepsilon \rangle \\
& + e_A e_B e'_C e_D {e_P}^2 \langle \{b(10)\}_\varepsilon | \{d(1'2)c(21)a(10)p(01)p(10)\}_\varepsilon \rangle \\
& + e_A e_B e_C e'_D {e_P}^2 \langle \{b(10)\}_\varepsilon | \{d(12)c(21')a(10)p(01)p(10)\}_\varepsilon \rangle \\
& + e_A e'_B e'_C e_D {e_P}^2 \langle \{b(1'0)\}_\varepsilon | \{d(1'2)c(21)a(10)p(01)p(10)\}_\varepsilon \rangle \\
& + e_A e_B e'_C e'_D {e_P}^2 \langle \{b(10)\}_\varepsilon | \{d(12)c(21)a(10)p(01)p(10)\}_\varepsilon \rangle \\
& + e_A e'_B e_C e'_D {e_P}^2 \langle \{b(1'0)\}_\varepsilon | \{d(12)c(21')a(10)p(01)p(10)\}_\varepsilon \rangle \\
& + e_A e'_B e'_C e'_D {e_P}^2 \langle \{b(1'0)\}_\varepsilon | \{d(12)c(21)a(10)p(01)p(10)\}_\varepsilon \rangle \\
& + e_A e_B e_C e_D {e'_P}^2 \langle \{b(10)\}_\varepsilon | \{d(1'2)c(21')a(10)p(01')p(1'0)\}_\varepsilon \rangle \\
& + e'_A e_B e_C e_D {e'_P}^2 \langle \{b(10)\}_\varepsilon | \{d(1'2)c(21')a(1'0)p(01')p(1'0)\}_\varepsilon \rangle \\
& + e'_A e'_B e_C e_D {e'_P}^2 \langle \{b(1'0)\}_\varepsilon | \{d(1'2)c(21')a(1'0)p(01')p(1'0)\}_\varepsilon \rangle \\
& + e'_A e_B e'_C e_D {e'_P}^2 \langle \{b(10)\}_\varepsilon | \{d(1'2)c(21)a(1'0)p(01')p(1'0)\}_\varepsilon \rangle \\
& + e'_A e_B e_C e'_D {e'_P}^2 \langle \{b(10)\}_\varepsilon | \{d(12)c(21')a(1'0)p(01')p(1'0)\}_\varepsilon \rangle \\
& + e'_A e'_B e'_C e_D {e'_P}^2 \langle \{b(1'0)\}_\varepsilon | \{d(1'2)c(21)a(1'0)p(01')p(1'0)\}_\varepsilon \rangle \\
& + e'_A e_B e'_C e'_D {e'_P}^2 \langle \{b(10)\}_\varepsilon | \{d(12)c(21)a(1'0)p(01')p(1'0)\}_\varepsilon \rangle \\
& + e'_A e'_B e_C e'_D {e'_P}^2 \langle \{b(1'0)\}_\varepsilon | \{d(12)c(21')a(1'0)p(01')p(1'0)\}_\varepsilon \rangle \\
& + e'_A e'_B e'_C e'_D {e'_P}^2 \langle \{b(1'0)\}_\varepsilon | \{d(12)c(21)a(1'0)p(01')p(1'0)\}_\varepsilon \rangle \\
& + e_A e'_B e_C e_D {e'_P}^2 \langle \{b(1'0)\}_\varepsilon | \{d(1'2)c(21')a(10)p(01')p(1'0)\}_\varepsilon \rangle \\
& + e_A e_B e'_C e_D {e'_P}^2 \langle \{b(10)\}_\varepsilon | \{d(1'2)c(21)a(10)p(01')p(1'0)\}_\varepsilon \rangle \\
& + e_A e_B e_C e'_D {e'_P}^2 \langle \{b(10)\}_\varepsilon | \{d(12)c(21')a(10)p(01')p(1'0)\}_\varepsilon \rangle \\
& + e_A e'_B e'_C e_D {e'_P}^2 \langle \{b(1'0)\}_\varepsilon | \{d(1'2)c(21)a(10)p(01')p(1'0)\}_\varepsilon \rangle \\
& + e_A e_B e'_C e'_D {e'_P}^2 \langle \{b(10)\}_\varepsilon | \{d(12)c(21)a(10)p(01')p(1'0)\}_\varepsilon \rangle \\
& + e_A e'_B e_C e'_D {e'_P}^2 \langle \{b(1'0)\}_\varepsilon | \{d(12)c(21')a(10)p(01')p(1'0)\}_\varepsilon \rangle \\
& + e_A e'_B e'_C e'_D {e'_P}^2 \langle \{b(1'0)\}_\varepsilon | \{d(12)c(21)a(10)p(01')p(1'0)\}_\varepsilon \rangle \qquad \text{(B.4)}
\end{aligned}
$$

$$
\begin{aligned}
\langle (B)_\varepsilon | (CDAPP)_\varepsilon \rangle^{++} = {} & e_A e_B e_C e_D e_P{}^2 \langle \{b(10)\}_\varepsilon | \{c(10)d(01)a(10)p(01)p(10)\}_\varepsilon \rangle \\
& + e'_A e_B e_C e_D e_P{}^2 \langle \{b(10)\}_\varepsilon | \{c(10)d(01)a(1'0)p(01)p(10)\}_\varepsilon \rangle \\
& + e'_A e'_B e_C e_D e_P{}^2 \langle \{b(1'0)\}_\varepsilon | \{c(10)d(01)a(1'0)p(01)p(10)\}_\varepsilon \rangle \\
& + e'_A e_B e'_C e_D e_P{}^2 \langle \{b(10)\}_\varepsilon | \{c(1'0)d(01)a(1'0)p(01)p(10)\}_\varepsilon \rangle \\
& + e'_A e_B e_C e'_D e_P{}^2 \langle \{b(10)\}_\varepsilon | \{c(10)d(01')a(1'0)p(01)p(10)\}_\varepsilon \rangle \\
& + e'_A e'_B e'_C e_D e_P{}^2 \langle \{b(1'0)\}_\varepsilon | \{c(1'0)d(01)a(1'0)p(01)p(10)\}_\varepsilon \rangle \\
& + e'_A e_B e'_C e'_D e_P{}^2 \langle \{b(10)\}_\varepsilon | \{c(1'0)d(01')a(1'0)p(01)p(10)\}_\varepsilon \rangle \\
& + e'_A e'_B e_C e'_D e_P{}^2 \langle \{b(1'0)\}_\varepsilon | \{c(10)d(01')a(1'0)p(01)p(10)\}_\varepsilon \rangle \\
& + e'_A e'_B e'_C e'_D e_P{}^2 \langle \{b(1'0)\}_\varepsilon | \{c(1'0)d(01')a(1'0)p(01)p(10)\}_\varepsilon \rangle \\
& + e_A e'_B e_C e_D e_P{}^2 \langle \{b(1'0)\}_\varepsilon | \{c(10)d(01)a(10)p(01)p(10)\}_\varepsilon \rangle \\
& + e_A e_B e'_C e_D e_P{}^2 \langle \{b(10)\}_\varepsilon | \{c(1'0)d(01)a(10)p(01)p(10)\}_\varepsilon \rangle \\
& + e_A e_B e_C e'_D e_P{}^2 \langle \{b(10)\}_\varepsilon | \{c(10)d(01')a(10)p(01)p(10)\}_\varepsilon \rangle \\
& + e_A e'_B e'_C e_D e_P{}^2 \langle \{b(1'0)\}_\varepsilon | \{c(1'0)d(01)a(10)p(01)p(10)\}_\varepsilon \rangle \\
& + e_A e_B e'_C e'_D e_P{}^2 \langle \{b(10)\}_\varepsilon | \{c(1'0)d(01')a(10)p(01)p(10)\}_\varepsilon \rangle \\
& + e_A e'_B e_C e'_D e_P{}^2 \langle \{b(1'0)\}_\varepsilon | \{c(10)d(01')a(10)p(01)p(10)\}_\varepsilon \rangle \\
& + e_A e'_B e'_C e'_D e_P{}^2 \langle \{b(1'0)\}_\varepsilon | \{c(1'0)d(01')a(10)p(01)p(10)\}_\varepsilon \rangle \\
& + e_A e_B e_C e_D e'_P{}^2 \langle \{b(10)\}_\varepsilon | \{c(10)d(01)a(10)p(01')p(1'0)\}_\varepsilon \rangle \\
& + e'_A e_B e_C e_D e'_P{}^2 \langle \{b(10)\}_\varepsilon | \{c(10)d(01)a(1'0)p(01')p(1'0)\}_\varepsilon \rangle \\
& + e'_A e'_B e_C e_D e'_P{}^2 \langle \{b(1'0)\}_\varepsilon | \{c(10)d(01)a(1'0)p(01')p(1'0)\}_\varepsilon \rangle \\
& + e'_A e_B e'_C e_D e'_P{}^2 \langle \{b(10)\}_\varepsilon | \{c(1'0)d(01)a(1'0)p(01')p(1'0)\}_\varepsilon \rangle \\
& + e'_A e_B e_C e'_D e'_P{}^2 \langle \{b(10)\}_\varepsilon | \{c(10)d(01')a(1'0)p(01')p(1'0)\}_\varepsilon \rangle \\
& + e'_A e'_B e'_C e_D e'_P{}^2 \langle \{b(1'0)\}_\varepsilon | \{c(1'0)d(01)a(1'0)p(01')p(1'0)\}_\varepsilon \rangle \\
& + e'_A e_B e'_C e'_D e'_P{}^2 \langle \{b(10)\}_\varepsilon | \{c(1'0)d(01')a(1'0)p(01')p(1'0)\}_\varepsilon \rangle \\
& + e'_A e'_B e_C e'_D e'_P{}^2 \langle \{b(1'0)\}_\varepsilon | \{c(10)d(01')a(1'0)p(01')p(1'0)\}_\varepsilon \rangle \\
& + e'_A e'_B e'_C e'_D e'_P{}^2 \langle \{b(1'0)\}_\varepsilon | \{c(1'0)d(01')a(1'0)p(01')p(1'0)\}_\varepsilon \rangle \\
& + e_A e'_B e_C e_D e'_P{}^2 \langle \{b(1'0)\}_\varepsilon | \{c(10)d(01)a(10)p(01')p(1'0)\}_\varepsilon \rangle \\
& + e_A e_B e'_C e_D e'_P{}^2 \langle \{b(10)\}_\varepsilon | \{c(1'0)d(01)a(10)p(01')p(1'0)\}_\varepsilon \rangle \\
& + e_A e_B e_C e'_D e'_P{}^2 \langle \{b(10)\}_\varepsilon | \{c(10)d(01')a(10)p(01')p(1'0)\}_\varepsilon \rangle \\
& + e_A e'_B e'_C e_D e'_P{}^2 \langle \{b(1'0)\}_\varepsilon | \{c(1'0)d(01)a(10)p(01')p(1'0)\}_\varepsilon \rangle \\
& + e_A e_B e'_C e'_D e'_P{}^2 \langle \{b(10)\}_\varepsilon | \{c(1'0)d(01')a(10)p(01')p(1'0)\}_\varepsilon \rangle \\
& + e_A e'_B e_C e'_D e'_P{}^2 \langle \{b(1'0)\}_\varepsilon | \{c(10)d(01')a(10)p(01')p(1'0)\}_\varepsilon \rangle \\
& + e_A e'_B e'_C e'_D e'_P{}^2 \langle \{b(1'0)\}_\varepsilon | \{c(1'0)d(01')a(10)p(01')p(1'0)\}_\varepsilon \rangle \qquad \text{(B.5)}
\end{aligned}
$$

$$
\begin{aligned}
\langle (BPP)_\varepsilon | (DCA)_\varepsilon \rangle^{++} = {} & e_A e_B e_C e_D e_P{}^2 \langle \{b(10)p(01)p(10)\}_\varepsilon | \{d(1'2)c(21')a(10)\}_\varepsilon \rangle \\
& + e'_A e_B e_C e_D e_P{}^2 \langle \{b(10)p(01)p(10)\}_\varepsilon | \{d(1'2)c(21')a(1'0)\}_\varepsilon \rangle \\
& + e'_A e'_B e_C e_D e_P{}^2 \langle \{b(1'0)p(01)p(10)\}_\varepsilon | \{d(1'2)c(21')a(1'0)\}_\varepsilon \rangle \\
& + e'_A e_B e'_C e_D e_P{}^2 \langle \{b(10)p(01)p(10)\}_\varepsilon | \{d(1'2)c(21)a(1'0)\}_\varepsilon \rangle \\
& + e'_A e_B e_C e'_D e_P{}^2 \langle \{b(10)p(01)p(10)\}_\varepsilon | \{d(12)c(21')a(1'0)\}_\varepsilon \rangle \\
& + e'_A e'_B e'_C e_D e_P{}^2 \langle \{b(1'0)p(01)p(10)\}_\varepsilon | \{d(1'2)c(21)a(1'0)\}_\varepsilon \rangle \\
& + e'_A e_B e'_C e'_D e_P{}^2 \langle \{b(10)p(01)p(10)\}_\varepsilon | \{d(12)c(21)a(1'0)\}_\varepsilon \rangle \\
& + e'_A e'_B e_C e'_D e_P{}^2 \langle \{b(1'0)p(01)p(10)\}_\varepsilon | \{d(12)c(21')a(1'0)\}_\varepsilon \rangle \\
& + e'_A e'_B e'_C e'_D e_P{}^2 \langle \{b(1'0)p(01)p(10)\}_\varepsilon | \{d(12)c(21)a(1'0)\}_\varepsilon \rangle \\
& + e_A e'_B e_C e_D e_P{}^2 \langle \{b(1'0)p(01)p(10)\}_\varepsilon | \{d(1'2)c(21')a(10)\}_\varepsilon \rangle \\
& + e_A e_B e'_C e_D e_P{}^2 \langle \{b(10)p(01)p(10)\}_\varepsilon | \{d(1'2)c(21)a(10)\}_\varepsilon \rangle \\
& + e_A e_B e_C e'_D e_P{}^2 \langle \{b(10)p(01)p(10)\}_\varepsilon | \{d(12)c(21')a(10)\}_\varepsilon \rangle \\
& + e_A e'_B e'_C e_D e_P{}^2 \langle \{b(1'0)p(01)p(10)\}_\varepsilon | \{d(1'2)c(21)a(10)\}_\varepsilon \rangle \\
& + e_A e_B e'_C e'_D e_P{}^2 \langle \{b(10)p(01)p(10)\}_\varepsilon | \{d(12)c(21)a(10)\}_\varepsilon \rangle \\
& + e_A e'_B e_C e'_D e_P{}^2 \langle \{b(1'0)p(01)p(10)\}_\varepsilon | \{d(12)c(21')a(10)\}_\varepsilon \rangle \\
& + e_A e'_B e'_C e'_D e_P{}^2 \langle \{b(1'0)p(01)p(10)\}_\varepsilon | \{d(12)c(21)a(10)\}_\varepsilon \rangle \\
& + e_A e_B e_C e_D e'_P{}^2 \langle \{b(10)p(01')p(1'0)\}_\varepsilon | \{d(1'2)c(21')a(10)\}_\varepsilon \rangle \\
& + e'_A e_B e_C e_D e'_P{}^2 \langle \{b(10)p(01')p(1'0)\}_\varepsilon | \{d(1'2)c(21')a(1'0)\}_\varepsilon \rangle \\
& + e'_A e'_B e_C e_D e'_P{}^2 \langle \{b(1'0)p(01')p(1'0)\}_\varepsilon | \{d(1'2)c(21')a(1'0)\}_\varepsilon \rangle \\
& + e'_A e_B e'_C e_D e'_P{}^2 \langle \{b(10)p(01')p(1'0)\}_\varepsilon | \{d(1'2)c(21)a(1'0)\}_\varepsilon \rangle \\
& + e'_A e_B e_C e'_D e'_P{}^2 \langle \{b(10)p(01')p(1'0)\}_\varepsilon | \{d(12)c(21')a(1'0)\}_\varepsilon \rangle \\
& + e'_A e'_B e'_C e_D e'_P{}^2 \langle \{b(1'0)p(01')p(1'0)\}_\varepsilon | \{d(1'2)c(21)a(1'0)\}_\varepsilon \rangle \\
& + e'_A e_B e'_C e'_D e'_P{}^2 \langle \{b(10)p(01')p(1'0)\}_\varepsilon | \{d(12)c(21)a(1'0)\}_\varepsilon \rangle \\
& + e'_A e'_B e_C e'_D e'_P{}^2 \langle \{b(1'0)p(01')p(1'0)\}_\varepsilon | \{d(12)c(21')a(1'0)\}_\varepsilon \rangle \\
& + e'_A e'_B e'_C e'_D e'_P{}^2 \langle \{b(1'0)p(01')p(1'0)\}_\varepsilon | \{d(12)c(21)a(1'0)\}_\varepsilon \rangle \\
& + e_A e'_B e_C e_D e'_P{}^2 \langle \{b(1'0)p(01')p(1'0)\}_\varepsilon | \{d(1'2)c(21')a(10)\}_\varepsilon \rangle \\
& + e_A e_B e'_C e_D e'_P{}^2 \langle \{b(10)p(01')p(1'0)\}_\varepsilon | \{d(1'2)c(21)a(10)\}_\varepsilon \rangle \\
& + e_A e_B e_C e'_D e'_P{}^2 \langle \{b(10)p(01')p(1'0)\}_\varepsilon | \{d(12)c(21')a(10)\}_\varepsilon \rangle \\
& + e_A e'_B e'_C e_D e'_P{}^2 \langle \{b(1'0)p(01')p(1'0)\}_\varepsilon | \{d(1'2)c(21)a(10)\}_\varepsilon \rangle \\
& + e_A e_B e'_C e'_D e'_P{}^2 \langle \{b(10)p(01')p(1'0)\}_\varepsilon | \{d(12)c(21)a(10)\}_\varepsilon \rangle \\
& + e_A e'_B e_C e'_D e'_P{}^2 \langle \{b(1'0)p(01')p(1'0)\}_\varepsilon | \{d(12)c(21')a(10)\}_\varepsilon \rangle \\
& + e_A e'_B e'_C e'_D e'_P{}^2 \langle \{b(1'0)p(01')p(1'0)\}_\varepsilon | \{d(12)c(21)a(10)\}_\varepsilon \rangle \qquad \text{(B.6)}
\end{aligned}
$$

$$
\begin{aligned}
\langle (BPP)_\varepsilon | (CDA)_\varepsilon \rangle^{++} = {} & e_A e_B e_C e_D {e_P}^2 \langle \{b(10)p(01)p(10)\}_\varepsilon | \{c(10)d(01)a(10)\}_\varepsilon \rangle \\
& + e'_A e_B e_C e_D {e_P}^2 \langle \{b(10)p(01)p(10)\}_\varepsilon | \{c(10)d(01)a(1'0)\}_\varepsilon \rangle \\
& + e'_A e'_B e_C e_D {e_P}^2 \langle \{b(1'0)p(01)p(10)\}_\varepsilon | \{c(10)d(01)a(1'0)\}_\varepsilon \rangle \\
& + e'_A e_B e'_C e_D {e_P}^2 \langle \{b(10)p(01)p(10)\}_\varepsilon | \{c(1'0)d(01)a(1'0)\}_\varepsilon \rangle \\
& + e'_A e_B e_C e'_D {e_P}^2 \langle \{b(10)p(01)p(10)\}_\varepsilon | \{c(10)d(01')a(1'0)\}_\varepsilon \rangle \\
& + e'_A e'_B e'_C e_D {e_P}^2 \langle \{b(1'0)p(01)p(10)\}_\varepsilon | \{c(1'0)d(01)a(1'0)\}_\varepsilon \rangle \\
& + e'_A e_B e'_C e'_D {e_P}^2 \langle \{b(10)p(01)p(10)\}_\varepsilon | \{c(1'0)d(01')a(1'0)\}_\varepsilon \rangle \\
& + e'_A e'_B e_C e'_D {e_P}^2 \langle \{b(1'0)p(01)p(10)\}_\varepsilon | \{c(10)d(01')a(1'0)\}_\varepsilon \rangle \\
& + e'_A e'_B e'_C e'_D {e_P}^2 \langle \{b(1'0)p(01)p(10)\}_\varepsilon | \{c(1'0)d(01')a(1'0)\}_\varepsilon \rangle \\
& + e_A e'_B e_C e_D {e_P}^2 \langle \{b(1'0)p(01)p(10)\}_\varepsilon | \{c(10)d(01)a(10)\}_\varepsilon \rangle \\
& + e_A e_B e'_C e_D {e_P}^2 \langle \{b(10)p(01)p(10)\}_\varepsilon | \{c(1'0)d(01)a(10)\}_\varepsilon \rangle \\
& + e_A e_B e_C e'_D {e_P}^2 \langle \{b(10)p(01)p(10)\}_\varepsilon | \{c(10)d(01')a(10)\}_\varepsilon \rangle \\
& + e_A e'_B e'_C e_D {e_P}^2 \langle \{b(1'0)p(01)p(10)\}_\varepsilon | \{c(1'0)d(01)a(10)\}_\varepsilon \rangle \\
& + e_A e_B e'_C e'_D {e_P}^2 \langle \{b(10)p(01)p(10)\}_\varepsilon | \{c(1'0)d(01')a(10)\}_\varepsilon \rangle \\
& + e_A e'_B e_C e'_D {e_P}^2 \langle \{b(1'0)p(01)p(10)\}_\varepsilon | \{c(10)d(01')a(10)\}_\varepsilon \rangle \\
& + e_A e'_B e'_C e'_D {e_P}^2 \langle \{b(1'0)p(01)p(10)\}_\varepsilon | \{c(1'0)d(01')a(10)\}_\varepsilon \rangle \\
& + e_A e_B e_C e_D {e'_P}^2 \langle \{b(10)p(01')p(1'0)\}_\varepsilon | \{c(10)d(01)a(10)\}_\varepsilon \rangle \\
& + e'_A e_B e_C e_D {e'_P}^2 \langle \{b(10)p(01')p(1'0)\}_\varepsilon | \{c(10)d(01)a(1'0)\}_\varepsilon \rangle \\
& + e'_A e'_B e_C e_D {e'_P}^2 \langle \{b(1'0)p(01')p(1'0)\}_\varepsilon | \{c(10)d(01)a(1'0)\}_\varepsilon \rangle \\
& + e'_A e_B e'_C e_D {e'_P}^2 \langle \{b(10)p(01')p(1'0)\}_\varepsilon | \{c(1'0)d(01)a(1'0)\}_\varepsilon \rangle \\
& + e'_A e_B e_C e'_D {e'_P}^2 \langle \{b(10)p(01')p(1'0)\}_\varepsilon | \{c(10)d(01')a(1'0)\}_\varepsilon \rangle \\
& + e'_A e'_B e'_C e_D {e'_P}^2 \langle \{b(1'0)p(01')p(1'0)\}_\varepsilon | \{c(1'0)d(01)a(1'0)\}_\varepsilon \rangle \\
& + e'_A e_B e'_C e'_D {e'_P}^2 \langle \{b(10)p(01')p(1'0)\}_\varepsilon | \{c(1'0)d(01')a(1'0)\}_\varepsilon \rangle \\
& + e'_A e'_B e_C e'_D {e'_P}^2 \langle \{b(1'0)p(01')p(1'0)\}_\varepsilon | \{c(10)d(01')a(1'0)\}_\varepsilon \rangle \\
& + e'_A e'_B e'_C e'_D {e'_P}^2 \langle \{b(1'0)p(01')p(1'0)\}_\varepsilon | \{c(1'0)d(01')a(1'0)\}_\varepsilon \rangle \\
& + e_A e'_B e_C e_D {e'_P}^2 \langle \{b(1'0)p(01')p(1'0)\}_\varepsilon | \{c(10)d(01)a(10)\}_\varepsilon \rangle \\
& + e_A e_B e'_C e_D {e'_P}^2 \langle \{b(10)p(01')p(1'0)\}_\varepsilon | \{c(1'0)d(01)a(10)\}_\varepsilon \rangle \\
& + e_A e_B e_C e'_D {e'_P}^2 \langle \{b(10)p(01')p(1'0)\}_\varepsilon | \{c(10)d(01')a(10)\}_\varepsilon \rangle \\
& + e_A e'_B e'_C e_D {e'_P}^2 \langle \{b(1'0)p(01')p(1'0)\}_\varepsilon | \{c(1'0)d(01)a(10)\}_\varepsilon \rangle \\
& + e_A e_B e'_C e'_D {e'_P}^2 \langle \{b(10)p(01')p(1'0)\}_\varepsilon | \{c(1'0)d(01')a(10)\}_\varepsilon \rangle \\
& + e_A e'_B e_C e'_D {e'_P}^2 \langle \{b(1'0)p(01')p(1'0)\}_\varepsilon | \{c(10)d(01')a(10)\}_\varepsilon \rangle \\
& + e_A e'_B e'_C e'_D {e'_P}^2 \langle \{b(1'0)p(01')p(1'0)\}_\varepsilon | \{c(1'0)d(01')a(10)\}_\varepsilon \rangle
\end{aligned}
\tag{B.7}
$$

$$
\begin{aligned}
\langle (DCBPP)_\varepsilon | (A)_\varepsilon \rangle^{++} = {} & e_A e_B e_C e_D {e_P}^2 \langle \{d(10)c(01)b(10)p(01)p(10)\}_\varepsilon | \{a(10)\}_\varepsilon \rangle \\
& + e'_A e_B e_C e_D {e_P}^2 \langle \{d(10)c(01)b(10)p(01)p(10)\}_\varepsilon | \{a(1'0)\}_\varepsilon \rangle \\
& + e'_A e'_B e_C e_D {e_P}^2 \langle \{d(10)c(01)b(1'0)p(01)p(10)\}_\varepsilon | \{a(1'0)\}_\varepsilon \rangle \\
& + e'_A e_B e'_C e_D {e_P}^2 \langle \{d(10)c(01')b(10)p(01)p(10)\}_\varepsilon | \{a(1'0)\}_\varepsilon \rangle \\
& + e'_A e_B e_C e'_D {e_P}^2 \langle \{d(1'0)c(01)b(10)p(01)p(10)\}_\varepsilon | \{a(1'0)\}_\varepsilon \rangle \\
& + e'_A e'_B e'_C e_D {e_P}^2 \langle \{d(10)c(01')b(1'0)p(01)p(10)\}_\varepsilon | \{a(1'0)\}_\varepsilon \rangle \\
& + e'_A e_B e'_C e'_D {e_P}^2 \langle \{d(1'0)c(01')b(10)p(01)p(10)\}_\varepsilon | \{a(1'0)\}_\varepsilon \rangle \\
& + e'_A e'_B e_C e'_D {e_P}^2 \langle \{d(1'0)c(01)b(1'0)p(01)p(10)\}_\varepsilon | \{a(1'0)\}_\varepsilon \rangle \\
& + e'_A e'_B e'_C e'_D {e_P}^2 \langle \{d(1'0)c(01')b(1'0)p(01)p(10)\}_\varepsilon | \{a(1'0)\}_\varepsilon \rangle \\
& + e_A e'_B e_C e_D {e_P}^2 \langle \{d(10)c(01)b(1'0)p(01)p(10)\}_\varepsilon | \{a(10)\}_\varepsilon \rangle \\
& + e_A e_B e'_C e_D {e_P}^2 \langle \{d(10)c(01')b(10)p(01)p(10)\}_\varepsilon | \{a(10)\}_\varepsilon \rangle \\
& + e_A e_B e_C e'_D {e_P}^2 \langle \{d(1'0)c(01)b(10)p(01)p(10)\}_\varepsilon | \{a(10)\}_\varepsilon \rangle \\
& + e_A e'_B e'_C e_D {e_P}^2 \langle \{d(10)c(01')b(1'0)p(01)p(10)\}_\varepsilon | \{a(10)\}_\varepsilon \rangle \\
& + e_A e_B e'_C e'_D {e_P}^2 \langle \{d(1'0)c(01')b(10)p(01)p(10)\}_\varepsilon | \{a(10)\}_\varepsilon \rangle \\
& + e_A e'_B e_C e'_D {e_P}^2 \langle \{d(1'0)c(01)b(1'0)p(01)p(10)\}_\varepsilon | \{a(10)\}_\varepsilon \rangle \\
& + e_A e'_B e'_C e'_D {e_P}^2 \langle \{d(1'0)c(01')b(1'0)p(01)p(10)\}_\varepsilon | \{a(10)\}_\varepsilon \rangle \\
& + e_A e_B e_C e_D {e'_P}^2 \langle \{d(10)c(01)b(10)p(01')p(1'0)\}_\varepsilon | \{a(10)\}_\varepsilon \rangle \\
& + e'_A e_B e_C e_D {e'_P}^2 \langle \{d(10)c(01)b(10)p(01')p(1'0)\}_\varepsilon | \{a(1'0)\}_\varepsilon \rangle \\
& + e'_A e'_B e_C e_D {e'_P}^2 \langle \{d(10)c(01)b(1'0)p(01')p(1'0)\}_\varepsilon | \{a(1'0)\}_\varepsilon \rangle \\
& + e'_A e_B e'_C e_D {e'_P}^2 \langle \{d(10)c(01')b(10)p(01')p(1'0)\}_\varepsilon | \{a(1'0)\}_\varepsilon \rangle \\
& + e'_A e_B e_C e'_D {e'_P}^2 \langle \{d(1'0)c(01)b(10)p(01')p(1'0)\}_\varepsilon | \{a(1'0)\}_\varepsilon \rangle \\
& + e'_A e'_B e'_C e_D {e'_P}^2 \langle \{d(10)c(01')b(1'0)p(01')p(1'0)\}_\varepsilon | \{a(1'0)\}_\varepsilon \rangle \\
& + e'_A e_B e'_C e'_D {e'_P}^2 \langle \{d(1'0)c(01')b(10)p(01')p(1'0)\}_\varepsilon | \{a(1'0)\}_\varepsilon \rangle \\
& + e'_A e'_B e_C e'_D {e'_P}^2 \langle \{d(1'0)c(01)b(1'0)p(01')p(1'0)\}_\varepsilon | \{a(1'0)\}_\varepsilon \rangle \\
& + e'_A e'_B e'_C e'_D {e'_P}^2 \langle \{d(1'0)c(01')b(1'0)p(01')p(1'0)\}_\varepsilon | \{a(1'0)\}_\varepsilon \rangle \\
& + e_A e'_B e_C e_D {e'_P}^2 \langle \{d(10)c(01)b(1'0)p(01')p(1'0)\}_\varepsilon | \{a(10)\}_\varepsilon \rangle \\
& + e_A e_B e'_C e_D {e'_P}^2 \langle \{d(10)c(01')b(10)p(01')p(1'0)\}_\varepsilon | \{a(10)\}_\varepsilon \rangle \\
& + e_A e_B e_C e'_D {e'_P}^2 \langle \{d(1'0)c(01)b(10)p(01')p(1'0)\}_\varepsilon | \{a(10)\}_\varepsilon \rangle \\
& + e_A e'_B e'_C e_D {e'_P}^2 \langle \{d(10)c(01')b(1'0)p(01')p(1'0)\}_\varepsilon | \{a(10)\}_\varepsilon \rangle \\
& + e_A e_B e'_C e'_D {e'_P}^2 \langle \{d(1'0)c(01')b(10)p(01')p(1'0)\}_\varepsilon | \{a(10)\}_\varepsilon \rangle \\
& + e_A e'_B e_C e'_D {e'_P}^2 \langle \{d(1'0)c(01)b(1'0)p(01')p(1'0)\}_\varepsilon | \{a(10)\}_\varepsilon \rangle \\
& + e_A e'_B e'_C e'_D {e'_P}^2 \langle \{d(1'0)c(01')b(1'0)p(01')p(1'0)\}_\varepsilon | \{a(10)\}_\varepsilon \rangle \qquad \text{(B.8)}
\end{aligned}
$$

$$
\begin{aligned}
\langle (CDBPP)_\varepsilon|(A)_\varepsilon\rangle^{++} &= e_A e_B e_C e_D {e_P}^2 \langle \{c(1'2)d(21')b(10)p(01)p(10)\}_\varepsilon|\{a(10)\}_\varepsilon\rangle \\
&+ e'_A e_B e_C e_D {e_P}^2 \langle \{c(1'2)d(21')b(10)p(01)p(10)\}_\varepsilon|\{a(1'0)\}_\varepsilon\rangle \\
&+ e'_A e'_B e_C e_D {e_P}^2 \langle \{c(1'2)d(21')b(1'0)p(01)p(10)\}_\varepsilon|\{a(1'0)\}_\varepsilon\rangle \\
&+ e'_A e_B e'_C e_D {e_P}^2 \langle \{c(12)d(21')b(10)p(01)p(10)\}_\varepsilon|\{a(1'0)\}_\varepsilon\rangle \\
&+ e'_A e_B e_C e'_D {e_P}^2 \langle \{c(1'2)d(21)b(10)p(01)p(10)\}_\varepsilon|\{a(1'0)\}_\varepsilon\rangle \\
&+ e'_A e'_B e'_C e_D {e_P}^2 \langle \{c(12)d(21')b(1'0)p(01)p(10)\}_\varepsilon|\{a(1'0)\}_\varepsilon\rangle \\
&+ e'_A e_B e'_C e'_D {e_P}^2 \langle \{c(12)d(21)b(10)p(01)p(10)\}_\varepsilon|\{a(1'0)\}_\varepsilon\rangle \\
&+ e'_A e'_B e_C e'_D {e_P}^2 \langle \{c(1'2)d(21)b(1'0)p(01)p(10)\}_\varepsilon|\{a(1'0)\}_\varepsilon\rangle \\
&+ e'_A e'_B e'_C e'_D {e_P}^2 \langle \{c(12)d(21)b(1'0)p(01)p(10)\}_\varepsilon|\{a(1'0)\}_\varepsilon\rangle \\
&+ e_A e'_B e_C e_D {e_P}^2 \langle \{c(1'2)d(21')b(1'0)p(01)p(10)\}_\varepsilon|\{a(10)\}_\varepsilon\rangle \\
&+ e_A e_B e'_C e_D {e_P}^2 \langle \{c(12)d(21')b(10)p(01)p(10)\}_\varepsilon|\{a(10)\}_\varepsilon\rangle \\
&+ e_A e_B e_C e'_D {e_P}^2 \langle \{c(1'2)d(21)b(10)p(01)p(10)\}_\varepsilon|\{a(10)\}_\varepsilon\rangle \\
&+ e_A e'_B e'_C e_D {e_P}^2 \langle \{c(12)d(21')b(1'0)p(01)p(10)\}_\varepsilon|\{a(10)\}_\varepsilon\rangle \\
&+ e_A e_B e'_C e'_D {e_P}^2 \langle \{c(12)d(21)b(10)p(01)p(10)\}_\varepsilon|\{a(10)\}_\varepsilon\rangle \\
&+ e_A e'_B e_C e'_D {e_P}^2 \langle \{c(1'2)d(21)b(1'0)p(01)p(10)\}_\varepsilon|\{a(10)\}_\varepsilon\rangle \\
&+ e_A e'_B e'_C e'_D {e_P}^2 \langle \{c(12)d(21)b(1'0)p(01)p(10)\}_\varepsilon|\{a(10)\}_\varepsilon\rangle \\
&+ e_A e_B e_C e_D {e'_P}^2 \langle \{c(1'2)d(21')b(10)p(01')p(1'0)\}_\varepsilon|\{a(10)\}_\varepsilon\rangle \\
&+ e'_A e_B e_C e_D {e'_P}^2 \langle \{c(1'2)d(21')b(10)p(01')p(1'0)\}_\varepsilon|\{a(1'0)\}_\varepsilon\rangle \\
&+ e'_A e'_B e_C e_D {e'_P}^2 \langle \{c(1'2)d(21')b(1'0)p(01')p(1'0)\}_\varepsilon|\{a(1'0)\}_\varepsilon\rangle \\
&+ e'_A e_B e'_C e_D {e'_P}^2 \langle \{c(12)d(21')b(10)p(01')p(1'0)\}_\varepsilon|\{a(1'0)\}_\varepsilon\rangle \\
&+ e'_A e_B e_C e'_D {e'_P}^2 \langle \{c(1'2)d(21)b(10)p(01')p(1'0)\}_\varepsilon|\{a(1'0)\}_\varepsilon\rangle \\
&+ e'_A e'_B e'_C e_D {e'_P}^2 \langle \{c(12)d(21')b(1'0)p(01')p(1'0)\}_\varepsilon|\{a(1'0)\}_\varepsilon\rangle \\
&+ e'_A e_B e'_C e'_D {e'_P}^2 \langle \{c(12)d(21)b(10)p(01')p(1'0)\}_\varepsilon|\{a(1'0)\}_\varepsilon\rangle \\
&+ e'_A e'_B e_C e'_D {e'_P}^2 \langle \{c(1'2)d(21)b(1'0)p(01')p(1'0)\}_\varepsilon|\{a(1'0)\}_\varepsilon\rangle \\
&+ e'_A e'_B e'_C e'_D {e'_P}^2 \langle \{c(12)d(21)b(1'0)p(01')p(1'0)\}_\varepsilon|\{a(1'0)\}_\varepsilon\rangle \\
&+ e_A e'_B e_C e_D {e'_P}^2 \langle \{c(1'2)d(21')b(1'0)p(01')p(1'0)\}_\varepsilon|\{a(10)\}_\varepsilon\rangle \\
&+ e_A e_B e'_C e_D {e'_P}^2 \langle \{c(12)d(21')b(10)p(01')p(1'0)\}_\varepsilon|\{a(10)\}_\varepsilon\rangle \\
&+ e_A e_B e_C e'_D {e'_P}^2 \langle \{c(1'2)d(21)b(10)p(01')p(1'0)\}_\varepsilon|\{a(10)\}_\varepsilon\rangle \\
&+ e_A e'_B e'_C e_D {e'_P}^2 \langle \{c(12)d(21')b(1'0)p(01')p(1'0)\}_\varepsilon|\{a(10)\}_\varepsilon\rangle \\
&+ e_A e_B e'_C e'_D {e'_P}^2 \langle \{c(12)d(21)b(10)p(01')p(1'0)\}_\varepsilon|\{a(10)\}_\varepsilon\rangle \\
&+ e_A e'_B e_C e'_D {e'_P}^2 \langle \{c(1'2)d(21)b(1'0)p(01')p(1'0)\}_\varepsilon|\{a(10)\}_\varepsilon\rangle \\
&+ e_A e'_B e'_C e'_D {e'_P}^2 \langle \{c(12)d(21)b(1'0)p(01')p(1'0)\}_\varepsilon|\{a(10)\}_\varepsilon\rangle \qquad \text{(B.9)}
\end{aligned}
$$

$$
\begin{aligned}
\langle (DCB)_\varepsilon | (APP)_\varepsilon \rangle^{++} = {} & e_A e_B e_C e_D {e_P}^2 \langle \{d(10)c(01)b(10)\}_\varepsilon | \{a(10)p(01)p(10)\}_\varepsilon \rangle \\
& + e'_A e_B e_C e_D {e_P}^2 \langle \{d(10)c(01)b(10)\}_\varepsilon | \{a(1'0)p(01)p(10)\}_\varepsilon \rangle \\
& + e'_A e'_B e_C e_D {e_P}^2 \langle \{d(10)c(01)b(1'0)\}_\varepsilon | \{a(1'0)p(01)p(10)\}_\varepsilon \rangle \\
& + e'_A e_B e'_C e_D {e_P}^2 \langle \{d(10)c(01')b(10)\}_\varepsilon | \{a(1'0)p(01)p(10)\}_\varepsilon \rangle \\
& + e'_A e_B e_C e'_D {e_P}^2 \langle \{d(1'0)c(01)b(10)\}_\varepsilon | \{a(1'0)p(01)p(10)\}_\varepsilon \rangle \\
& + e'_A e'_B e'_C e_D {e_P}^2 \langle \{d(10)c(01')b(1'0)\}_\varepsilon | \{a(1'0)p(01)p(10)\}_\varepsilon \rangle \\
& + e'_A e_B e'_C e'_D {e_P}^2 \langle \{d(1'0)c(01')b(10)\}_\varepsilon | \{a(1'0)p(01)p(10)\}_\varepsilon \rangle \\
& + e'_A e'_B e_C e'_D {e_P}^2 \langle \{d(1'0)c(01)b(1'0)\}_\varepsilon | \{a(1'0)p(01)p(10)\}_\varepsilon \rangle \\
& + e'_A e'_B e'_C e'_D {e_P}^2 \langle \{d(1'0)c(01')b(1'0)\}_\varepsilon | \{a(1'0)p(01)p(10)\}_\varepsilon \rangle \\
& + e_A e'_B e_C e_D {e_P}^2 \langle \{d(10)c(01)b(1'0)\}_\varepsilon | \{a(10)p(01)p(10)\}_\varepsilon \rangle \\
& + e_A e_B e'_C e_D {e_P}^2 \langle \{d(10)c(01')b(10)\}_\varepsilon | \{a(10)p(01)p(10)\}_\varepsilon \rangle \\
& + e_A e_B e_C e'_D {e_P}^2 \langle \{d(1'0)c(01)b(10)\}_\varepsilon | \{a(10)p(01)p(10)\}_\varepsilon \rangle \\
& + e_A e'_B e'_C e_D {e_P}^2 \langle \{d(10)c(01')b(1'0)\}_\varepsilon | \{a(10)p(01)p(10)\}_\varepsilon \rangle \\
& + e_A e_B e'_C e'_D {e_P}^2 \langle \{d(1'0)c(01')b(10)\}_\varepsilon | \{a(10)p(01)p(10)\}_\varepsilon \rangle \\
& + e_A e'_B e_C e'_D {e_P}^2 \langle \{d(1'0)c(01)b(1'0)\}_\varepsilon | \{a(10)p(01)p(10)\}_\varepsilon \rangle \\
& + e_A e'_B e'_C e'_D {e_P}^2 \langle \{d(1'0)c(01')b(1'0)\}_\varepsilon | \{a(10)p(01)p(10)\}_\varepsilon \rangle \\
& + e_A e_B e_C e_D {e'_P}^2 \langle \{d(10)c(01)b(10)\}_\varepsilon | \{a(10)p(01')p(1'0)\}_\varepsilon \rangle \\
& + e'_A e_B e_C e_D {e'_P}^2 \langle \{d(10)c(01)b(10)\}_\varepsilon | \{a(1'0)p(01')p(1'0)\}_\varepsilon \rangle \\
& + e'_A e'_B e_C e_D {e'_P}^2 \langle \{d(10)c(01)b(1'0)\}_\varepsilon | \{a(1'0)p(01')p(1'0)\}_\varepsilon \rangle \\
& + e'_A e_B e'_C e_D {e'_P}^2 \langle \{d(10)c(01')b(10)\}_\varepsilon | \{a(1'0)p(01')p(1'0)\}_\varepsilon \rangle \\
& + e'_A e_B e_C e'_D {e'_P}^2 \langle \{d(1'0)c(01)b(10)\}_\varepsilon | \{a(1'0)p(01')p(1'0)\}_\varepsilon \rangle \\
& + e'_A e'_B e'_C e_D {e'_P}^2 \langle \{d(10)c(01')b(1'0)\}_\varepsilon | \{a(1'0)p(01')p(1'0)\}_\varepsilon \rangle \\
& + e'_A e_B e'_C e'_D {e'_P}^2 \langle \{d(1'0)c(01')b(10)\}_\varepsilon | \{a(1'0)p(01')p(1'0)\}_\varepsilon \rangle \\
& + e'_A e'_B e_C e'_D {e'_P}^2 \langle \{d(1'0)c(01)b(1'0)\}_\varepsilon | \{a(1'0)p(01')p(1'0)\}_\varepsilon \rangle \\
& + e'_A e'_B e'_C e'_D {e'_P}^2 \langle \{d(1'0)c(01')b(1'0)\}_\varepsilon | \{a(1'0)p(01')p(1'0)\}_\varepsilon \rangle \\
& + e_A e'_B e_C e_D {e'_P}^2 \langle \{d(10)c(01)b(1'0)\}_\varepsilon | \{a(10)p(01')p(1'0)\}_\varepsilon \rangle \\
& + e_A e_B e'_C e_D {e'_P}^2 \langle \{d(10)c(01')b(10)\}_\varepsilon | \{a(10)p(01')p(1'0)\}_\varepsilon \rangle \\
& + e_A e_B e_C e'_D {e'_P}^2 \langle \{d(1'0)c(01)b(10)\}_\varepsilon | \{a(10)p(01')p(1'0)\}_\varepsilon \rangle \\
& + e_A e'_B e'_C e_D {e'_P}^2 \langle \{d(10)c(01')b(1'0)\}_\varepsilon | \{a(10)p(01')p(1'0)\}_\varepsilon \rangle \\
& + e_A e_B e'_C e'_D {e'_P}^2 \langle \{d(1'0)c(01')b(10)\}_\varepsilon | \{a(10)p(01')p(1'0)\}_\varepsilon \rangle \\
& + e_A e'_B e_C e'_D {e'_P}^2 \langle \{d(1'0)c(01)b(1'0)\}_\varepsilon | \{a(10)p(01')p(1'0)\}_\varepsilon \rangle \\
& + e_A e'_B e'_C e'_D {e'_P}^2 \langle \{d(1'0)c(01')b(1'0)\}_\varepsilon | \{a(10)p(01')p(1'0)\}_\varepsilon \rangle \qquad \text{(B.10)}
\end{aligned}
$$

$$
\begin{aligned}
\langle (CDB)_\varepsilon | (APP)_\varepsilon \rangle^{++} = {} & e_A e_B e_C e_D {e_P}^2 \langle \{c(1'2)d(21')b(10)\}_\varepsilon | \{a(10)p(01)p(10)\}_\varepsilon \rangle \\
& + e'_A e_B e_C e_D {e_P}^2 \langle \{c(1'2)d(21')b(10)\}_\varepsilon | \{a(1'0)p(01)p(10)\}_\varepsilon \rangle \\
& + e'_A e'_B e_C e_D {e_P}^2 \langle \{c(1'2)d(21')b(1'0)\}_\varepsilon | \{a(1'0)p(01)p(10)\}_\varepsilon \rangle \\
& + e'_A e_B e'_C e_D {e_P}^2 \langle \{c(12)d(21')b(10)\}_\varepsilon | \{a(1'0)p(01)p(10)\}_\varepsilon \rangle \\
& + e'_A e_B e_C e'_D {e_P}^2 \langle \{c(1'2)d(21)b(10)\}_\varepsilon | \{a(1'0)p(01)p(10)\}_\varepsilon \rangle \\
& + e'_A e'_B e'_C e_D {e_P}^2 \langle \{c(12)d(21')b(1'0)\}_\varepsilon | \{a(1'0)p(01)p(10)\}_\varepsilon \rangle \\
& + e'_A e_B e'_C e'_D {e_P}^2 \langle \{c(12)d(21)b(10)\}_\varepsilon | \{a(1'0)p(01)p(10)\}_\varepsilon \rangle \\
& + e'_A e'_B e_C e'_D {e_P}^2 \langle \{c(1'2)d(21)b(1'0)\}_\varepsilon | \{a(1'0)p(01)p(10)\}_\varepsilon \rangle \\
& + e'_A e'_B e'_C e'_D {e_P}^2 \langle \{c(12)d(21)b(1'0)\}_\varepsilon | \{a(1'0)p(01)p(10)\}_\varepsilon \rangle \\
& + e_A e'_B e_C e_D {e_P}^2 \langle \{c(1'2)d(21')b(1'0)\}_\varepsilon | \{a(10)p(01)p(10)\}_\varepsilon \rangle \\
& + e_A e_B e'_C e_D {e_P}^2 \langle \{c(12)d(21')b(10)\}_\varepsilon | \{a(10)p(01)p(10)\}_\varepsilon \rangle \\
& + e_A e_B e_C e'_D {e_P}^2 \langle \{c(1'2)d(21)b(10)\}_\varepsilon | \{a(10)p(01)p(10)\}_\varepsilon \rangle \\
& + e_A e'_B e'_C e_D {e_P}^2 \langle \{c(12)d(21')b(1'0)\}_\varepsilon | \{a(10)p(01)p(10)\}_\varepsilon \rangle \\
& + e_A e_B e'_C e'_D {e_P}^2 \langle \{c(12)d(21)b(10)\}_\varepsilon | \{a(10)p(01)p(10)\}_\varepsilon \rangle \\
& + e_A e'_B e_C e'_D {e_P}^2 \langle \{c(1'2)d(21)b(1'0)\}_\varepsilon | \{a(10)p(01)p(10)\}_\varepsilon \rangle \\
& + e_A e'_B e'_C e'_D {e_P}^2 \langle \{c(12)d(21)b(1'0)\}_\varepsilon | \{a(10)p(01)p(10)\}_\varepsilon \rangle \\
& + e_A e_B e_C e_D {e'_P}^2 \langle \{c(1'2)d(21')b(10)\}_\varepsilon | \{a(10)p(01')p(1'0)\}_\varepsilon \rangle \\
& + e'_A e_B e_C e_D {e'_P}^2 \langle \{c(1'2)d(21')b(10)\}_\varepsilon | \{a(1'0)p(01')p(1'0)\}_\varepsilon \rangle \\
& + e'_A e'_B e_C e_D {e'_P}^2 \langle \{c(1'2)d(21')b(1'0)\}_\varepsilon | \{a(1'0)p(01')p(1'0)\}_\varepsilon \rangle \\
& + e'_A e_B e'_C e_D {e'_P}^2 \langle \{c(12)d(21')b(10)\}_\varepsilon | \{a(1'0)p(01')p(1'0)\}_\varepsilon \rangle \\
& + e'_A e_B e_C e'_D {e'_P}^2 \langle \{c(1'2)d(21)b(10)\}_\varepsilon | \{a(1'0)p(01')p(1'0)\}_\varepsilon \rangle \\
& + e'_A e'_B e'_C e_D {e'_P}^2 \langle \{c(12)d(21')b(1'0)\}_\varepsilon | \{a(1'0)p(01')p(1'0)\}_\varepsilon \rangle \\
& + e'_A e_B e'_C e'_D {e'_P}^2 \langle \{c(12)d(21)b(10)\}_\varepsilon | \{a(1'0)p(01')p(1'0)\}_\varepsilon \rangle \\
& + e'_A e'_B e_C e'_D {e'_P}^2 \langle \{c(1'2)d(21)b(1'0)\}_\varepsilon | \{a(1'0)p(01')p(1'0)\}_\varepsilon \rangle \\
& + e'_A e'_B e'_C e'_D {e'_P}^2 \langle \{c(12)d(21)b(1'0)\}_\varepsilon | \{a(1'0)p(01')p(1'0)\}_\varepsilon \rangle \\
& + e_A e'_B e_C e_D {e'_P}^2 \langle \{c(1'2)d(21')b(1'0)\}_\varepsilon | \{a(10)p(01')p(1'0)\}_\varepsilon \rangle \\
& + e_A e_B e'_C e_D {e'_P}^2 \langle \{c(12)d(21')b(10)\}_\varepsilon | \{a(10)p(01')p(1'0)\}_\varepsilon \rangle \\
& + e_A e_B e_C e'_D {e'_P}^2 \langle \{c(1'2)d(21)b(10)\}_\varepsilon | \{a(10)p(01')p(1'0)\}_\varepsilon \rangle \\
& + e_A e'_B e'_C e_D {e'_P}^2 \langle \{c(12)d(21')b(1'0)\}_\varepsilon | \{a(10)p(01')p(1'0)\}_\varepsilon \rangle \\
& + e_A e_B e'_C e'_D {e'_P}^2 \langle \{c(12)d(21)b(10)\}_\varepsilon | \{a(10)p(01')p(1'0)\}_\varepsilon \rangle \\
& + e_A e'_B e_C e'_D {e'_P}^2 \langle \{c(1'2)d(21)b(1'0)\}_\varepsilon | \{a(10)p(01')p(1'0)\}_\varepsilon \rangle \\
& + e_A e'_B e'_C e'_D {e'_P}^2 \langle \{c(12)d(21)b(1'0)\}_\varepsilon | \{a(10)p(01')p(1'0)\}_\varepsilon \rangle
\end{aligned}
\tag{B.11}
$$

$$
\begin{aligned}
\langle (D)_\varepsilon | (CBAPP)_\varepsilon \rangle^{++} = {} & e_A e_B e_C e_D {e_P}^2 \langle \{d(10)\}_\varepsilon | \{c(10)b(01)a(10)p(01)p(10)\}_\varepsilon \rangle \\
& + e'_A e_B e_C e_D {e_P}^2 \langle \{d(10)\}_\varepsilon | \{c(10)b(01)a(1'0)p(01)p(10)\}_\varepsilon \rangle \\
& + e'_A e'_B e_C e_D {e_P}^2 \langle \{d(10)\}_\varepsilon | \{c(10)b(01')a(1'0)p(01)p(10)\}_\varepsilon \rangle \\
& + e'_A e_B e'_C e_D {e_P}^2 \langle \{d(10)\}_\varepsilon | \{c(1'0)b(01)a(1'0)p(01)p(10)\}_\varepsilon \rangle \\
& + e'_A e_B e_C e'_D {e_P}^2 \langle \{d(1'0)\}_\varepsilon | \{c(10)b(01)a(1'0)p(01)p(10)\}_\varepsilon \rangle \\
& + e'_A e'_B e'_C e_D {e_P}^2 \langle \{d(10)\}_\varepsilon | \{c(1'0)b(01')a(1'0)p(01)p(10)\}_\varepsilon \rangle \\
& + e'_A e_B e'_C e'_D {e_P}^2 \langle \{d(1'0)\}_\varepsilon | \{c(1'0)b(01)a(1'0)p(01)p(10)\}_\varepsilon \rangle \\
& + e'_A e'_B e_C e'_D {e_P}^2 \langle \{d(1'0)\}_\varepsilon | \{c(10)b(01')a(1'0)p(01)p(10)\}_\varepsilon \rangle \\
& + e'_A e'_B e'_C e'_D {e_P}^2 \langle \{d(1'0)\}_\varepsilon | \{c(1'0)b(01')a(1'0)p(01)p(10)\}_\varepsilon \rangle \\
& + e_A e'_B e_C e_D {e_P}^2 \langle \{d(10)\}_\varepsilon | \{c(10)b(01')a(10)p(01)p(10)\}_\varepsilon \rangle \\
& + e_A e_B e'_C e_D {e_P}^2 \langle \{d(10)\}_\varepsilon | \{c(1'0)b(01)a(10)p(01)p(10)\}_\varepsilon \rangle \\
& + e_A e_B e_C e'_D {e_P}^2 \langle \{d(1'0)\}_\varepsilon | \{c(10)b(01)a(10)p(01)p(10)\}_\varepsilon \rangle \\
& + e_A e'_B e'_C e_D {e_P}^2 \langle \{d(10)\}_\varepsilon | \{c(1'0)b(01')a(10)p(01)p(10)\}_\varepsilon \rangle \\
& + e_A e_B e'_C e'_D {e_P}^2 \langle \{d(1'0)\}_\varepsilon | \{c(1'0)b(01)a(10)p(01)p(10)\}_\varepsilon \rangle \\
& + e_A e'_B e_C e'_D {e_P}^2 \langle \{d(1'0)\}_\varepsilon | \{c(10)b(01')a(10)p(01)p(10)\}_\varepsilon \rangle \\
& + e_A e'_B e'_C e'_D {e_P}^2 \langle \{d(1'0)\}_\varepsilon | \{c(1'0)b(01')a(10)p(01)p(10)\}_\varepsilon \rangle \\
& + e_A e_B e_C e_D {e'_P}^2 \langle \{d(10)\}_\varepsilon | \{c(10)b(01)a(10)p(01')p(1'0)\}_\varepsilon \rangle \\
& + e'_A e_B e_C e_D {e'_P}^2 \langle \{d(10)\}_\varepsilon | \{c(10)b(01)a(1'0)p(01')p(1'0)\}_\varepsilon \rangle \\
& + e'_A e'_B e_C e_D {e'_P}^2 \langle \{d(10)\}_\varepsilon | \{c(10)b(01')a(1'0)p(01')p(1'0)\}_\varepsilon \rangle \\
& + e'_A e_B e'_C e_D {e'_P}^2 \langle \{d(10)\}_\varepsilon | \{c(1'0)b(01)a(1'0)p(01')p(1'0)\}_\varepsilon \rangle \\
& + e'_A e_B e_C e'_D {e'_P}^2 \langle \{d(1'0)\}_\varepsilon | \{c(10)b(01)a(1'0)p(01')p(1'0)\}_\varepsilon \rangle \\
& + e'_A e'_B e'_C e_D {e'_P}^2 \langle \{d(10)\}_\varepsilon | \{c(1'0)b(01')a(1'0)p(01')p(1'0)\}_\varepsilon \rangle \\
& + e'_A e_B e'_C e'_D {e'_P}^2 \langle \{d(1'0)\}_\varepsilon | \{c(1'0)b(01)a(1'0)p(01')p(1'0)\}_\varepsilon \rangle \\
& + e'_A e'_B e_C e'_D {e'_P}^2 \langle \{d(1'0)\}_\varepsilon | \{c(10)b(01')a(1'0)p(01')p(1'0)\}_\varepsilon \rangle \\
& + e'_A e'_B e'_C e'_D {e'_P}^2 \langle \{d(1'0)\}_\varepsilon | \{c(1'0)b(01')a(1'0)p(01')p(1'0)\}_\varepsilon \rangle \\
& + e_A e'_B e_C e_D {e'_P}^2 \langle \{d(10)\}_\varepsilon | \{c(10)b(01')a(10)p(01')p(1'0)\}_\varepsilon \rangle \\
& + e_A e_B e'_C e_D {e'_P}^2 \langle \{d(10)\}_\varepsilon | \{c(1'0)b(01)a(10)p(01')p(1'0)\}_\varepsilon \rangle \\
& + e_A e_B e_C e'_D {e'_P}^2 \langle \{d(1'0)\}_\varepsilon | \{c(10)b(01)a(10)p(01')p(1'0)\}_\varepsilon \rangle \\
& + e_A e'_B e'_C e_D {e'_P}^2 \langle \{d(10)\}_\varepsilon | \{c(1'0)b(01')a(10)p(01')p(1'0)\}_\varepsilon \rangle \\
& + e_A e_B e'_C e'_D {e'_P}^2 \langle \{d(1'0)\}_\varepsilon | \{c(1'0)b(01)a(10)p(01')p(1'0)\}_\varepsilon \rangle \\
& + e_A e'_B e_C e'_D {e'_P}^2 \langle \{d(1'0)\}_\varepsilon | \{c(10)b(01')a(10)p(01')p(1'0)\}_\varepsilon \rangle \\
& + e_A e'_B e'_C e'_D {e'_P}^2 \langle \{d(1'0)\}_\varepsilon | \{c(1'0)b(01')a(10)p(01')p(1'0)\}_\varepsilon \rangle
\end{aligned}
\tag{B.12}
$$

$$
\begin{aligned}
\langle (DPP)_\varepsilon | (CBA)_\varepsilon \rangle^{++} = {} & e_A e_B e_C e_D {e_P}^2 \langle \{d(10)p(01)p(10)\}_\varepsilon | \{c(10)b(01)a(10)\}_\varepsilon \rangle \\
& + e'_A e_B e_C e_D {e_P}^2 \langle \{d(10)p(01)p(10)\}_\varepsilon | \{c(10)b(01)a(1'0)\}_\varepsilon \rangle \\
& + e'_A e'_B e_C e_D {e_P}^2 \langle \{d(10)p(01)p(10)\}_\varepsilon | \{c(10)b(01')a(1'0)\}_\varepsilon \rangle \\
& + e'_A e_B e'_C e_D {e_P}^2 \langle \{d(10)p(01)p(10)\}_\varepsilon | \{c(1'0)b(01)a(1'0)\}_\varepsilon \rangle \\
& + e'_A e_B e_C e'_D {e_P}^2 \langle \{d(1'0)p(01)p(10)\}_\varepsilon | \{c(10)b(01)a(1'0)\}_\varepsilon \rangle \\
& + e'_A e'_B e'_C e_D {e_P}^2 \langle \{d(10)p(01)p(10)\}_\varepsilon | \{c(1'0)b(01')a(1'0)\}_\varepsilon \rangle \\
& + e'_A e_B e'_C e'_D {e_P}^2 \langle \{d(1'0)p(01)p(10)\}_\varepsilon | \{c(1'0)b(01)a(1'0)\}_\varepsilon \rangle \\
& + e'_A e'_B e_C e'_D {e_P}^2 \langle \{d(1'0)p(01)p(10)\}_\varepsilon | \{c(10)b(01')a(1'0)\}_\varepsilon \rangle \\
& + e'_A e'_B e'_C e'_D {e_P}^2 \langle \{d(1'0)p(01)p(10)\}_\varepsilon | \{c(1'0)b(01')a(1'0)\}_\varepsilon \rangle \\
& + e_A e'_B e_C e_D {e_P}^2 \langle \{d(10)p(01)p(10)\}_\varepsilon | \{c(10)b(01')a(10)\}_\varepsilon \rangle \\
& + e_A e_B e'_C e_D {e_P}^2 \langle \{d(10)p(01)p(10)\}_\varepsilon | \{c(1'0)b(01)a(10)\}_\varepsilon \rangle \\
& + e_A e_B e_C e'_D {e_P}^2 \langle \{d(1'0)p(01)p(10)\}_\varepsilon | \{c(10)b(01)a(10)\}_\varepsilon \rangle \\
& + e_A e'_B e'_C e_D {e_P}^2 \langle \{d(10)p(01)p(10)\}_\varepsilon | \{c(1'0)b(01')a(10)\}_\varepsilon \rangle \\
& + e_A e_B e'_C e'_D {e_P}^2 \langle \{d(1'0)p(01)p(10)\}_\varepsilon | \{c(1'0)b(01)a(10)\}_\varepsilon \rangle \\
& + e_A e'_B e_C e'_D {e_P}^2 \langle \{d(1'0)p(01)p(10)\}_\varepsilon | \{c(10)b(01')a(10)\}_\varepsilon \rangle \\
& + e_A e'_B e'_C e'_D {e_P}^2 \langle \{d(1'0)p(01)p(10)\}_\varepsilon | \{c(1'0)b(01')a(10)\}_\varepsilon \rangle \\
& + e_A e_B e_C e_D {e'_P}^2 \langle \{d(10)p(01')p(1'0)\}_\varepsilon | \{c(10)b(01)a(10)\}_\varepsilon \rangle \\
& + e'_A e_B e_C e_D {e'_P}^2 \langle \{d(10)p(01')p(1'0)\}_\varepsilon | \{c(10)b(01)a(1'0)\}_\varepsilon \rangle \\
& + e'_A e'_B e_C e_D {e'_P}^2 \langle \{d(10)p(01')p(1'0)\}_\varepsilon | \{c(10)b(01')a(1'0)\}_\varepsilon \rangle \\
& + e'_A e_B e'_C e_D {e'_P}^2 \langle \{d(10)p(01')p(1'0)\}_\varepsilon | \{c(1'0)b(01)a(1'0)\}_\varepsilon \rangle \\
& + e'_A e_B e_C e'_D {e'_P}^2 \langle \{d(1'0)p(01')p(1'0)\}_\varepsilon | \{c(10)b(01)a(1'0)\}_\varepsilon \rangle \\
& + e'_A e'_B e'_C e_D {e'_P}^2 \langle \{d(10)p(01')p(1'0)\}_\varepsilon | \{c(1'0)b(01')a(1'0)\}_\varepsilon \rangle \\
& + e'_A e_B e'_C e'_D {e'_P}^2 \langle \{d(1'0)p(01')p(1'0)\}_\varepsilon | \{c(1'0)b(01)a(1'0)\}_\varepsilon \rangle \\
& + e'_A e'_B e_C e'_D {e'_P}^2 \langle \{d(1'0)p(01')p(1'0)\}_\varepsilon | \{c(10)b(01')a(1'0)\}_\varepsilon \rangle \\
& + e'_A e'_B e'_C e'_D {e'_P}^2 \langle \{d(1'0)p(01')p(1'0)\}_\varepsilon | \{c(1'0)b(01')a(1'0)\}_\varepsilon \rangle \\
& + e_A e'_B e_C e_D {e'_P}^2 \langle \{d(10)p(01')p(1'0)\}_\varepsilon | \{c(10)b(01')a(10)\}_\varepsilon \rangle \\
& + e_A e_B e'_C e_D {e'_P}^2 \langle \{d(10)p(01')p(1'0)\}_\varepsilon | \{c(1'0)b(01)a(10)\}_\varepsilon \rangle \\
& + e_A e_B e_C e'_D {e'_P}^2 \langle \{d(1'0)p(01')p(1'0)\}_\varepsilon | \{c(10)b(01)a(10)\}_\varepsilon \rangle \\
& + e_A e'_B e'_C e_D {e'_P}^2 \langle \{d(10)p(01')p(1'0)\}_\varepsilon | \{c(1'0)b(01')a(10)\}_\varepsilon \rangle \\
& + e_A e_B e'_C e'_D {e'_P}^2 \langle \{d(1'0)p(01')p(1'0)\}_\varepsilon | \{c(1'0)b(01)a(10)\}_\varepsilon \rangle \\
& + e_A e'_B e_C e'_D {e'_P}^2 \langle \{d(1'0)p(01')p(1'0)\}_\varepsilon | \{c(10)b(01')a(10)\}_\varepsilon \rangle \\
& + e_A e'_B e'_C e'_D {e'_P}^2 \langle \{d(1'0)p(01')p(1'0)\}_\varepsilon | \{c(1'0)b(01')a(10)\}_\varepsilon \rangle
\end{aligned} \tag{B.13}
$$

$$\begin{aligned}
\langle (DABPP)_\varepsilon|(C)_\varepsilon\rangle^{++} &= e_A e_B e_C e_D {e_P}^2 \langle\{d(10)a(01)b(10)p(01)p(10)\}_\varepsilon|\{c(10)\}_\varepsilon\rangle \\
&+ e'_A e'_B e_C e_D {e_P}^2 \langle\{d(10)a(01')b(1'0)p(01)p(10)\}_\varepsilon|\{c(10)\}_\varepsilon\rangle \\
&+ e'_A e'_B e'_C e_D {e_P}^2 \langle\{d(10)a(01')b(1'0)p(01)p(10)\}_\varepsilon|\{c(1'0)\}_\varepsilon\rangle \\
&+ e'_A e'_B e_C e'_D {e_P}^2 \langle\{d(1'0)a(01')b(1'0)p(01)p(10)\}_\varepsilon|\{c(10)\}_\varepsilon\rangle \\
&+ e'_A e'_B e'_C e'_D {e_P}^2 \langle\{d(1'0)a(01')b(1'0)p(01)p(10)\}_\varepsilon|\{c(1'0)\}_\varepsilon\rangle \\
&+ e_A e_B e'_C e_D {e_P}^2 \langle\{d(10)a(01)b(10)p(01)p(10)\}_\varepsilon|\{c(1'0)\}_\varepsilon\rangle \\
&+ e_A e_B e_C e'_D {e_P}^2 \langle\{d(1'0)a(01)b(10)p(01)p(10)\}_\varepsilon|\{c(10)\}_\varepsilon\rangle \\
&+ e_A e_B e'_C e'_D {e_P}^2 \langle\{d(1'0)a(01)b(10)p(01)p(10)\}_\varepsilon|\{c(1'0)\}_\varepsilon\rangle \\
&+ e_A e_B e_C e_D {e'_P}^2 \langle\{d(10)a(01)b(10)p(01')p(1'0)\}_\varepsilon|\{c(10)\}_\varepsilon\rangle \\
&+ e'_A e'_B e_C e_D {e'_P}^2 \langle\{d(10)a(01')b(1'0)p(01')p(1'0)\}_\varepsilon|\{c(10)\}_\varepsilon\rangle \\
&+ e'_A e'_B e'_C e_D {e'_P}^2 \langle\{d(10)a(01')b(1'0)p(01')p(1'0)\}_\varepsilon|\{c(1'0)\}_\varepsilon\rangle \\
&+ e'_A e'_B e_C e'_D {e'_P}^2 \langle\{d(1'0)a(01')b(1'0)p(01')p(1'0)\}_\varepsilon|\{c(10)\}_\varepsilon\rangle \\
&+ e'_A e'_B e'_C e'_D {e'_P}^2 \langle\{d(1'0)a(01')b(1'0)p(01')p(1'0)\}_\varepsilon|\{c(1'0)\}_\varepsilon\rangle \\
&+ e_A e_B e'_C e_D {e'_P}^2 \langle\{d(10)a(01)b(10)p(01')p(1'0)\}_\varepsilon|\{c(1'0)\}_\varepsilon\rangle \\
&+ e_A e_B e_C e'_D {e'_P}^2 \langle\{d(1'0)a(01)b(10)p(01')p(1'0)\}_\varepsilon|\{c(10)\}_\varepsilon\rangle \\
&+ e_A e_B e'_C e'_D {e'_P}^2 \langle\{d(1'0)a(01)b(10)p(01')p(1'0)\}_\varepsilon|\{c(1'0)\}_\varepsilon\rangle
\end{aligned} \tag{B.14}$$

$$\begin{aligned}
\langle (DAB)_\varepsilon|(CPP)_\varepsilon\rangle^{++} &= e_A e_B e_C e_D {e_P}^2 \langle\{d(10)a(01)b(10)\}_\varepsilon|\{c(10)p(01)p(10)\}_\varepsilon\rangle \\
&+ e'_A e'_B e_C e_D {e_P}^2 \langle\{d(10)a(01')b(1'0)\}_\varepsilon|\{c(10)p(01)p(10)\}_\varepsilon\rangle \\
&+ e'_A e_B e'_C e_D {e_P}^2 \langle\{d(10)a(01')b(10)\}_\varepsilon|\{c(1'0)p(01)p(10)\}_\varepsilon\rangle \\
&+ e'_A e'_B e'_C e_D {e_P}^2 \langle\{d(10)a(01')b(1'0)\}_\varepsilon|\{c(1'0)p(01)p(10)\}_\varepsilon\rangle \\
&+ e'_A e'_B e_C e'_D {e_P}^2 \langle\{d(1'0)a(01')b(1'0)\}_\varepsilon|\{c(10)p(01)p(10)\}_\varepsilon\rangle \\
&+ e'_A e'_B e'_C e'_D {e_P}^2 \langle\{d(1'0)a(01')b(1'0)\}_\varepsilon|\{c(1'0)p(01)p(10)\}_\varepsilon\rangle \\
&+ e_A e_B e'_C e_D {e_P}^2 \langle\{d(10)a(01)b(10)\}_\varepsilon|\{c(1'0)p(01)p(10)\}_\varepsilon\rangle \\
&+ e_A e_B e_C e'_D {e_P}^2 \langle\{d(1'0)a(01)b(10)\}_\varepsilon|\{c(10)p(01)p(10)\}_\varepsilon\rangle \\
&+ e_A e_B e'_C e'_D {e_P}^2 \langle\{d(1'0)a(01)b(10)\}_\varepsilon|\{c(1'0)p(01)p(10)\}_\varepsilon\rangle \\
&+ e_A e_B e_C e_D {e'_P}^2 \langle\{d(10)a(01)b(10)\}_\varepsilon|\{c(10)p(01')p(1'0)\}_\varepsilon\rangle \\
&+ e'_A e'_B e_C e_D {e'_P}^2 \langle\{d(10)a(01')b(1'0)\}_\varepsilon|\{c(10)p(01')p(1'0)\}_\varepsilon\rangle \\
&+ e'_A e'_B e'_C e_D {e'_P}^2 \langle\{d(10)a(01')b(1'0)\}_\varepsilon|\{c(1'0)p(01')p(1'0)\}_\varepsilon\rangle \\
&+ e'_A e'_B e_C e'_D {e'_P}^2 \langle\{d(1'0)a(01')b(1'0)\}_\varepsilon|\{c(10)p(01')p(1'0)\}_\varepsilon\rangle \\
&+ e'_A e'_B e'_C e'_D {e'_P}^2 \langle\{d(1'0)a(01')b(1'0)\}_\varepsilon|\{c(1'0)p(01')p(1'0)\}_\varepsilon\rangle \\
&+ e_A e_B e'_C e_D {e'_P}^2 \langle\{d(10)a(01)b(10)\}_\varepsilon|\{c(1'0)p(01')p(1'0)\}_\varepsilon\rangle \\
&+ e_A e_B e_C e'_D {e'_P}^2 \langle\{d(1'0)a(01)b(10)\}_\varepsilon|\{c(10)p(01')p(1'0)\}_\varepsilon\rangle \\
&+ e_A e_B e'_C e'_D {e'_P}^2 \langle\{d(1'0)a(01)b(10)\}_\varepsilon|\{c(1'0)p(01')p(1'0)\}_\varepsilon\rangle
\end{aligned} \tag{B.15}$$

$$
\begin{aligned}
\langle (B)_\varepsilon|(DCAPP)_\varepsilon\rangle^{+-} &= e_Ae_Be_Ce_De_P{}^2\langle\{b(10)\}_\varepsilon|\{d(10)c(01)a(10)p(01)p(10)\}_\varepsilon\rangle\\
&+ e'_Ae_Be_Ce_De_P{}^2\langle\{b(10)\}_\varepsilon|\{d(10)c(01)a(1'0)p(01)p(10)\}_\varepsilon\rangle\\
&+ e'_Ae'_Be_Ce_De_P{}^2\langle\{b(1'0)\}_\varepsilon|\{d(10)c(01)a(1'0)p(01)p(10)\}_\varepsilon\rangle\\
&+ e'_Ae_Be'_Ce_De_P{}^2\langle\{b(10)\}_\varepsilon|\{d(10)c(01')a(1'0)p(01)p(10)\}_\varepsilon\rangle\\
&+ e'_Ae_Be_Ce'_De_P{}^2\langle\{b(10)\}_\varepsilon|\{d(1'0)c(01)a(1'0)p(01)p(10)\}_\varepsilon\rangle\\
&+ e'_Ae'_Be'_Ce_De_P{}^2\langle\{b(1'0)\}_\varepsilon|\{d(10)c(01')a(1'0)p(01)p(10)\}_\varepsilon\rangle\\
&+ e'_Ae_Be'_Ce'_De_P{}^2\langle\{b(10)\}_\varepsilon|\{d(1'0)c(01')a(1'0)p(01)p(10)\}_\varepsilon\rangle\\
&+ e'_Ae'_Be_Ce'_De_P{}^2\langle\{b(1'0)\}_\varepsilon|\{d(1'0)c(01)a(1'0)p(01)p(10)\}_\varepsilon\rangle\\
&+ e'_Ae'_Be'_Ce'_De_P{}^2\langle\{b(1'0)\}_\varepsilon|\{d(1'0)c(01')a(1'0)p(01)p(10)\}_\varepsilon\rangle\\
&+ e_Ae'_Be_Ce_De_P{}^2\langle\{b(1'0)\}_\varepsilon|\{d(10)c(01)a(10)p(01)p(10)\}_\varepsilon\rangle\\
&+ e_Ae_Be'_Ce_De_P{}^2\langle\{b(10)\}_\varepsilon|\{d(10)c(01')a(10)p(01)p(10)\}_\varepsilon\rangle\\
&+ e_Ae_Be_Ce'_De_P{}^2\langle\{b(10)\}_\varepsilon|\{d(1'0)c(01)a(10)p(01)p(10)\}_\varepsilon\rangle\\
&+ e_Ae'_Be'_Ce_De_P{}^2\langle\{b(1'0)\}_\varepsilon|\{d(10)c(01')a(10)p(01)p(10)\}_\varepsilon\rangle\\
&+ e_Ae_Be'_Ce'_De_P{}^2\langle\{b(10)\}_\varepsilon|\{d(1'0)c(01')a(10)p(01)p(10)\}_\varepsilon\rangle\\
&+ e_Ae'_Be_Ce'_De_P{}^2\langle\{b(1'0)\}_\varepsilon|\{d(1'0)c(01)a(10)p(01)p(10)\}_\varepsilon\rangle\\
&+ e_Ae'_Be'_Ce'_De_P{}^2\langle\{b(1'0)\}_\varepsilon|\{d(1'0)c(01')a(10)p(01)p(10)\}_\varepsilon\rangle\\
&+ e_Ae_Be_Ce_De'_P{}^2\langle\{b(10)\}_\varepsilon|\{d(10)c(01)a(10)p(01')p(1'0)\}_\varepsilon\rangle\\
&+ e'_Ae_Be_Ce_De'_P{}^2\langle\{b(10)\}_\varepsilon|\{d(10)c(01)a(1'0)p(01')p(1'0)\}_\varepsilon\rangle\\
&+ e'_Ae'_Be_Ce_De'_P{}^2\langle\{b(1'0)\}_\varepsilon|\{d(10)c(01)a(1'0)p(01')p(1'0)\}_\varepsilon\rangle\\
&+ e'_Ae_Be'_Ce_De'_P{}^2\langle\{b(10)\}_\varepsilon|\{d(10)c(01')a(1'0)p(01')p(1'0)\}_\varepsilon\rangle\\
&+ e'_Ae_Be_Ce'_De'_P{}^2\langle\{b(10)\}_\varepsilon|\{d(1'0)c(01)a(1'0)p(01')p(1'0)\}_\varepsilon\rangle\\
&+ e'_Ae'_Be'_Ce_De'_P{}^2\langle\{b(1'0)\}_\varepsilon|\{d(10)c(01')a(1'0)p(01')p(1'0)\}_\varepsilon\rangle\\
&+ e'_Ae_Be'_Ce'_De'_P{}^2\langle\{b(10)\}_\varepsilon|\{d(1'0)c(01')a(1'0)p(01')p(1'0)\}_\varepsilon\rangle\\
&+ e'_Ae'_Be_Ce'_De'_P{}^2\langle\{b(1'0)\}_\varepsilon|\{d(1'0)c(01)a(1'0)p(01')p(1'0)\}_\varepsilon\rangle\\
&+ e'_Ae'_Be'_Ce'_De'_P{}^2\langle\{b(1'0)\}_\varepsilon|\{d(1'0)c(01')a(1'0)p(01')p(1'0)\}_\varepsilon\rangle\\
&+ e_Ae'_Be_Ce_De'_P{}^2\langle\{b(1'0)\}_\varepsilon|\{d(10)c(01)a(10)p(01')p(1'0)\}_\varepsilon\rangle\\
&+ e_Ae_Be'_Ce_De'_P{}^2\langle\{b(10)\}_\varepsilon|\{d(10)c(01')a(10)p(01')p(1'0)\}_\varepsilon\rangle\\
&+ e_Ae_Be_Ce'_De'_P{}^2\langle\{b(10)\}_\varepsilon|\{d(1'0)c(01)a(10)p(01')p(1'0)\}_\varepsilon\rangle\\
&+ e_Ae'_Be'_Ce_De'_P{}^2\langle\{b(1'0)\}_\varepsilon|\{d(10)c(01')a(10)p(01')p(1'0)\}_\varepsilon\rangle\\
&+ e_Ae_Be'_Ce'_De'_P{}^2\langle\{b(10)\}_\varepsilon|\{d(1'0)c(01')a(10)p(01')p(1'0)\}_\varepsilon\rangle\\
&+ e_Ae'_Be_Ce'_De'_P{}^2\langle\{b(1'0)\}_\varepsilon|\{d(1'0)c(01)a(10)p(01')p(1'0)\}_\varepsilon\rangle\\
&+ e_Ae'_Be'_Ce'_De'_P{}^2\langle\{b(1'0)\}_\varepsilon|\{d(1'0)c(01')a(10)p(01')p(1'0)\}_\varepsilon\rangle \qquad \text{(B.16)}
\end{aligned}
$$

$$
\begin{aligned}
\langle (B)_\varepsilon | (CDAPP)_\varepsilon \rangle^{+-} = {} & e_A e_B e_C e_D {e_P}^2 \langle \{b(10)\}_\varepsilon | \{c(1'2)d(21')a(10)p(01)p(10)\}_\varepsilon \rangle \\
& + e'_A e_B e_C e_D {e_P}^2 \langle \{b(10)\}_\varepsilon | \{c(1'2)d(21')a(1'0)p(01)p(10)\}_\varepsilon \rangle \\
& + e'_A e'_B e_C e_D {e_P}^2 \langle \{b(1'0)\}_\varepsilon | \{c(1'2)d(21')a(1'0)p(01)p(10)\}_\varepsilon \rangle \\
& + e'_A e_B e'_C e_D {e_P}^2 \langle \{b(10)\}_\varepsilon | \{c(12)d(21')a(1'0)p(01)p(10)\}_\varepsilon \rangle \\
& + e'_A e_B e_C e'_D {e_P}^2 \langle \{b(10)\}_\varepsilon | \{c(1'2)d(21)a(1'0)p(01)p(10)\}_\varepsilon \rangle \\
& + e'_A e'_B e'_C e_D {e_P}^2 \langle \{b(1'0)\}_\varepsilon | \{c(12)d(21')a(1'0)p(01)p(10)\}_\varepsilon \rangle \\
& + e'_A e_B e'_C e'_D {e_P}^2 \langle \{b(10)\}_\varepsilon | \{c(12)d(21)a(1'0)p(01)p(10)\}_\varepsilon \rangle \\
& + e'_A e'_B e_C e'_D {e_P}^2 \langle \{b(1'0)\}_\varepsilon | \{c(1'2)d(21)a(1'0)p(01)p(10)\}_\varepsilon \rangle \\
& + e'_A e'_B e'_C e'_D {e_P}^2 \langle \{b(1'0)\}_\varepsilon | \{c(12)d(21)a(1'0)p(01)p(10)\}_\varepsilon \rangle \\
& + e_A e'_B e_C e_D {e_P}^2 \langle \{b(1'0)\}_\varepsilon | \{c(1'2)d(21')a(10)p(01)p(10)\}_\varepsilon \rangle \\
& + e_A e_B e'_C e_D {e_P}^2 \langle \{b(10)\}_\varepsilon | \{c(12)d(21')a(10)p(01)p(10)\}_\varepsilon \rangle \\
& + e_A e_B e_C e'_D {e_P}^2 \langle \{b(10)\}_\varepsilon | \{c(1'2)d(21)a(10)p(01)p(10)\}_\varepsilon \rangle \\
& + e_A e'_B e'_C e_D {e_P}^2 \langle \{b(1'0)\}_\varepsilon | \{c(12)d(21')a(10)p(01)p(10)\}_\varepsilon \rangle \\
& + e_A e_B e'_C e'_D {e_P}^2 \langle \{b(10)\}_\varepsilon | \{c(12)d(21)a(10)p(01)p(10)\}_\varepsilon \rangle \\
& + e_A e'_B e_C e'_D {e_P}^2 \langle \{b(1'0)\}_\varepsilon | \{c(1'2)d(21)a(10)p(01)p(10)\}_\varepsilon \rangle \\
& + e_A e'_B e'_C e'_D {e_P}^2 \langle \{b(1'0)\}_\varepsilon | \{c(12)d(21)a(10)p(01)p(10)\}_\varepsilon \rangle \\
& + e_A e_B e_C e_D {e'_P}^2 \langle \{b(10)\}_\varepsilon | \{c(1'2)d(21')a(10)p(01')p(1'0)\}_\varepsilon \rangle \\
& + e'_A e_B e_C e_D {e'_P}^2 \langle \{b(10)\}_\varepsilon | \{c(1'2)d(21')a(1'0)p(01')p(1'0)\}_\varepsilon \rangle \\
& + e'_A e'_B e_C e_D {e'_P}^2 \langle \{b(1'0)\}_\varepsilon | \{c(1'2)d(21')a(1'0)p(01')p(1'0)\}_\varepsilon \rangle \\
& + e'_A e_B e'_C e_D {e'_P}^2 \langle \{b(10)\}_\varepsilon | \{c(12)d(21')a(1'0)p(01')p(1'0)\}_\varepsilon \rangle \\
& + e'_A e_B e_C e'_D {e'_P}^2 \langle \{b(10)\}_\varepsilon | \{c(1'2)d(21)a(1'0)p(01')p(1'0)\}_\varepsilon \rangle \\
& + e'_A e'_B e'_C e_D {e'_P}^2 \langle \{b(1'0)\}_\varepsilon | \{c(12)d(21')a(1'0)p(01')p(1'0)\}_\varepsilon \rangle \\
& + e'_A e_B e'_C e'_D {e'_P}^2 \langle \{b(10)\}_\varepsilon | \{c(12)d(21)a(1'0)p(01')p(1'0)\}_\varepsilon \rangle \\
& + e'_A e'_B e_C e'_D {e'_P}^2 \langle \{b(1'0)\}_\varepsilon | \{c(1'2)d(21)a(1'0)p(01')p(1'0)\}_\varepsilon \rangle \\
& + e'_A e'_B e'_C e'_D {e'_P}^2 \langle \{b(1'0)\}_\varepsilon | \{c(12)d(21)a(1'0)p(01')p(1'0)\}_\varepsilon \rangle \\
& + e_A e'_B e_C e_D {e'_P}^2 \langle \{b(1'0)\}_\varepsilon | \{c(1'2)d(21')a(10)p(01')p(1'0)\}_\varepsilon \rangle \\
& + e_A e_B e'_C e_D {e'_P}^2 \langle \{b(10)\}_\varepsilon | \{c(12)d(21')a(10)p(01')p(1'0)\}_\varepsilon \rangle \\
& + e_A e_B e_C e'_D {e'_P}^2 \langle \{b(10)\}_\varepsilon | \{c(1'2)d(21)a(10)p(01')p(1'0)\}_\varepsilon \rangle \\
& + e_A e'_B e'_C e_D {e'_P}^2 \langle \{b(1'0)\}_\varepsilon | \{c(12)d(21')a(10)p(01')p(1'0)\}_\varepsilon \rangle \\
& + e_A e_B e'_C e'_D {e'_P}^2 \langle \{b(10)\}_\varepsilon | \{c(12)d(21)a(10)p(01')p(1'0)\}_\varepsilon \rangle \\
& + e_A e'_B e_C e'_D {e'_P}^2 \langle \{b(1'0)\}_\varepsilon | \{c(1'2)d(21)a(10)p(01')p(1'0)\}_\varepsilon \rangle \\
& + e_A e'_B e'_C e'_D {e'_P}^2 \langle \{b(1'0)\}_\varepsilon | \{c(12)d(21)a(10)p(01')p(1'0)\}_\varepsilon \rangle
\end{aligned}
\tag{B.17}
$$

$$
\begin{aligned}
\langle (BPP)_\varepsilon | (DCA)_\varepsilon \rangle^{+-} &= e_A e_B e_C e_D {e_P}^2 \langle \{b(10)p(01)p(10)\}_\varepsilon | \{d(10)c(01)a(10)\}_\varepsilon \rangle \\
&+ e'_A e_B e_C e_D {e_P}^2 \langle \{b(10)p(01)p(10)\}_\varepsilon | \{d(10)c(01)a(1'0)\}_\varepsilon \rangle \\
&+ e'_A e'_B e_C e_D {e_P}^2 \langle \{b(1'0)p(01)p(10)\}_\varepsilon | \{d(10)c(01)a(1'0)\}_\varepsilon \rangle \\
&+ e'_A e_B e'_C e_D {e_P}^2 \langle \{b(10)p(01)p(10)\}_\varepsilon | \{d(10)c(01')a(1'0)\}_\varepsilon \rangle \\
&+ e'_A e_B e_C e'_D {e_P}^2 \langle \{b(10)p(01)p(10)\}_\varepsilon | \{d(1'0)c(01)a(1'0)\}_\varepsilon \rangle \\
&+ e'_A e'_B e'_C e_D {e_P}^2 \langle \{b(1'0)p(01)p(10)\}_\varepsilon | \{d(10)c(01')a(1'0)\}_\varepsilon \rangle \\
&+ e'_A e_B e'_C e'_D {e_P}^2 \langle \{b(10)p(01)p(10)\}_\varepsilon | \{d(1'0)c(01')a(1'0)\}_\varepsilon \rangle \\
&+ e'_A e'_B e_C e'_D {e_P}^2 \langle \{b(1'0)p(01)p(10)\}_\varepsilon | \{d(1'0)c(01)a(1'0)\}_\varepsilon \rangle \\
&+ e'_A e'_B e'_C e'_D {e_P}^2 \langle \{b(1'0)p(01)p(10)\}_\varepsilon | \{d(1'0)c(01')a(1'0)\}_\varepsilon \rangle \\
&+ e_A e'_B e_C e_D {e_P}^2 \langle \{b(1'0)p(01)p(10)\}_\varepsilon | \{d(10)c(01)a(10)\}_\varepsilon \rangle \\
&+ e_A e_B e'_C e_D {e_P}^2 \langle \{b(10)p(01)p(10)\}_\varepsilon | \{d(10)c(01')a(10)\}_\varepsilon \rangle \\
&+ e_A e_B e_C e'_D {e_P}^2 \langle \{b(10)p(01)p(10)\}_\varepsilon | \{d(1'0)c(01)a(10)\}_\varepsilon \rangle \\
&+ e_A e'_B e'_C e_D {e_P}^2 \langle \{b(1'0)p(01)p(10)\}_\varepsilon | \{d(10)c(01')a(10)\}_\varepsilon \rangle \\
&+ e_A e_B e'_C e'_D {e_P}^2 \langle \{b(10)p(01)p(10)\}_\varepsilon | \{d(1'0)c(01')a(10)\}_\varepsilon \rangle \\
&+ e_A e'_B e_C e'_D {e_P}^2 \langle \{b(1'0)p(01)p(10)\}_\varepsilon | \{d(1'0)c(01)a(10)\}_\varepsilon \rangle \\
&+ e_A e'_B e'_C e'_D {e_P}^2 \langle \{b(1'0)p(01)p(10)\}_\varepsilon | \{d(1'0)c(01')a(10)\}_\varepsilon \rangle \\
&+ e_A e_B e_C e_D {e'_P}^2 \langle \{b(10)p(01')p(1'0)\}_\varepsilon | \{d(10)c(01)a(10)\}_\varepsilon \rangle \\
&+ e'_A e_B e_C e_D {e'_P}^2 \langle \{b(10)p(01')p(1'0)\}_\varepsilon | \{d(10)c(01)a(1'0)\}_\varepsilon \rangle \\
&+ e'_A e'_B e_C e_D {e'_P}^2 \langle \{b(1'0)p(01')p(1'0)\}_\varepsilon | \{d(10)c(01)a(1'0)\}_\varepsilon \rangle \\
&+ e'_A e_B e'_C e_D {e'_P}^2 \langle \{b(10)p(01')p(1'0)\}_\varepsilon | \{d(10)c(01')a(1'0)\}_\varepsilon \rangle \\
&+ e'_A e_B e_C e'_D {e'_P}^2 \langle \{b(10)p(01')p(1'0)\}_\varepsilon | \{d(1'0)c(01)a(1'0)\}_\varepsilon \rangle \\
&+ e'_A e'_B e'_C e_D {e'_P}^2 \langle \{b(1'0)p(01')p(1'0)\}_\varepsilon | \{d(10)c(01')a(1'0)\}_\varepsilon \rangle \\
&+ e'_A e_B e'_C e'_D {e'_P}^2 \langle \{b(10)p(01')p(1'0)\}_\varepsilon | \{d(1'0)c(01')a(1'0)\}_\varepsilon \rangle \\
&+ e'_A e'_B e_C e'_D {e'_P}^2 \langle \{b(1'0)p(01')p(1'0)\}_\varepsilon | \{d(1'0)c(01)a(1'0)\}_\varepsilon \rangle \\
&+ e'_A e'_B e'_C e'_D {e'_P}^2 \langle \{b(1'0)p(01')p(1'0)\}_\varepsilon | \{d(1'0)c(01')a(1'0)\}_\varepsilon \rangle \\
&+ e_A e'_B e_C e_D {e'_P}^2 \langle \{b(1'0)p(01')p(1'0)\}_\varepsilon | \{d(10)c(01)a(10)\}_\varepsilon \rangle \\
&+ e_A e_B e'_C e_D {e'_P}^2 \langle \{b(10)p(01')p(1'0)\}_\varepsilon | \{d(10)c(01')a(10)\}_\varepsilon \rangle \\
&+ e_A e_B e_C e'_D {e'_P}^2 \langle \{b(10)p(01')p(1'0)\}_\varepsilon | \{d(1'0)c(01)a(10)\}_\varepsilon \rangle \\
&+ e_A e'_B e'_C e_D {e'_P}^2 \langle \{b(1'0)p(01')p(1'0)\}_\varepsilon | \{d(10)c(01')a(10)\}_\varepsilon \rangle \\
&+ e_A e_B e'_C e'_D {e'_P}^2 \langle \{b(10)p(01')p(1'0)\}_\varepsilon | \{d(1'0)c(01')a(10)\}_\varepsilon \rangle \\
&+ e_A e'_B e_C e'_D {e'_P}^2 \langle \{b(1'0)p(01')p(1'0)\}_\varepsilon | \{d(1'0)c(01)a(10)\}_\varepsilon \rangle \\
&+ e_A e'_B e'_C e'_D {e'_P}^2 \langle \{b(1'0)p(01')p(1'0)\}_\varepsilon | \{d(1'0)c(01')a(10)\}_\varepsilon \rangle \qquad \text{(B.18)}
\end{aligned}
$$

$$
\begin{aligned}
\langle (BPP)_\varepsilon | (CDA)_\varepsilon \rangle^{+-} &= e_A e_B e_C e_D {e_P}^2 \langle \{ b(10) p(01) p(10) \}_\varepsilon | \{ c(1'2) d(21') a(10) \}_\varepsilon \rangle \\
&+ e'_A e_B e_C e_D {e_P}^2 \langle \{ b(10) p(01) p(10) \}_\varepsilon | \{ c(1'2) d(21') a(1'0) \}_\varepsilon \rangle \\
&+ e'_A e'_B e_C e_D {e_P}^2 \langle \{ b(1'0) p(01) p(10) \}_\varepsilon | \{ c(1'2) d(21') a(1'0) \}_\varepsilon \rangle \\
&+ e'_A e_B e'_C e_D {e_P}^2 \langle \{ b(10) p(01) p(10) \}_\varepsilon | \{ c(12) d(21') a(1'0) \}_\varepsilon \rangle \\
&+ e'_A e_B e_C e'_D {e_P}^2 \langle \{ b(10) p(01) p(10) \}_\varepsilon | \{ c(1'2) d(21) a(1'0) \}_\varepsilon \rangle \\
&+ e'_A e'_B e'_C e_D {e_P}^2 \langle \{ b(1'0) p(01) p(10) \}_\varepsilon | \{ c(12) d(21') a(1'0) \}_\varepsilon \rangle \\
&+ e'_A e_B e'_C e'_D {e_P}^2 \langle \{ b(10) p(01) p(10) \}_\varepsilon | \{ c(12) d(21) a(1'0) \}_\varepsilon \rangle \\
&+ e'_A e'_B e_C e'_D {e_P}^2 \langle \{ b(1'0) p(01) p(10) \}_\varepsilon | \{ c(1'2) d(21) a(1'0) \}_\varepsilon \rangle \\
&+ e'_A e'_B e'_C e'_D {e_P}^2 \langle \{ b(1'0) p(01) p(10) \}_\varepsilon | \{ c(12) d(21) a(1'0) \}_\varepsilon \rangle \\
&+ e_A e'_B e_C e_D {e_P}^2 \langle \{ b(1'0) p(01) p(10) \}_\varepsilon | \{ c(1'2) d(21') a(10) \}_\varepsilon \rangle \\
&+ e_A e_B e'_C e_D {e_P}^2 \langle \{ b(10) p(01) p(10) \}_\varepsilon | \{ c(12) d(21') a(10) \}_\varepsilon \rangle \\
&+ e_A e_B e_C e'_D {e_P}^2 \langle \{ b(10) p(01) p(10) \}_\varepsilon | \{ c(1'2) d(21) a(10) \}_\varepsilon \rangle \\
&+ e_A e'_B e'_C e_D {e_P}^2 \langle \{ b(1'0) p(01) p(10) \}_\varepsilon | \{ c(12) d(21') a(10) \}_\varepsilon \rangle \\
&+ e_A e_B e'_C e'_D {e_P}^2 \langle \{ b(10) p(01) p(10) \}_\varepsilon | \{ c(12) d(21) a(10) \}_\varepsilon \rangle \\
&+ e_A e'_B e_C e'_D {e_P}^2 \langle \{ b(1'0) p(01) p(10) \}_\varepsilon | \{ c(1'2) d(21) a(10) \}_\varepsilon \rangle \\
&+ e_A e'_B e'_C e'_D {e_P}^2 \langle \{ b(1'0) p(01) p(10) \}_\varepsilon | \{ c(12) d(21) a(10) \}_\varepsilon \rangle \\
&+ e_A e_B e_C e_D {e'_P}^2 \langle \{ b(10) p(01') p(1'0) \}_\varepsilon | \{ c(1'2) d(21') a(10) \}_\varepsilon \rangle \\
&+ e'_A e_B e_C e_D {e'_P}^2 \langle \{ b(10) p(01') p(1'0) \}_\varepsilon | \{ c(1'2) d(21') a(1'0) \}_\varepsilon \rangle \\
&+ e'_A e'_B e_C e_D {e'_P}^2 \langle \{ b(1'0) p(01') p(1'0) \}_\varepsilon | \{ c(1'2) d(21') a(1'0) \}_\varepsilon \rangle \\
&+ e'_A e_B e'_C e_D {e'_P}^2 \langle \{ b(10) p(01') p(1'0) \}_\varepsilon | \{ c(12) d(21') a(1'0) \}_\varepsilon \rangle \\
&+ e'_A e_B e_C e'_D {e'_P}^2 \langle \{ b(10) p(01') p(1'0) \}_\varepsilon | \{ c(1'2) d(21) a(1'0) \}_\varepsilon \rangle \\
&+ e'_A e'_B e'_C e_D {e'_P}^2 \langle \{ b(1'0) p(01') p(1'0) \}_\varepsilon | \{ c(12) d(21') a(1'0) \}_\varepsilon \rangle \\
&+ e'_A e_B e'_C e'_D {e'_P}^2 \langle \{ b(10) p(01') p(1'0) \}_\varepsilon | \{ c(12) d(21) a(1'0) \}_\varepsilon \rangle \\
&+ e'_A e'_B e_C e'_D {e'_P}^2 \langle \{ b(1'0) p(01') p(1'0) \}_\varepsilon | \{ c(1'2) d(21) a(1'0) \}_\varepsilon \rangle \\
&+ e'_A e'_B e'_C e'_D {e'_P}^2 \langle \{ b(1'0) p(01') p(1'0) \}_\varepsilon | \{ c(12) d(21) a(1'0) \}_\varepsilon \rangle \\
&+ e_A e'_B e_C e_D {e'_P}^2 \langle \{ b(1'0) p(01') p(1'0) \}_\varepsilon | \{ c(1'2) d(21') a(10) \}_\varepsilon \rangle \\
&+ e_A e_B e'_C e_D {e'_P}^2 \langle \{ b(10) p(01') p(1'0) \}_\varepsilon | \{ c(12) d(21') a(10) \}_\varepsilon \rangle \\
&+ e_A e_B e_C e'_D {e'_P}^2 \langle \{ b(10) p(01') p(1'0) \}_\varepsilon | \{ c(1'2) d(21) a(10) \}_\varepsilon \rangle \\
&+ e_A e'_B e'_C e_D {e'_P}^2 \langle \{ b(1'0) p(01') p(1'0) \}_\varepsilon | \{ c(12) d(21') a(10) \}_\varepsilon \rangle \\
&+ e_A e_B e'_C e'_D {e'_P}^2 \langle \{ b(10) p(01') p(1'0) \}_\varepsilon | \{ c(12) d(21) a(10) \}_\varepsilon \rangle \\
&+ e_A e'_B e_C e'_D {e'_P}^2 \langle \{ b(1'0) p(01') p(1'0) \}_\varepsilon | \{ c(1'2) d(21) a(10) \}_\varepsilon \rangle \\
&+ e_A e'_B e'_C e'_D {e'_P}^2 \langle \{ b(1'0) p(01') p(1'0) \}_\varepsilon | \{ c(12) d(21) a(10) \}_\varepsilon \rangle \qquad \text{(B.19)}
\end{aligned}
$$

$$
\begin{aligned}
\langle (DCBPP)_\varepsilon|(A)_\varepsilon\rangle^{+-} = {} & e_A e_B e_C e_D {e_P}^2 \langle\{d(1'2)c(21')b(10)p(01)p(10)\}_\varepsilon|\{a(10)\}_\varepsilon\rangle \\
& + e'_A e_B e_C e_D {e_P}^2 \langle\{d(1'2)c(21')b(10)p(01)p(10)\}_\varepsilon|\{a(1'0)\}_\varepsilon\rangle \\
& + e'_A e'_B e_C e_D {e_P}^2 \langle\{d(1'2)c(21')b(1'0)p(01)p(10)\}_\varepsilon|\{a(1'0)\}_\varepsilon\rangle \\
& + e'_A e_B e'_C e_D {e_P}^2 \langle\{d(1'2)c(21)b(10)p(01)p(10)\}_\varepsilon|\{a(1'0)\}_\varepsilon\rangle \\
& + e'_A e_B e_C e'_D {e_P}^2 \langle\{d(12)c(21')b(10)p(01)p(10)\}_\varepsilon|\{a(1'0)\}_\varepsilon\rangle \\
& + e'_A e'_B e'_C e_D {e_P}^2 \langle\{d(1'2)c(21)b(1'0)p(01)p(10)\}_\varepsilon|\{a(1'0)\}_\varepsilon\rangle \\
& + e'_A e_B e'_C e'_D {e_P}^2 \langle\{d(12)c(21)b(10)p(01)p(10)\}_\varepsilon|\{a(1'0)\}_\varepsilon\rangle \\
& + e'_A e'_B e_C e'_D {e_P}^2 \langle\{d(12)c(21')b(1'0)p(01)p(10)\}_\varepsilon|\{a(1'0)\}_\varepsilon\rangle \\
& + e'_A e'_B e'_C e'_D {e_P}^2 \langle\{d(12)c(21)b(1'0)p(01)p(10)\}_\varepsilon|\{a(1'0)\}_\varepsilon\rangle \\
& + e_A e'_B e_C e_D {e_P}^2 \langle\{d(1'2)c(21')b(1'0)p(01)p(10)\}_\varepsilon|\{a(10)\}_\varepsilon\rangle \\
& + e_A e_B e'_C e_D {e_P}^2 \langle\{d(1'2)c(21)b(10)p(01)p(10)\}_\varepsilon|\{a(10)\}_\varepsilon\rangle \\
& + e_A e_B e_C e'_D {e_P}^2 \langle\{d(12)c(21')b(10)p(01)p(10)\}_\varepsilon|\{a(10)\}_\varepsilon\rangle \\
& + e_A e'_B e'_C e_D {e_P}^2 \langle\{d(1'2)c(21)b(1'0)p(01)p(10)\}_\varepsilon|\{a(10)\}_\varepsilon\rangle \\
& + e_A e_B e'_C e'_D {e_P}^2 \langle\{d(12)c(21)b(10)p(01)p(10)\}_\varepsilon|\{a(10)\}_\varepsilon\rangle \\
& + e_A e'_B e_C e'_D {e_P}^2 \langle\{d(12)c(21')b(1'0)p(01)p(10)\}_\varepsilon|\{a(10)\}_\varepsilon\rangle \\
& + e_A e'_B e'_C e'_D {e_P}^2 \langle\{d(12)c(21)b(1'0)p(01)p(10)\}_\varepsilon|\{a(10)\}_\varepsilon\rangle \\
& + e_A e_B e_C e_D {e'_P}^2 \langle\{d(1'2)c(21')b(10)p(01')p(1'0)\}_\varepsilon|\{a(10)\}_\varepsilon\rangle \\
& + e'_A e_B e_C e_D {e'_P}^2 \langle\{d(1'2)c(21')b(10)p(01')p(1'0)\}_\varepsilon|\{a(1'0)\}_\varepsilon\rangle \\
& + e'_A e'_B e_C e_D {e'_P}^2 \langle\{d(1'2)c(21')b(1'0)p(01')p(1'0)\}_\varepsilon|\{a(1'0)\}_\varepsilon\rangle \\
& + e'_A e_B e'_C e_D {e'_P}^2 \langle\{d(1'2)c(21)b(10)p(01')p(1'0)\}_\varepsilon|\{a(1'0)\}_\varepsilon\rangle \\
& + e'_A e_B e_C e'_D {e'_P}^2 \langle\{d(12)c(21')b(10)p(01')p(1'0)\}_\varepsilon|\{a(1'0)\}_\varepsilon\rangle \\
& + e'_A e'_B e'_C e_D {e'_P}^2 \langle\{d(1'2)c(21)b(1'0)p(01')p(1'0)\}_\varepsilon|\{a(1'0)\}_\varepsilon\rangle \\
& + e'_A e_B e'_C e'_D {e'_P}^2 \langle\{d(12)c(21)b(10)p(01')p(1'0)\}_\varepsilon|\{a(1'0)\}_\varepsilon\rangle \\
& + e'_A e'_B e_C e'_D {e'_P}^2 \langle\{d(12)c(21')b(1'0)p(01')p(1'0)\}_\varepsilon|\{a(1'0)\}_\varepsilon\rangle \\
& + e'_A e'_B e'_C e'_D {e'_P}^2 \langle\{d(12)c(21)b(1'0)p(01')p(1'0)\}_\varepsilon|\{a(1'0)\}_\varepsilon\rangle \\
& + e_A e'_B e_C e_D {e'_P}^2 \langle\{d(1'2)c(21')b(1'0)p(01')p(1'0)\}_\varepsilon|\{a(10)\}_\varepsilon\rangle \\
& + e_A e_B e'_C e_D {e'_P}^2 \langle\{d(1'2)c(21)b(10)p(01')p(1'0)\}_\varepsilon|\{a(10)\}_\varepsilon\rangle \\
& + e_A e_B e_C e'_D {e'_P}^2 \langle\{d(12)c(21')b(10)p(01')p(1'0)\}_\varepsilon|\{a(10)\}_\varepsilon\rangle \\
& + e_A e'_B e'_C e_D {e'_P}^2 \langle\{d(1'2)c(21)b(1'0)p(01')p(1'0)\}_\varepsilon|\{a(10)\}_\varepsilon\rangle \\
& + e_A e_B e'_C e'_D {e'_P}^2 \langle\{d(12)c(21)b(10)p(01')p(1'0)\}_\varepsilon|\{a(10)\}_\varepsilon\rangle \\
& + e_A e'_B e_C e'_D {e'_P}^2 \langle\{d(12)c(21')b(1'0)p(01')p(1'0)\}_\varepsilon|\{a(10)\}_\varepsilon\rangle \\
& + e_A e'_B e'_C e'_D {e'_P}^2 \langle\{d(12)c(21)b(1'0)p(01')p(1'0)\}_\varepsilon|\{a(10)\}_\varepsilon\rangle \qquad \text{(B.20)}
\end{aligned}
$$

$$
\begin{aligned}
\langle (CDBPP)_\varepsilon | (A)_\varepsilon \rangle^{+-} = {} & e_A e_B e_C e_D {e_P}^2 \langle \{c(10)d(01)b(10)p(01)p(10)\}_\varepsilon | \{a(10)\}_\varepsilon \rangle \\
& + e'_A e_B e_C e_D {e_P}^2 \langle \{c(10)d(01)b(10)p(01)p(10)\}_\varepsilon | \{a(1'0)\}_\varepsilon \rangle \\
& + e'_A e'_B e_C e_D {e_P}^2 \langle \{c(10)d(01)b(1'0)p(01)p(10)\}_\varepsilon | \{a(1'0)\}_\varepsilon \rangle \\
& + e'_A e_B e'_C e_D {e_P}^2 \langle \{c(1'0)d(01)b(10)p(01)p(10)\}_\varepsilon | \{a(1'0)\}_\varepsilon \rangle \\
& + e'_A e_B e_C e'_D {e_P}^2 \langle \{c(10)d(01')b(10)p(01)p(10)\}_\varepsilon | \{a(1'0)\}_\varepsilon \rangle \\
& + e'_A e'_B e'_C e_D {e_P}^2 \langle \{c(1'0)d(01)b(1'0)p(01)p(10)\}_\varepsilon | \{a(1'0)\}_\varepsilon \rangle \\
& + e'_A e_B e'_C e'_D {e_P}^2 \langle \{c(1'0)d(01')b(10)p(01)p(10)\}_\varepsilon | \{a(1'0)\}_\varepsilon \rangle \\
& + e'_A e'_B e_C e'_D {e_P}^2 \langle \{c(10)d(01')b(1'0)p(01)p(10)\}_\varepsilon | \{a(1'0)\}_\varepsilon \rangle \\
& + e'_A e'_B e'_C e'_D {e_P}^2 \langle \{c(1'0)d(01')b(1'0)p(01)p(10)\}_\varepsilon | \{a(1'0)\}_\varepsilon \rangle \\
& + e_A e'_B e_C e_D {e_P}^2 \langle \{c(10)d(01)b(1'0)p(01)p(10)\}_\varepsilon | \{a(10)\}_\varepsilon \rangle \\
& + e_A e_B e'_C e_D {e_P}^2 \langle \{c(1'0)d(01)b(10)p(01)p(10)\}_\varepsilon | \{a(10)\}_\varepsilon \rangle \\
& + e_A e_B e_C e'_D {e_P}^2 \langle \{c(10)d(01')b(10)p(01)p(10)\}_\varepsilon | \{a(10)\}_\varepsilon \rangle \\
& + e_A e'_B e'_C e_D {e_P}^2 \langle \{c(1'0)d(01)b(1'0)p(01)p(10)\}_\varepsilon | \{a(10)\}_\varepsilon \rangle \\
& + e_A e_B e'_C e'_D {e_P}^2 \langle \{c(1'0)d(01')b(10)p(01)p(10)\}_\varepsilon | \{a(10)\}_\varepsilon \rangle \\
& + e_A e'_B e_C e'_D {e_P}^2 \langle \{c(10)d(01')b(1'0)p(01)p(10)\}_\varepsilon | \{a(10)\}_\varepsilon \rangle \\
& + e_A e'_B e'_C e'_D {e_P}^2 \langle \{c(1'0)d(01')b(1'0)p(01)p(10)\}_\varepsilon | \{a(10)\}_\varepsilon \rangle \\
& + e_A e_B e_C e_D {e'_P}^2 \langle \{c(10)d(01)b(10)p(01')p(1'0)\}_\varepsilon | \{a(10)\}_\varepsilon \rangle \\
& + e'_A e_B e_C e_D {e'_P}^2 \langle \{c(10)d(01)b(10)p(01')p(1'0)\}_\varepsilon | \{a(1'0)\}_\varepsilon \rangle \\
& + e'_A e'_B e_C e_D {e'_P}^2 \langle \{c(10)d(01)b(1'0)p(01')p(1'0)\}_\varepsilon | \{a(1'0)\}_\varepsilon \rangle \\
& + e'_A e_B e'_C e_D {e'_P}^2 \langle \{c(1'0)d(01)b(10)p(01')p(1'0)\}_\varepsilon | \{a(1'0)\}_\varepsilon \rangle \\
& + e'_A e_B e_C e'_D {e'_P}^2 \langle \{c(10)d(01')b(10)p(01')p(1'0)\}_\varepsilon | \{a(1'0)\}_\varepsilon \rangle \\
& + e'_A e'_B e'_C e_D {e'_P}^2 \langle \{c(1'0)d(01)b(1'0)p(01')p(1'0)\}_\varepsilon | \{a(1'0)\}_\varepsilon \rangle \\
& + e'_A e_B e'_C e'_D {e'_P}^2 \langle \{c(1'0)d(01')b(10)p(01')p(1'0)\}_\varepsilon | \{a(1'0)\}_\varepsilon \rangle \\
& + e'_A e'_B e_C e'_D {e'_P}^2 \langle \{c(10)d(01')b(1'0)p(01')p(1'0)\}_\varepsilon | \{a(1'0)\}_\varepsilon \rangle \\
& + e'_A e'_B e'_C e'_D {e'_P}^2 \langle \{c(1'0)d(01')b(1'0)p(01')p(1'0)\}_\varepsilon | \{a(1'0)\}_\varepsilon \rangle \\
& + e_A e'_B e_C e_D {e'_P}^2 \langle \{c(10)d(01)b(1'0)p(01')p(1'0)\}_\varepsilon | \{a(10)\}_\varepsilon \rangle \\
& + e_A e_B e'_C e_D {e'_P}^2 \langle \{c(1'0)d(01)b(10)p(01')p(1'0)\}_\varepsilon | \{a(10)\}_\varepsilon \rangle \\
& + e_A e_B e_C e'_D {e'_P}^2 \langle \{c(10)d(01')b(10)p(01')p(1'0)\}_\varepsilon | \{a(10)\}_\varepsilon \rangle \\
& + e_A e'_B e'_C e_D {e'_P}^2 \langle \{c(1'0)d(01)b(1'0)p(01')p(1'0)\}_\varepsilon | \{a(10)\}_\varepsilon \rangle \\
& + e_A e_B e'_C e'_D {e'_P}^2 \langle \{c(1'0)d(01')b(10)p(01')p(1'0)\}_\varepsilon | \{a(10)\}_\varepsilon \rangle \\
& + e_A e'_B e_C e'_D {e'_P}^2 \langle \{c(10)d(01')b(1'0)p(01')p(1'0)\}_\varepsilon | \{a(10)\}_\varepsilon \rangle \\
& + e_A e'_B e'_C e'_D {e'_P}^2 \langle \{c(1'0)d(01')b(1'0)p(01')p(1'0)\}_\varepsilon | \{a(10)\}_\varepsilon \rangle \qquad \text{(B.21)}
\end{aligned}
$$

$$
\begin{aligned}
\langle (DCB)_\varepsilon | (APP)_\varepsilon \rangle^{+-} &= e_A e_B e_C e_D {e_P}^2 \langle \{d(1'2)c(21')b(10)\}_\varepsilon | \{a(10)p(01)p(10)\}_\varepsilon \rangle \\
&+ e'_A e_B e_C e_D {e_P}^2 \langle \{d(1'2)c(21')b(10)\}_\varepsilon | \{a(1'0)p(01)p(10)\}_\varepsilon \rangle \\
&+ e'_A e'_B e_C e_D {e_P}^2 \langle \{d(1'2)c(21')b(1'0)\}_\varepsilon | \{a(1'0)p(01)p(10)\}_\varepsilon \rangle \\
&+ e'_A e_B e'_C e_D {e_P}^2 \langle \{d(1'2)c(21)b(10)\}_\varepsilon | \{a(1'0)p(01)p(10)\}_\varepsilon \rangle \\
&+ e'_A e_B e_C e'_D {e_P}^2 \langle \{d(12)c(21')b(10)\}_\varepsilon | \{a(1'0)p(01)p(10)\}_\varepsilon \rangle \\
&+ e'_A e'_B e'_C e_D {e_P}^2 \langle \{d(1'2)c(21)b(1'0)\}_\varepsilon | \{a(1'0)p(01)p(10)\}_\varepsilon \rangle \\
&+ e'_A e_B e'_C e'_D {e_P}^2 \langle \{d(12)c(21)b(10)\}_\varepsilon | \{a(1'0)p(01)p(10)\}_\varepsilon \rangle \\
&+ e'_A e'_B e_C e'_D {e_P}^2 \langle \{d(12)c(21')b(1'0)\}_\varepsilon | \{a(1'0)p(01)p(10)\}_\varepsilon \rangle \\
&+ e'_A e'_B e'_C e'_D {e_P}^2 \langle \{d(12)c(21)b(1'0)\}_\varepsilon | \{a(1'0)p(01)p(10)\}_\varepsilon \rangle \\
&+ e_A e'_B e_C e_D {e_P}^2 \langle \{d(1'2)c(21')b(1'0)\}_\varepsilon | \{a(10)p(01)p(10)\}_\varepsilon \rangle \\
&+ e_A e_B e'_C e_D {e_P}^2 \langle \{d(1'2)c(21)b(10)\}_\varepsilon | \{a(10)p(01)p(10)\}_\varepsilon \rangle \\
&+ e_A e_B e_C e'_D {e_P}^2 \langle \{d(12)c(21')b(10)\}_\varepsilon | \{a(10)p(01)p(10)\}_\varepsilon \rangle \\
&+ e_A e'_B e'_C e_D {e_P}^2 \langle \{d(1'2)c(21)b(1'0)\}_\varepsilon | \{a(10)p(01)p(10)\}_\varepsilon \rangle \\
&+ e_A e_B e'_C e'_D {e_P}^2 \langle \{d(12)c(21)b(10)\}_\varepsilon | \{a(10)p(01)p(10)\}_\varepsilon \rangle \\
&+ e_A e'_B e_C e'_D {e_P}^2 \langle \{d(12)c(21')b(1'0)\}_\varepsilon | \{a(10)p(01)p(10)\}_\varepsilon \rangle \\
&+ e_A e'_B e'_C e'_D {e_P}^2 \langle \{d(12)c(21)b(1'0)\}_\varepsilon | \{a(10)p(01)p(10)\}_\varepsilon \rangle \\
&+ e_A e_B e_C e_D {e'_P}^2 \langle \{d(1'2)c(21')b(10)\}_\varepsilon | \{a(10)p(01')p(1'0)\}_\varepsilon \rangle \\
&+ e'_A e_B e_C e_D {e'_P}^2 \langle \{d(1'2)c(21')b(10)\}_\varepsilon | \{a(1'0)p(01')p(1'0)\}_\varepsilon \rangle \\
&+ e'_A e'_B e_C e_D {e'_P}^2 \langle \{d(1'2)c(21')b(1'0)\}_\varepsilon | \{a(1'0)p(01')p(1'0)\}_\varepsilon \rangle \\
&+ e'_A e_B e'_C e_D {e'_P}^2 \langle \{d(1'2)c(21)b(10)\}_\varepsilon | \{a(1'0)p(01')p(1'0)\}_\varepsilon \rangle \\
&+ e'_A e_B e_C e'_D {e'_P}^2 \langle \{d(12)c(21')b(10)\}_\varepsilon | \{a(1'0)p(01')p(1'0)\}_\varepsilon \rangle \\
&+ e'_A e'_B e'_C e_D {e'_P}^2 \langle \{d(1'2)c(21)b(1'0)\}_\varepsilon | \{a(1'0)p(01')p(1'0)\}_\varepsilon \rangle \\
&+ e'_A e_B e'_C e'_D {e'_P}^2 \langle \{d(12)c(21)b(10)\}_\varepsilon | \{a(1'0)p(01')p(1'0)\}_\varepsilon \rangle \\
&+ e'_A e'_B e_C e'_D {e'_P}^2 \langle \{d(12)c(21')b(1'0)\}_\varepsilon | \{a(1'0)p(01')p(1'0)\}_\varepsilon \rangle \\
&+ e'_A e'_B e'_C e'_D {e'_P}^2 \langle \{d(12)c(21)b(1'0)\}_\varepsilon | \{a(1'0)p(01')p(1'0)\}_\varepsilon \rangle \\
&+ e_A e'_B e_C e_D {e'_P}^2 \langle \{d(1'2)c(21')b(1'0)\}_\varepsilon | \{a(10)p(01')p(1'0)\}_\varepsilon \rangle \\
&+ e_A e_B e'_C e_D {e'_P}^2 \langle \{d(1'2)c(21)b(10)\}_\varepsilon | \{a(10)p(01')p(1'0)\}_\varepsilon \rangle \\
&+ e_A e_B e_C e'_D {e'_P}^2 \langle \{d(12)c(21')b(10)\}_\varepsilon | \{a(10)p(01')p(1'0)\}_\varepsilon \rangle \\
&+ e_A e'_B e'_C e_D {e'_P}^2 \langle \{d(1'2)c(21)b(1'0)\}_\varepsilon | \{a(10)p(01')p(1'0)\}_\varepsilon \rangle \\
&+ e_A e_B e'_C e'_D {e'_P}^2 \langle \{d(12)c(21)b(10)\}_\varepsilon | \{a(10)p(01')p(1'0)\}_\varepsilon \rangle \\
&+ e_A e'_B e_C e'_D {e'_P}^2 \langle \{d(12)c(21')b(1'0)\}_\varepsilon | \{a(10)p(01')p(1'0)\}_\varepsilon \rangle \\
&+ e_A e'_B e'_C e'_D {e'_P}^2 \langle \{d(12)c(21)b(1'0)\}_\varepsilon | \{a(10)p(01')p(1'0)\}_\varepsilon \rangle
\end{aligned}
\tag{B.22}
$$

$$
\begin{aligned}
\langle (CDB)_\varepsilon | (APP)_\varepsilon \rangle^{+-} &= e_A e_B e_C e_D {e_P}^2 \langle \{c(10)d(01)b(10)\}_\varepsilon | \{a(10)p(01)p(10)\}_\varepsilon \rangle \\
&+ e'_A e_B e_C e_D {e_P}^2 \langle \{c(10)d(01)b(10)\}_\varepsilon | \{a(1'0)p(01)p(10)\}_\varepsilon \rangle \\
&+ e'_A e'_B e_C e_D {e_P}^2 \langle \{c(10)d(01)b(1'0)\}_\varepsilon | \{a(1'0)p(01)p(10)\}_\varepsilon \rangle \\
&+ e'_A e_B e'_C e_D {e_P}^2 \langle \{c(1'0)d(01)b(10)\}_\varepsilon | \{a(1'0)p(01)p(10)\}_\varepsilon \rangle \\
&+ e'_A e_B e_C e'_D {e_P}^2 \langle \{c(10)d(01')b(10)\}_\varepsilon | \{a(1'0)p(01)p(10)\}_\varepsilon \rangle \\
&+ e'_A e'_B e'_C e_D {e_P}^2 \langle \{c(1'0)d(01)b(1'0)\}_\varepsilon | \{a(1'0)p(01)p(10)\}_\varepsilon \rangle \\
&+ e'_A e_B e'_C e'_D {e_P}^2 \langle \{c(1'0)d(01')b(10)\}_\varepsilon | \{a(1'0)p(01)p(10)\}_\varepsilon \rangle \\
&+ e'_A e'_B e_C e'_D {e_P}^2 \langle \{c(10)d(01')b(1'0)\}_\varepsilon | \{a(1'0)p(01)p(10)\}_\varepsilon \rangle \\
&+ e'_A e'_B e'_C e'_D {e_P}^2 \langle \{c(1'0)d(01')b(1'0)\}_\varepsilon | \{a(1'0)p(01)p(10)\}_\varepsilon \rangle \\
&+ e_A e'_B e_C e_D {e_P}^2 \langle \{c(10)d(01)b(1'0)\}_\varepsilon | \{a(10)p(01)p(10)\}_\varepsilon \rangle \\
&+ e_A e_B e'_C e_D {e_P}^2 \langle \{c(1'0)d(01)b(10)\}_\varepsilon | \{a(10)p(01)p(10)\}_\varepsilon \rangle \\
&+ e_A e_B e_C e'_D {e_P}^2 \langle \{c(10)d(01')b(10)\}_\varepsilon | \{a(10)p(01)p(10)\}_\varepsilon \rangle \\
&+ e_A e'_B e'_C e_D {e_P}^2 \langle \{c(1'0)d(01)b(1'0)\}_\varepsilon | \{a(10)p(01)p(10)\}_\varepsilon \rangle \\
&+ e_A e_B e'_C e'_D {e_P}^2 \langle \{c(1'0)d(01')b(10)\}_\varepsilon | \{a(10)p(01)p(10)\}_\varepsilon \rangle \\
&+ e_A e'_B e_C e'_D {e_P}^2 \langle \{c(10)d(01')b(1'0)\}_\varepsilon | \{a(10)p(01)p(10)\}_\varepsilon \rangle \\
&+ e_A e'_B e'_C e'_D {e_P}^2 \langle \{c(1'0)d(01')b(1'0)\}_\varepsilon | \{a(10)p(01)p(10)\}_\varepsilon \rangle \\
&+ e_A e_B e_C e_D {e'_P}^2 \langle \{c(10)d(01)b(10)\}_\varepsilon | \{a(10)p(01')p(1'0)\}_\varepsilon \rangle \\
&+ e'_A e_B e_C e_D {e'_P}^2 \langle \{c(10)d(01)b(10)\}_\varepsilon | \{a(1'0)p(01')p(1'0)\}_\varepsilon \rangle \\
&+ e'_A e'_B e_C e_D {e'_P}^2 \langle \{c(10)d(01)b(1'0)\}_\varepsilon | \{a(1'0)p(01')p(1'0)\}_\varepsilon \rangle \\
&+ e'_A e_B e'_C e_D {e'_P}^2 \langle \{c(1'0)d(01)b(10)\}_\varepsilon | \{a(1'0)p(01')p(1'0)\}_\varepsilon \rangle \\
&+ e'_A e_B e_C e'_D {e'_P}^2 \langle \{c(10)d(01')b(10)\}_\varepsilon | \{a(1'0)p(01')p(1'0)\}_\varepsilon \rangle \\
&+ e'_A e'_B e'_C e_D {e'_P}^2 \langle \{c(1'0)d(01)b(1'0)\}_\varepsilon | \{a(1'0)p(01')p(1'0)\}_\varepsilon \rangle \\
&+ e'_A e_B e'_C e'_D {e'_P}^2 \langle \{c(1'0)d(01')b(10)\}_\varepsilon | \{a(1'0)p(01')p(1'0)\}_\varepsilon \rangle \\
&+ e'_A e'_B e_C e'_D {e'_P}^2 \langle \{c(10)d(01')b(1'0)\}_\varepsilon | \{a(1'0)p(01')p(1'0)\}_\varepsilon \rangle \\
&+ e'_A e'_B e'_C e'_D {e'_P}^2 \langle \{c(1'0)d(01')b(1'0)\}_\varepsilon | \{a(1'0)p(01')p(1'0)\}_\varepsilon \rangle \\
&+ e_A e'_B e_C e_D {e'_P}^2 \langle \{c(10)d(01)b(1'0)\}_\varepsilon | \{a(10)p(01')p(1'0)\}_\varepsilon \rangle \\
&+ e_A e_B e'_C e_D {e'_P}^2 \langle \{c(1'0)d(01)b(10)\}_\varepsilon | \{a(10)p(01')p(1'0)\}_\varepsilon \rangle \\
&+ e_A e_B e_C e'_D {e'_P}^2 \langle \{c(10)d(01')b(10)\}_\varepsilon | \{a(10)p(01')p(1'0)\}_\varepsilon \rangle \\
&+ e_A e'_B e'_C e_D {e'_P}^2 \langle \{c(1'0)d(01)b(1'0)\}_\varepsilon | \{a(10)p(01')p(1'0)\}_\varepsilon \rangle \\
&+ e_A e_B e'_C e'_D {e'_P}^2 \langle \{c(1'0)d(01')b(10)\}_\varepsilon | \{a(10)p(01')p(1'0)\}_\varepsilon \rangle \\
&+ e_A e'_B e_C e'_D {e'_P}^2 \langle \{c(10)d(01')b(1'0)\}_\varepsilon | \{a(10)p(01')p(1'0)\}_\varepsilon \rangle \\
&+ e_A e'_B e'_C e'_D {e'_P}^2 \langle \{c(1'0)d(01')b(1'0)\}_\varepsilon | \{a(10)p(01')p(1'0)\}_\varepsilon \rangle \qquad \text{(B.23)}
\end{aligned}
$$

$$
\begin{aligned}
\langle (C)_\varepsilon | (DBAPP)_\varepsilon \rangle^{+-} &= e_A e_B e_C e_D {e_P}^2 \langle \{c(10)\}_\varepsilon | \{d(10)b(01)a(10)p(01)p(10)\}_\varepsilon \rangle \\
&+ e'_A e_B e_C e_D {e_P}^2 \langle \{c(10)\}_\varepsilon | \{d(10)b(01)a(1'0)p(01)p(10)\}_\varepsilon \rangle \\
&+ e'_A e'_B e_C e_D {e_P}^2 \langle \{c(10)\}_\varepsilon | \{d(10)b(01')a(1'0)p(01)p(10)\}_\varepsilon \rangle \\
&+ e'_A e_B e'_C e_D {e_P}^2 \langle \{c(1'0)\}_\varepsilon | \{d(10)b(01)a(1'0)p(01)p(10)\}_\varepsilon \rangle \\
&+ e'_A e_B e_C e'_D {e_P}^2 \langle \{c(10)\}_\varepsilon | \{d(1'0)b(01)a(1'0)p(01)p(10)\}_\varepsilon \rangle \\
&+ e'_A e'_B e'_C e_D {e_P}^2 \langle \{c(1'0)\}_\varepsilon | \{d(10)b(01')a(1'0)p(01)p(10)\}_\varepsilon \rangle \\
&+ e'_A e_B e'_C e'_D {e_P}^2 \langle \{c(1'0)\}_\varepsilon | \{d(1'0)b(01)a(1'0)p(01)p(10)\}_\varepsilon \rangle \\
&+ e'_A e'_B e_C e'_D {e_P}^2 \langle \{c(10)\}_\varepsilon | \{d(1'0)b(01')a(1'0)p(01)p(10)\}_\varepsilon \rangle \\
&+ e'_A e'_B e'_C e'_D {e_P}^2 \langle \{c(1'0)\}_\varepsilon | \{d(1'0)b(01')a(1'0)p(01)p(10)\}_\varepsilon \rangle \\
&+ e_A e'_B e_C e_D {e_P}^2 \langle \{c(10)\}_\varepsilon | \{d(10)b(01')a(10)p(01)p(10)\}_\varepsilon \rangle \\
&+ e_A e_B e'_C e_D {e_P}^2 \langle \{c(1'0)\}_\varepsilon | \{d(10)b(01)a(10)p(01)p(10)\}_\varepsilon \rangle \\
&+ e_A e_B e_C e'_D {e_P}^2 \langle \{c(10)\}_\varepsilon | \{d(1'0)b(01)a(10)p(01)p(10)\}_\varepsilon \rangle \\
&+ e_A e'_B e'_C e_D {e_P}^2 \langle \{c(1'0)\}_\varepsilon | \{d(10)b(01')a(10)p(01)p(10)\}_\varepsilon \rangle \\
&+ e_A e_B e'_C e'_D {e_P}^2 \langle \{c(1'0)\}_\varepsilon | \{d(1'0)b(01)a(10)p(01)p(10)\}_\varepsilon \rangle \\
&+ e_A e'_B e_C e'_D {e_P}^2 \langle \{c(10)\}_\varepsilon | \{d(1'0)b(01')a(10)p(01)p(10)\}_\varepsilon \rangle \\
&+ e_A e'_B e'_C e'_D {e_P}^2 \langle \{c(1'0)\}_\varepsilon | \{d(1'0)b(01')a(10)p(01)p(10)\}_\varepsilon \rangle \\
&+ e_A e_B e_C e_D {e'_P}^2 \langle \{c(10)\}_\varepsilon | \{d(10)b(01)a(10)p(01')p(1'0)\}_\varepsilon \rangle \\
&+ e'_A e_B e_C e_D {e'_P}^2 \langle \{c(10)\}_\varepsilon | \{d(10)b(01)a(1'0)p(01')p(1'0)\}_\varepsilon \rangle \\
&+ e'_A e'_B e_C e_D {e'_P}^2 \langle \{c(10)\}_\varepsilon | \{d(10)b(01')a(1'0)p(01')p(1'0)\}_\varepsilon \rangle \\
&+ e'_A e_B e'_C e_D {e'_P}^2 \langle \{c(1'0)\}_\varepsilon | \{d(10)b(01)a(1'0)p(01')p(1'0)\}_\varepsilon \rangle \\
&+ e'_A e_B e_C e'_D {e'_P}^2 \langle \{c(10)\}_\varepsilon | \{d(1'0)b(01)a(1'0)p(01')p(1'0)\}_\varepsilon \rangle \\
&+ e'_A e'_B e'_C e_D {e'_P}^2 \langle \{c(1'0)\}_\varepsilon | \{d(10)b(01')a(1'0)p(01')p(1'0)\}_\varepsilon \rangle \\
&+ e'_A e_B e'_C e'_D {e'_P}^2 \langle \{c(1'0)\}_\varepsilon | \{d(1'0)b(01)a(1'0)p(01')p(1'0)\}_\varepsilon \rangle \\
&+ e'_A e'_B e_C e'_D {e'_P}^2 \langle \{c(10)\}_\varepsilon | \{d(1'0)b(01')a(1'0)p(01')p(1'0)\}_\varepsilon \rangle \\
&+ e'_A e'_B e'_C e'_D {e'_P}^2 \langle \{c(1'0)\}_\varepsilon | \{d(1'0)b(01')a(1'0)p(01')p(1'0)\}_\varepsilon \rangle \\
&+ e_A e'_B e_C e_D {e'_P}^2 \langle \{c(10)\}_\varepsilon | \{d(10)b(01')a(10)p(01')p(1'0)\}_\varepsilon \rangle \\
&+ e_A e_B e'_C e_D {e'_P}^2 \langle \{c(1'0)\}_\varepsilon | \{d(10)b(01)a(10)p(01')p(1'0)\}_\varepsilon \rangle \\
&+ e_A e_B e_C e'_D {e'_P}^2 \langle \{c(10)\}_\varepsilon | \{d(1'0)b(01)a(10)p(01')p(1'0)\}_\varepsilon \rangle \\
&+ e_A e'_B e'_C e_D {e'_P}^2 \langle \{c(1'0)\}_\varepsilon | \{d(10)b(01')a(10)p(01')p(1'0)\}_\varepsilon \rangle \\
&+ e_A e_B e'_C e'_D {e'_P}^2 \langle \{c(1'0)\}_\varepsilon | \{d(1'0)b(01)a(10)p(01')p(1'0)\}_\varepsilon \rangle \\
&+ e_A e'_B e_C e'_D {e'_P}^2 \langle \{c(10)\}_\varepsilon | \{d(1'0)b(01')a(10)p(01')p(1'0)\}_\varepsilon \rangle \\
&+ e_A e'_B e'_C e'_D {e'_P}^2 \langle \{c(1'0)\}_\varepsilon | \{d(1'0)b(01')a(10)p(01')p(1'0)\}_\varepsilon \rangle \qquad \text{(B.24)}
\end{aligned}
$$

$$
\begin{aligned}
\langle (CPP)_\varepsilon|(DBA)_\varepsilon\rangle^{+-} &= e_A e_B e_C e_D {e_P}^2 \langle \{c(10)p(01)p(10)\}_\varepsilon|\{d(10)b(01)a(10)\}_\varepsilon\rangle \\
&+ e'_A e_B e_C e_D {e_P}^2 \langle \{c(10)p(01)p(10)\}_\varepsilon|\{d(10)b(01)a(1'0)\}_\varepsilon\rangle \\
&+ e'_A e'_B e_C e_D {e_P}^2 \langle \{c(10)p(01)p(10)\}_\varepsilon|\{d(10)b(01')a(1'0)\}_\varepsilon\rangle \\
&+ e'_A e_B e'_C e_D {e_P}^2 \langle \{c(1'0)p(01)p(10)\}_\varepsilon|\{d(10)b(01)a(1'0)\}_\varepsilon\rangle \\
&+ e'_A e_B e_C e'_D {e_P}^2 \langle \{c(10)p(01)p(10)\}_\varepsilon|\{d(1'0)b(01)a(1'0)\}_\varepsilon\rangle \\
&+ e'_A e'_B e'_C e_D {e_P}^2 \langle \{c(1'0)p(01)p(10)\}_\varepsilon|\{d(10)b(01')a(1'0)\}_\varepsilon\rangle \\
&+ e'_A e_B e'_C e'_D {e_P}^2 \langle \{c(1'0)p(01)p(10)\}_\varepsilon|\{d(1'0)b(01)a(1'0)\}_\varepsilon\rangle \\
&+ e'_A e'_B e_C e'_D {e_P}^2 \langle \{c(10)p(01)p(10)\}_\varepsilon|\{d(1'0)b(01')a(1'0)\}_\varepsilon\rangle \\
&+ e'_A e'_B e'_C e'_D {e_P}^2 \langle \{c(1'0)p(01)p(10)\}_\varepsilon|\{d(1'0)b(01')a(1'0)\}_\varepsilon\rangle \\
&+ e_A e'_B e_C e_D {e_P}^2 \langle \{c(10)p(01)p(10)\}_\varepsilon|\{d(10)b(01')a(10)\}_\varepsilon\rangle \\
&+ e_A e_B e'_C e_D {e_P}^2 \langle \{c(1'0)p(01)p(10)\}_\varepsilon|\{d(10)b(01)a(10)\}_\varepsilon\rangle \\
&+ e_A e_B e_C e'_D {e_P}^2 \langle \{c(10)p(01)p(10)\}_\varepsilon|\{d(1'0)b(01)a(10)\}_\varepsilon\rangle \\
&+ e_A e'_B e'_C e_D {e_P}^2 \langle \{c(1'0)p(01)p(10)\}_\varepsilon|\{d(10)b(01')a(10)\}_\varepsilon\rangle \\
&+ e_A e_B e'_C e'_D {e_P}^2 \langle \{c(1'0)p(01)p(10)\}_\varepsilon|\{d(1'0)b(01)a(10)\}_\varepsilon\rangle \\
&+ e_A e'_B e_C e'_D {e_P}^2 \langle \{c(10)p(01)p(10)\}_\varepsilon|\{d(1'0)b(01')a(10)\}_\varepsilon\rangle \\
&+ e_A e'_B e'_C e'_D {e_P}^2 \langle \{c(1'0)p(01)p(10)\}_\varepsilon|\{d(1'0)b(01')a(10)\}_\varepsilon\rangle \\
&+ e_A e_B e_C e_D {e'_P}^2 \langle \{c(10)p(01')p(1'0)\}_\varepsilon|\{d(10)b(01)a(10)\}_\varepsilon\rangle \\
&+ e'_A e_B e_C e_D {e'_P}^2 \langle \{c(10)p(01')p(1'0)\}_\varepsilon|\{d(10)b(01)a(1'0)\}_\varepsilon\rangle \\
&+ e'_A e'_B e_C e_D {e'_P}^2 \langle \{c(10)p(01')p(1'0)\}_\varepsilon|\{d(10)b(01')a(1'0)\}_\varepsilon\rangle \\
&+ e'_A e_B e'_C e_D {e'_P}^2 \langle \{c(1'0)p(01')p(1'0)\}_\varepsilon|\{d(10)b(01)a(1'0)\}_\varepsilon\rangle \\
&+ e'_A e_B e_C e'_D {e'_P}^2 \langle \{c(10)p(01')p(1'0)\}_\varepsilon|\{d(1'0)b(01)a(1'0)\}_\varepsilon\rangle \\
&+ e'_A e'_B e'_C e_D {e'_P}^2 \langle \{c(1'0)p(01')p(1'0)\}_\varepsilon|\{d(10)b(01')a(1'0)\}_\varepsilon\rangle \\
&+ e'_A e_B e'_C e'_D {e'_P}^2 \langle \{c(1'0)p(01')p(1'0)\}_\varepsilon|\{d(1'0)b(01)a(1'0)\}_\varepsilon\rangle \\
&+ e'_A e'_B e_C e'_D {e'_P}^2 \langle \{c(10)p(01')p(1'0)\}_\varepsilon|\{d(1'0)b(01')a(1'0)\}_\varepsilon\rangle \\
&+ e'_A e'_B e'_C e'_D {e'_P}^2 \langle \{c(1'0)p(01')p(1'0)\}_\varepsilon|\{d(1'0)b(01')a(1'0)\}_\varepsilon\rangle \\
&+ e_A e'_B e_C e_D {e'_P}^2 \langle \{c(10)p(01')p(1'0)\}_\varepsilon|\{d(10)b(01')a(10)\}_\varepsilon\rangle \\
&+ e_A e_B e'_C e_D {e'_P}^2 \langle \{c(1'0)p(01')p(1'0)\}_\varepsilon|\{d(10)b(01)a(10)\}_\varepsilon\rangle \\
&+ e_A e_B e_C e'_D {e'_P}^2 \langle \{c(10)p(01')p(1'0)\}_\varepsilon|\{d(1'0)b(01)a(10)\}_\varepsilon\rangle \\
&+ e_A e'_B e'_C e_D {e'_P}^2 \langle \{c(1'0)p(01')p(1'0)\}_\varepsilon|\{d(10)b(01')a(10)\}_\varepsilon\rangle \\
&+ e_A e_B e'_C e'_D {e'_P}^2 \langle \{c(1'0)p(01')p(1'0)\}_\varepsilon|\{d(1'0)b(01)a(10)\}_\varepsilon\rangle \\
&+ e_A e'_B e_C e'_D {e'_P}^2 \langle \{c(10)p(01')p(1'0)\}_\varepsilon|\{d(1'0)b(01')a(10)\}_\varepsilon\rangle \\
&+ e_A e'_B e'_C e'_D {e'_P}^2 \langle \{c(1'0)p(01')p(1'0)\}_\varepsilon|\{d(1'0)b(01')a(10)\}_\varepsilon\rangle \qquad \text{(B.25)}
\end{aligned}
$$

$$
\begin{aligned}
\langle (CABPP)_\varepsilon|(D)_\varepsilon\rangle^{+-} &= e_Ae_Be_Ce_De_P{}^2\langle\{c(10)a(01)b(10)p(01)p(10)\}_\varepsilon|\{d(10)\}_\varepsilon\rangle \\
&+ e'_Ae'_Be_Ce_De_P{}^2\langle\{c(10)a(01')b(1'0)p(01)p(10)\}_\varepsilon|\{d(10)\}_\varepsilon\rangle \\
&+ e'_Ae'_Be'_Ce_De_P{}^2\langle\{c(1'0)a(01')b(1'0)p(01)p(10)\}_\varepsilon|\{d(10)\}_\varepsilon\rangle \\
&+ e'_Ae'_Be_Ce'_De_P{}^2\langle\{c(10)a(01')b(1'0)p(01)p(10)\}_\varepsilon|\{d(1'0)\}_\varepsilon\rangle \\
&+ e'_Ae'_Be'_Ce'_De_P{}^2\langle\{c(1'0)a(01')b(1'0)p(01)p(10)\}_\varepsilon|\{d(1'0)\}_\varepsilon\rangle \\
&+ e_Ae_Be'_Ce_De_P{}^2\langle\{c(1'0)a(01)b(10)p(01)p(10)\}_\varepsilon|\{d(10)\}_\varepsilon\rangle \\
&+ e_Ae_Be_Ce'_De_P{}^2\langle\{c(10)a(01)b(10)p(01)p(10)\}_\varepsilon|\{d(1'0)\}_\varepsilon\rangle \\
&+ e_Ae_Be'_Ce'_De_P{}^2\langle\{c(1'0)a(01)b(10)p(01)p(10)\}_\varepsilon|\{d(1'0)\}_\varepsilon\rangle \\
&+ e_Ae_Be_Ce_De'_P{}^2\langle\{c(10)a(01)b(10)p(01')p(1'0)\}_\varepsilon|\{d(10)\}_\varepsilon\rangle \\
&+ e'_Ae'_Be_Ce_De'_P{}^2\langle\{c(10)a(01')b(1'0)p(01')p(1'0)\}_\varepsilon|\{d(10)\}_\varepsilon\rangle \\
&+ e'_Ae'_Be'_Ce_De'_P{}^2\langle\{c(1'0)a(01')b(1'0)p(01')p(1'0)\}_\varepsilon|\{d(10)\}_\varepsilon\rangle \\
&+ e'_Ae'_Be_Ce'_De'_P{}^2\langle\{c(10)a(01')b(1'0)p(01')p(1'0)\}_\varepsilon|\{d(1'0)\}_\varepsilon\rangle \\
&+ e'_Ae'_Be'_Ce'_De'_P{}^2\langle\{c(1'0)a(01')b(1'0)p(01')p(1'0)\}_\varepsilon|\{d(1'0)\}_\varepsilon\rangle \\
&+ e_Ae_Be'_Ce_De'_P{}^2\langle\{c(1'0)a(01)b(10)p(01')p(1'0)\}_\varepsilon|\{d(10)\}_\varepsilon\rangle \\
&+ e_Ae_Be_Ce'_De'_P{}^2\langle\{c(10)a(01)b(10)p(01')p(1'0)\}_\varepsilon|\{d(1'0)\}_\varepsilon\rangle \\
&+ e_Ae_Be'_Ce'_De'_P{}^2\langle\{c(1'0)a(01)b(10)p(01')p(1'0)\}_\varepsilon|\{d(1'0)\}_\varepsilon\rangle \qquad \text{(B.26)}
\end{aligned}
$$

$$
\begin{aligned}
\langle (CAB)_\varepsilon|(DPP)_\varepsilon\rangle^{+-} &= e_Ae_Be_Ce_De_P{}^2\langle\{c(10)a(01)b(10)\}_\varepsilon|\{d(10)p(01)p(10)\}_\varepsilon\rangle \\
&+ e'_Ae'_Be_Ce_De_P{}^2\langle\{c(10)a(01')b(1'0)\}_\varepsilon|\{d(10)p(01)p(10)\}_\varepsilon\rangle \\
&+ e'_Ae'_Be'_Ce_De_P{}^2\langle\{c(1'0)a(01')b(1'0)\}_\varepsilon|\{d(10)p(01)p(10)\}_\varepsilon\rangle \\
&+ e'_Ae'_Be_Ce'_De_P{}^2\langle\{c(10)a(01')b(1'0)\}_\varepsilon|\{d(1'0)p(01)p(10)\}_\varepsilon\rangle \\
&+ e'_Ae'_Be'_Ce'_De_P{}^2\langle\{c(1'0)a(01')b(1'0)\}_\varepsilon|\{d(1'0)p(01)p(10)\}_\varepsilon\rangle \\
&+ e_Ae_Be'_Ce_De_P{}^2\langle\{c(1'0)a(01)b(10)\}_\varepsilon|\{d(10)p(01)p(10)\}_\varepsilon\rangle \\
&+ e_Ae_Be_Ce'_De_P{}^2\langle\{c(10)a(01)b(10)\}_\varepsilon|\{d(1'0)p(01)p(10)\}_\varepsilon\rangle \\
&+ e_Ae_Be'_Ce'_De_P{}^2\langle\{c(1'0)a(01)b(10)\}_\varepsilon|\{d(1'0)p(01)p(10)\}_\varepsilon\rangle \\
&+ e_Ae_Be_Ce_De'_P{}^2\langle\{c(10)a(01)b(10)\}_\varepsilon|\{d(10)p(01')p(1'0)\}_\varepsilon\rangle \\
&+ e'_Ae'_Be_Ce_De'_P{}^2\langle\{c(10)a(01')b(1'0)\}_\varepsilon|\{d(10)p(01')p(1'0)\}_\varepsilon\rangle \\
&+ e'_Ae'_Be'_Ce_De'_P{}^2\langle\{c(1'0)a(01')b(1'0)\}_\varepsilon|\{d(10)p(01')p(1'0)\}_\varepsilon\rangle \\
&+ e'_Ae'_Be_Ce'_De'_P{}^2\langle\{c(10)a(01')b(1'0)\}_\varepsilon|\{d(1'0)p(01')p(1'0)\}_\varepsilon\rangle \\
&+ e'_Ae'_Be'_Ce'_De'_P{}^2\langle\{c(1'0)a(01')b(1'0)\}_\varepsilon|\{d(1'0)p(01')p(1'0)\}_\varepsilon\rangle \\
&+ e_Ae_Be'_Ce_De'_P{}^2\langle\{c(1'0)a(01)b(10)\}_\varepsilon|\{d(10)p(01')p(1'0)\}_\varepsilon\rangle \\
&+ e_Ae_Be_Ce'_De'_P{}^2\langle\{c(10)a(01)b(10)\}_\varepsilon|\{d(1'0)p(01')p(1'0)\}_\varepsilon\rangle \\
&+ e_Ae_Be'_Ce'_De'_P{}^2\langle\{c(1'0)a(01)b(10)\}_\varepsilon|\{d(1'0)p(01')p(1'0)\}_\varepsilon\rangle \qquad \text{(B.27)}
\end{aligned}
$$

## APPENDIX C: PUMP-PROBE LIMIT OF NL-WPI SIGNAL FOLLOWING CONTROL PULSE

Here we provide detailed expressions for the pump-probe limit to the nonlinear wave-packet interferometry signal from an energy transfer complex subjected to a pre-resonant control pulse (derived in *Using Wave-Packet Interferometry to Monitor the External Vibrational Control of Electronic Excitation Transfer*, Section 4). The signal is composed of three components

$$\Theta_\varepsilon = \Theta_\varepsilon^{GSB} + \Theta_\varepsilon^{ESA} + \Theta_\varepsilon^{SE} \ . \tag{C.1}$$

$$\Theta_\varepsilon^{GSB} = 8\mathrm{Re}\left(\langle (C)_\varepsilon | (DBAPP)_\varepsilon \rangle^{+-} + \langle (CPP)_\varepsilon | (DBA)_\varepsilon \rangle^{+-}\right) \tag{C.2}$$

$$\Theta_\varepsilon^{ESA} = 8\mathrm{Re}\left(\langle (B)_\varepsilon | (CDAPP)_\varepsilon \rangle^{+-} + \langle (BPP)_\varepsilon | (CDA)_\varepsilon \rangle^{+-}\right) \tag{C.3}$$

$$\Theta_\varepsilon^{SE} = 8\mathrm{Re}\left(\langle (B)_\varepsilon | (DCAPP)_\varepsilon \rangle^{+-} + \langle (BPP)_\varepsilon | (DCA)_\varepsilon \rangle^{+-}\right) \tag{C.4}$$

In the ground-state bleach terms, the $A$-$B$ pulse acts twice to return amplitude to the ground electronic state. In the remaining terms, the $C$-$D$ pulse acts twice on a wave packet already prepared in an excited electronic state by the $A$-$B$ pulse ($\varepsilon \leftarrow 2 \leftarrow \varepsilon$ for excited-state absorption and $\varepsilon \leftarrow 0 \leftarrow \varepsilon$ for stimulated emission).

In the GSB terms the contributing amplitudes are in the electronic-ground state during the inter-pulse delay $t_{CA}$. Since we ignore energy transfer on the short timescale of interaction with the pulse, these contributions are therefore insensitive to EET (i.e. they do not change in going to the $J \rightarrow 0$ limit). In the SE and ESA terms, however, both the bra and ket evolve in the one-exciton manifold during $t_{CA}$. Accounting for the possibility of energy transfer during this interval, these terms therefore contain four times as many contributions as the GSB terms.

$$
\begin{aligned}
\langle (C)_\varepsilon | (DBAPP)_\varepsilon \rangle^{+-} &= e_A e_B e_C e_D {e_P}^2 \langle \{c(10)\}_\varepsilon | \{c(10)a(01)a(10)p(01)p(10)\}_\varepsilon \rangle \\
&+ e'_A e'_B e_C e_D {e_P}^2 \langle \{c(10)\}_\varepsilon | \{c(10)a(01')a(1'0)p(01)p(10)\}_\varepsilon \rangle \\
&+ e'_A e'_B e'_C e'_D {e_P}^2 \langle \{c(1'0)\}_\varepsilon | \{c(1'0)a(01')a(1'0)p(01)p(10)\}_\varepsilon \rangle \\
&+ e_A e_B e'_C e'_D {e_P}^2 \langle \{c(1'0)\}_\varepsilon | \{c(1'0)a(01)a(10)p(01)p(10)\}_\varepsilon \rangle \\
&+ e_A e_B e_C e_D {e'_P}^2 \langle \{c(10)\}_\varepsilon | \{c(10)a(01)a(10)p(01')p(1'0)\}_\varepsilon \rangle \\
&+ e'_A e'_B e_C e_D {e'_P}^2 \langle \{c(10)\}_\varepsilon | \{c(10)a(01')a(1'0)p(01')p(1'0)\}_\varepsilon \rangle \\
&+ e'_A e'_B e'_C e'_D {e'_P}^2 \langle \{c(1'0)\}_\varepsilon | \{c(1'0)a(01')a(1'0)p(01')p(1'0)\}_\varepsilon \rangle \\
&+ e_A e_B e'_C e'_D {e'_P}^2 \langle \{c(1'0)\}_\varepsilon | \{c(1'0)a(01)a(10)p(01')p(1'0)\}_\varepsilon \rangle \qquad \text{(C.5)}
\end{aligned}
$$

$$
\begin{aligned}
\langle (CPP)_\varepsilon | (DBA)_\varepsilon \rangle^{+-} &= e_A e_B e_C e_D {e_P}^2 \langle \{c(10)p(01)p(10)\}_\varepsilon | \{c(10)a(01)a(10)\}_\varepsilon \rangle \\
&+ e'_A e'_B e_C e_D {e_P}^2 \langle \{c(10)p(01)p(10)\}_\varepsilon | \{c(10)a(01')a(1'0)\}_\varepsilon \rangle \\
&+ e'_A e'_B e'_C e'_D {e_P}^2 \langle \{c(1'0)p(01)p(10)\}_\varepsilon | \{c(1'0)a(01')a(1'0)\}_\varepsilon \rangle \\
&+ e_A e_B e'_C e'_D {e_P}^2 \langle \{c(1'0)p(01)p(10)\}_\varepsilon | \{c(1'0)a(01)a(10)\}_\varepsilon \rangle \\
&+ e_A e_B e_C e_D {e'_P}^2 \langle \{c(10)p(01')p(1'0)\}_\varepsilon | \{c(10)a(01)a(10)\}_\varepsilon \rangle \\
&+ e'_A e'_B e_C e_D {e'_P}^2 \langle \{c(10)p(01')p(1'0)\}_\varepsilon | \{c(10)a(01')a(1'0)\}_\varepsilon \rangle \\
&+ e'_A e'_B e'_C e'_D {e'_P}^2 \langle \{c(1'0)p(01')p(1'0)\}_\varepsilon | \{c(1'0)a(01')a(1'0)\}_\varepsilon \rangle \\
&+ e_A e_B e'_C e'_D {e'_P}^2 \langle \{c(1'0)p(01')p(1'0)\}_\varepsilon | \{c(1'0)a(01)a(10)\}_\varepsilon \rangle \qquad \text{(C.6)}
\end{aligned}
$$

$$
\begin{aligned}
\langle (B)_\varepsilon | (CDAPP)_\varepsilon \rangle^{+-} = {} & e_A e_B e_C e_D {e_P}^2 \langle \{a(10)\}_\varepsilon | \{c(1'2)c(21')a(10)p(01)p(10)\}_\varepsilon \rangle \\
& + e'_A e_B e_C e_D {e_P}^2 \langle \{a(10)\}_\varepsilon | \{c(1'2)c(21')a(1'0)p(01)p(10)\}_\varepsilon \rangle \\
& + e'_A e'_B e_C e_D {e_P}^2 \langle \{a(1'0)\}_\varepsilon | \{c(1'2)c(21')a(1'0)p(01)p(10)\}_\varepsilon \rangle \\
& + e'_A e_B e'_C e_D {e_P}^2 \langle \{a(10)\}_\varepsilon | \{c(12)c(21')a(1'0)p(01)p(10)\}_\varepsilon \rangle \\
& + e'_A e_B e_C e'_D {e_P}^2 \langle \{a(10)\}_\varepsilon | \{c(1'2)c(21)a(1'0)p(01)p(10)\}_\varepsilon \rangle \\
& + e'_A e'_B e'_C e_D {e_P}^2 \langle \{a(1'0)\}_\varepsilon | \{c(12)c(21')a(1'0)p(01)p(10)\}_\varepsilon \rangle \\
& + e'_A e_B e'_C e'_D {e_P}^2 \langle \{a(10)\}_\varepsilon | \{c(12)c(21)a(1'0)p(01)p(10)\}_\varepsilon \rangle \\
& + e'_A e'_B e_C e'_D {e_P}^2 \langle \{a(1'0)\}_\varepsilon | \{c(1'2)c(21)a(1'0)p(01)p(10)\}_\varepsilon \rangle \\
& + e'_A e'_B e'_C e'_D {e_P}^2 \langle \{a(1'0)\}_\varepsilon | \{c(12)c(21)a(1'0)p(01)p(10)\}_\varepsilon \rangle \\
& + e_A e'_B e_C e_D {e_P}^2 \langle \{a(1'0)\}_\varepsilon | \{c(1'2)c(21')a(10)p(01)p(10)\}_\varepsilon \rangle \\
& + e_A e_B e'_C e_D {e_P}^2 \langle \{a(10)\}_\varepsilon | \{c(12)c(21')a(10)p(01)p(10)\}_\varepsilon \rangle \\
& + e_A e_B e_C e'_D {e_P}^2 \langle \{a(10)\}_\varepsilon | \{c(1'2)c(21)a(10)p(01)p(10)\}_\varepsilon \rangle \\
& + e_A e'_B e'_C e_D {e_P}^2 \langle \{a(1'0)\}_\varepsilon | \{c(12)c(21')a(10)p(01)p(10)\}_\varepsilon \rangle \\
& + e_A e_B e'_C e'_D {e_P}^2 \langle \{a(10)\}_\varepsilon | \{c(12)c(21)a(10)p(01)p(10)\}_\varepsilon \rangle \\
& + e_A e'_B e_C e'_D {e_P}^2 \langle \{a(1'0)\}_\varepsilon | \{c(1'2)c(21)a(10)p(01)p(10)\}_\varepsilon \rangle \\
& + e_A e'_B e'_C e'_D {e_P}^2 \langle \{a(1'0)\}_\varepsilon | \{c(12)c(21)a(10)p(01)p(10)\}_\varepsilon \rangle \\
& + e_A e_B e_C e_D {e'_P}^2 \langle \{a(10)\}_\varepsilon | \{c(1'2)c(21')a(10)p(01')p(1'0)\}_\varepsilon \rangle \\
& + e'_A e_B e_C e_D {e'_P}^2 \langle \{a(10)\}_\varepsilon | \{c(1'2)c(21')a(1'0)p(01')p(1'0)\}_\varepsilon \rangle \\
& + e'_A e'_B e_C e_D {e'_P}^2 \langle \{a(1'0)\}_\varepsilon | \{c(1'2)c(21')a(1'0)p(01')p(1'0)\}_\varepsilon \rangle \\
& + e'_A e_B e'_C e_D {e'_P}^2 \langle \{a(10)\}_\varepsilon | \{c(12)c(21')a(1'0)p(01')p(1'0)\}_\varepsilon \rangle \\
& + e'_A e_B e_C e'_D {e'_P}^2 \langle \{a(10)\}_\varepsilon | \{c(1'2)c(21)a(1'0)p(01')p(1'0)\}_\varepsilon \rangle \\
& + e'_A e'_B e'_C e_D {e'_P}^2 \langle \{a(1'0)\}_\varepsilon | \{c(12)c(21')a(1'0)p(01')p(1'0)\}_\varepsilon \rangle \\
& + e'_A e_B e'_C e'_D {e'_P}^2 \langle \{a(10)\}_\varepsilon | \{c(12)c(21)a(1'0)p(01')p(1'0)\}_\varepsilon \rangle \\
& + e'_A e'_B e_C e'_D {e'_P}^2 \langle \{a(1'0)\}_\varepsilon | \{c(1'2)c(21)a(1'0)p(01')p(1'0)\}_\varepsilon \rangle \\
& + e'_A e'_B e'_C e'_D {e'_P}^2 \langle \{a(1'0)\}_\varepsilon | \{c(12)c(21)a(1'0)p(01')p(1'0)\}_\varepsilon \rangle \\
& + e_A e'_B e_C e_D {e'_P}^2 \langle \{a(1'0)\}_\varepsilon | \{c(1'2)c(21')a(10)p(01')p(1'0)\}_\varepsilon \rangle \\
& + e_A e_B e'_C e_D {e'_P}^2 \langle \{a(10)\}_\varepsilon | \{c(12)c(21')a(10)p(01')p(1'0)\}_\varepsilon \rangle \\
& + e_A e_B e_C e'_D {e'_P}^2 \langle \{a(10)\}_\varepsilon | \{c(1'2)c(21)a(10)p(01')p(1'0)\}_\varepsilon \rangle \\
& + e_A e'_B e'_C e_D {e'_P}^2 \langle \{a(1'0)\}_\varepsilon | \{c(12)c(21')a(10)p(01')p(1'0)\}_\varepsilon \rangle \\
& + e_A e_B e'_C e'_D {e'_P}^2 \langle \{a(10)\}_\varepsilon | \{c(12)c(21)a(10)p(01')p(1'0)\}_\varepsilon \rangle \\
& + e_A e'_B e_C e'_D {e'_P}^2 \langle \{a(1'0)\}_\varepsilon | \{c(1'2)c(21)a(10)p(01')p(1'0)\}_\varepsilon \rangle \\
& + e_A e'_B e'_C e'_D {e'_P}^2 \langle \{a(1'0)\}_\varepsilon | \{c(12)c(21)a(10)p(01')p(1'0)\}_\varepsilon \rangle \qquad \text{(C.7)}
\end{aligned}
$$

$$
\begin{aligned}
\langle (BPP)_\varepsilon | (CDA)_\varepsilon \rangle^{+-} = & \; e_A e_B e_C e_D {e_P}^2 \langle \{a(10)p(01)p(10)\}_\varepsilon | \{c(1'2)c(21')a(10)\}_\varepsilon \rangle \\
& + e'_A e_B e_C e_D {e_P}^2 \langle \{a(10)p(01)p(10)\}_\varepsilon | \{c(1'2)c(21')a(1'0)\}_\varepsilon \rangle \\
& + e'_A e'_B e_C e_D {e_P}^2 \langle \{a(1'0)p(01)p(10)\}_\varepsilon | \{c(1'2)c(21')a(1'0)\}_\varepsilon \rangle \\
& + e'_A e_B e'_C e_D {e_P}^2 \langle \{a(10)p(01)p(10)\}_\varepsilon | \{c(12)c(21')a(1'0)\}_\varepsilon \rangle \\
& + e'_A e_B e_C e'_D {e_P}^2 \langle \{a(10)p(01)p(10)\}_\varepsilon | \{c(1'2)c(21)a(1'0)\}_\varepsilon \rangle \\
& + e'_A e'_B e'_C e_D {e_P}^2 \langle \{a(1'0)p(01)p(10)\}_\varepsilon | \{c(12)c(21')a(1'0)\}_\varepsilon \rangle \\
& + e'_A e_B e'_C e'_D {e_P}^2 \langle \{a(10)p(01)p(10)\}_\varepsilon | \{c(12)c(21)a(1'0)\}_\varepsilon \rangle \\
& + e'_A e'_B e_C e'_D {e_P}^2 \langle \{a(1'0)p(01)p(10)\}_\varepsilon | \{c(1'2)c(21)a(1'0)\}_\varepsilon \rangle \\
& + e'_A e'_B e'_C e'_D {e_P}^2 \langle \{a(1'0)p(01)p(10)\}_\varepsilon | \{c(12)c(21)a(1'0)\}_\varepsilon \rangle \\
& + e_A e'_B e_C e_D {e_P}^2 \langle \{a(1'0)p(01)p(10)\}_\varepsilon | \{c(1'2)c(21')a(10)\}_\varepsilon \rangle \\
& + e_A e_B e'_C e_D {e_P}^2 \langle \{a(10)p(01)p(10)\}_\varepsilon | \{c(12)c(21')a(10)\}_\varepsilon \rangle \\
& + e_A e_B e_C e'_D {e_P}^2 \langle \{a(10)p(01)p(10)\}_\varepsilon | \{c(1'2)c(21)a(10)\}_\varepsilon \rangle \\
& + e_A e'_B e'_C e_D {e_P}^2 \langle \{a(1'0)p(01)p(10)\}_\varepsilon | \{c(12)c(21')a(10)\}_\varepsilon \rangle \\
& + e_A e_B e'_C e'_D {e_P}^2 \langle \{a(10)p(01)p(10)\}_\varepsilon | \{c(12)c(21)a(10)\}_\varepsilon \rangle \\
& + e_A e'_B e_C e'_D {e_P}^2 \langle \{a(1'0)p(01)p(10)\}_\varepsilon | \{c(1'2)c(21)a(10)\}_\varepsilon \rangle \\
& + e_A e'_B e'_C e'_D {e_P}^2 \langle \{a(1'0)p(01)p(10)\}_\varepsilon | \{c(12)c(21)a(10)\}_\varepsilon \rangle \\
& + e_A e_B e_C e_D {e'_P}^2 \langle \{a(10)p(01')p(1'0)\}_\varepsilon | \{c(1'2)c(21')a(10)\}_\varepsilon \rangle \\
& + e'_A e_B e_C e_D {e'_P}^2 \langle \{a(10)p(01')p(1'0)\}_\varepsilon | \{c(1'2)c(21')a(1'0)\}_\varepsilon \rangle \\
& + e'_A e'_B e_C e_D {e'_P}^2 \langle \{a(1'0)p(01')p(1'0)\}_\varepsilon | \{c(1'2)c(21')a(1'0)\}_\varepsilon \rangle \\
& + e'_A e_B e'_C e_D {e'_P}^2 \langle \{a(10)p(01')p(1'0)\}_\varepsilon | \{c(12)c(21')a(1'0)\}_\varepsilon \rangle \\
& + e'_A e_B e_C e'_D {e'_P}^2 \langle \{a(10)p(01')p(1'0)\}_\varepsilon | \{c(1'2)c(21)a(1'0)\}_\varepsilon \rangle \\
& + e'_A e'_B e'_C e_D {e'_P}^2 \langle \{a(1'0)p(01')p(1'0)\}_\varepsilon | \{c(12)c(21')a(1'0)\}_\varepsilon \rangle \\
& + e'_A e_B e'_C e'_D {e'_P}^2 \langle \{a(10)p(01')p(1'0)\}_\varepsilon | \{c(12)c(21)a(1'0)\}_\varepsilon \rangle \\
& + e'_A e'_B e_C e'_D {e'_P}^2 \langle \{a(1'0)p(01')p(1'0)\}_\varepsilon | \{c(1'2)c(21)a(1'0)\}_\varepsilon \rangle \\
& + e'_A e'_B e'_C e'_D {e'_P}^2 \langle \{a(1'0)p(01')p(1'0)\}_\varepsilon | \{c(12)c(21)a(1'0)\}_\varepsilon \rangle \\
& + e_A e'_B e_C e_D {e'_P}^2 \langle \{a(1'0)p(01')p(1'0)\}_\varepsilon | \{c(1'2)c(21')a(10)\}_\varepsilon \rangle \\
& + e_A e_B e'_C e_D {e'_P}^2 \langle \{a(10)p(01')p(1'0)\}_\varepsilon | \{c(12)c(21')a(10)\}_\varepsilon \rangle \\
& + e_A e_B e_C e'_D {e'_P}^2 \langle \{a(10)p(01')p(1'0)\}_\varepsilon | \{c(1'2)c(21)a(10)\}_\varepsilon \rangle \\
& + e_A e'_B e'_C e_D {e'_P}^2 \langle \{a(1'0)p(01')p(1'0)\}_\varepsilon | \{c(12)c(21')a(10)\}_\varepsilon \rangle \\
& + e_A e_B e'_C e'_D {e'_P}^2 \langle \{a(10)p(01')p(1'0)\}_\varepsilon | \{c(12)c(21)a(10)\}_\varepsilon \rangle \\
& + e_A e'_B e_C e'_D {e'_P}^2 \langle \{a(1'0)p(01')p(1'0)\}_\varepsilon | \{c(1'2)c(21)a(10)\}_\varepsilon \rangle \\
& + e_A e'_B e'_C e'_D {e'_P}^2 \langle \{a(1'0)p(01')p(1'0)\}_\varepsilon | \{c(12)c(21)a(10)\}_\varepsilon \rangle
\end{aligned} \tag{C.8}
$$

$$
\begin{aligned}
\langle (B)_\varepsilon | (DCAPP)_\varepsilon \rangle^{+-} = \; & e_A e_B e_C e_D {e_P}^2 \langle \{a(10)\}_\varepsilon | \{c(10)c(01)a(10)p(01)p(10)\}_\varepsilon \rangle \\
& + e'_A e_B e_C e_D {e_P}^2 \langle \{a(10)\}_\varepsilon | \{c(10)c(01)a(1'0)p(01)p(10)\}_\varepsilon \rangle \\
& + e'_A e'_B e_C e_D {e_P}^2 \langle \{a(1'0)\}_\varepsilon | \{c(10)c(01)a(1'0)p(01)p(10)\}_\varepsilon \rangle \\
& + e'_A e_B e'_C e_D {e_P}^2 \langle \{a(10)\}_\varepsilon | \{c(10)c(01')a(1'0)p(01)p(10)\}_\varepsilon \rangle \\
& + e'_A e_B e_C e'_D {e_P}^2 \langle \{a(10)\}_\varepsilon | \{c(1'0)c(01)a(1'0)p(01)p(10)\}_\varepsilon \rangle \\
& + e'_A e'_B e'_C e_D {e_P}^2 \langle \{a(1'0)\}_\varepsilon | \{c(10)c(01')a(1'0)p(01)p(10)\}_\varepsilon \rangle \\
& + e'_A e_B e'_C e'_D {e_P}^2 \langle \{a(10)\}_\varepsilon | \{c(1'0)c(01')a(1'0)p(01)p(10)\}_\varepsilon \rangle \\
& + e'_A e'_B e_C e'_D {e_P}^2 \langle \{a(1'0)\}_\varepsilon | \{c(1'0)c(01)a(1'0)p(01)p(10)\}_\varepsilon \rangle \\
& + e'_A e'_B e'_C e'_D {e_P}^2 \langle \{a(1'0)\}_\varepsilon | \{c(1'0)c(01')a(1'0)p(01)p(10)\}_\varepsilon \rangle \\
& + e_A e'_B e_C e_D {e_P}^2 \langle \{a(1'0)\}_\varepsilon | \{c(10)c(01)a(10)p(01)p(10)\}_\varepsilon \rangle \\
& + e_A e_B e'_C e_D {e_P}^2 \langle \{a(10)\}_\varepsilon | \{c(10)c(01')a(10)p(01)p(10)\}_\varepsilon \rangle \\
& + e_A e_B e_C e'_D {e_P}^2 \langle \{a(10)\}_\varepsilon | \{c(1'0)c(01)a(10)p(01)p(10)\}_\varepsilon \rangle \\
& + e_A e'_B e'_C e_D {e_P}^2 \langle \{a(1'0)\}_\varepsilon | \{c(10)c(01')a(10)p(01)p(10)\}_\varepsilon \rangle \\
& + e_A e_B e'_C e'_D {e_P}^2 \langle \{a(10)\}_\varepsilon | \{c(1'0)c(01')a(10)p(01)p(10)\}_\varepsilon \rangle \\
& + e_A e'_B e_C e'_D {e_P}^2 \langle \{a(1'0)\}_\varepsilon | \{c(1'0)c(01)a(10)p(01)p(10)\}_\varepsilon \rangle \\
& + e_A e'_B e'_C e'_D {e_P}^2 \langle \{a(1'0)\}_\varepsilon | \{c(1'0)c(01')a(10)p(01)p(10)\}_\varepsilon \rangle \\
& + e_A e_B e_C e_D {e'_P}^2 \langle \{a(10)\}_\varepsilon | \{c(10)c(01)a(10)p(01')p(1'0)\}_\varepsilon \rangle \\
& + e'_A e_B e_C e_D {e'_P}^2 \langle \{a(10)\}_\varepsilon | \{c(10)c(01)a(1'0)p(01')p(1'0)\}_\varepsilon \rangle \\
& + e'_A e'_B e_C e_D {e'_P}^2 \langle \{a(1'0)\}_\varepsilon | \{c(10)c(01)a(1'0)p(01')p(1'0)\}_\varepsilon \rangle \\
& + e'_A e_B e'_C e_D {e'_P}^2 \langle \{a(10)\}_\varepsilon | \{c(10)c(01')a(1'0)p(01')p(1'0)\}_\varepsilon \rangle \\
& + e'_A e_B e_C e'_D {e'_P}^2 \langle \{a(10)\}_\varepsilon | \{c(1'0)c(01)a(1'0)p(01')p(1'0)\}_\varepsilon \rangle \\
& + e'_A e'_B e'_C e_D {e'_P}^2 \langle \{a(1'0)\}_\varepsilon | \{c(10)c(01')a(1'0)p(01')p(1'0)\}_\varepsilon \rangle \\
& + e'_A e_B e'_C e'_D {e'_P}^2 \langle \{a(10)\}_\varepsilon | \{c(1'0)c(01')a(1'0)p(01')p(1'0)\}_\varepsilon \rangle \\
& + e'_A e'_B e_C e'_D {e'_P}^2 \langle \{a(1'0)\}_\varepsilon | \{c(1'0)c(01)a(1'0)p(01')p(1'0)\}_\varepsilon \rangle \\
& + e'_A e'_B e'_C e'_D {e'_P}^2 \langle \{a(1'0)\}_\varepsilon | \{c(1'0)c(01')a(1'0)p(01')p(1'0)\}_\varepsilon \rangle \\
& + e_A e'_B e_C e_D {e'_P}^2 \langle \{a(1'0)\}_\varepsilon | \{c(10)c(01)a(10)p(01')p(1'0)\}_\varepsilon \rangle \\
& + e_A e_B e'_C e_D {e'_P}^2 \langle \{a(10)\}_\varepsilon | \{c(10)c(01')a(10)p(01')p(1'0)\}_\varepsilon \rangle \\
& + e_A e_B e_C e'_D {e'_P}^2 \langle \{a(10)\}_\varepsilon | \{c(1'0)c(01)a(10)p(01')p(1'0)\}_\varepsilon \rangle \\
& + e_A e'_B e'_C e_D {e'_P}^2 \langle \{a(1'0)\}_\varepsilon | \{c(10)c(01')a(10)p(01')p(1'0)\}_\varepsilon \rangle \\
& + e_A e_B e'_C e'_D {e'_P}^2 \langle \{a(10)\}_\varepsilon | \{c(1'0)c(01')a(10)p(01')p(1'0)\}_\varepsilon \rangle \\
& + e_A e'_B e_C e'_D {e'_P}^2 \langle \{a(1'0)\}_\varepsilon | \{c(1'0)c(01)a(10)p(01')p(1'0)\}_\varepsilon \rangle \\
& + e_A e'_B e'_C e'_D {e'_P}^2 \langle \{a(1'0)\}_\varepsilon | \{c(1'0)c(01')a(10)p(01')p(1'0)\}_\varepsilon \rangle \qquad \text{(C.9)}
\end{aligned}
$$

$$
\begin{aligned}
\langle (BPP)_\varepsilon | (DCA)_\varepsilon \rangle^{+-} = {} & e_A e_B e_C e_D {e_P}^2 \langle \{a(10)p(01)p(10)\}_\varepsilon | \{c(10)c(01)a(10)\}_\varepsilon \rangle \\
& + e'_A e_B e_C e_D {e_P}^2 \langle \{a(10)p(01)p(10)\}_\varepsilon | \{c(10)c(01)a(1'0)\}_\varepsilon \rangle \\
& + e'_A e'_B e_C e_D {e_P}^2 \langle \{a(1'0)p(01)p(10)\}_\varepsilon | \{c(10)c(01)a(1'0)\}_\varepsilon \rangle \\
& + e'_A e_B e'_C e_D {e_P}^2 \langle \{a(10)p(01)p(10)\}_\varepsilon | \{c(10)c(01')a(1'0)\}_\varepsilon \rangle \\
& + e'_A e_B e_C e'_D {e_P}^2 \langle \{a(10)p(01)p(10)\}_\varepsilon | \{c(1'0)c(01)a(1'0)\}_\varepsilon \rangle \\
& + e'_A e'_B e'_C e_D {e_P}^2 \langle \{a(1'0)p(01)p(10)\}_\varepsilon | \{c(10)c(01')a(1'0)\}_\varepsilon \rangle \\
& + e'_A e_B e'_C e'_D {e_P}^2 \langle \{a(10)p(01)p(10)\}_\varepsilon | \{c(1'0)c(01')a(1'0)\}_\varepsilon \rangle \\
& + e'_A e'_B e_C e'_D {e_P}^2 \langle \{a(1'0)p(01)p(10)\}_\varepsilon | \{c(1'0)c(01)a(1'0)\}_\varepsilon \rangle \\
& + e'_A e'_B e'_C e'_D {e_P}^2 \langle \{a(1'0)p(01)p(10)\}_\varepsilon | \{c(1'0)c(01')a(1'0)\}_\varepsilon \rangle \\
& + e_A e'_B e_C e_D {e_P}^2 \langle \{a(1'0)p(01)p(10)\}_\varepsilon | \{c(10)c(01)a(10)\}_\varepsilon \rangle \\
& + e_A e_B e'_C e_D {e_P}^2 \langle \{a(10)p(01)p(10)\}_\varepsilon | \{c(10)c(01')a(10)\}_\varepsilon \rangle \\
& + e_A e_B e_C e'_D {e_P}^2 \langle \{a(10)p(01)p(10)\}_\varepsilon | \{c(1'0)c(01)a(10)\}_\varepsilon \rangle \\
& + e_A e'_B e'_C e_D {e_P}^2 \langle \{a(1'0)p(01)p(10)\}_\varepsilon | \{c(10)c(01')a(10)\}_\varepsilon \rangle \\
& + e_A e_B e'_C e'_D {e_P}^2 \langle \{a(10)p(01)p(10)\}_\varepsilon | \{c(1'0)c(01')a(10)\}_\varepsilon \rangle \\
& + e_A e'_B e_C e'_D {e_P}^2 \langle \{a(1'0)p(01)p(10)\}_\varepsilon | \{c(1'0)c(01)a(10)\}_\varepsilon \rangle \\
& + e_A e'_B e'_C e'_D {e_P}^2 \langle \{a(1'0)p(01)p(10)\}_\varepsilon | \{c(1'0)c(01')a(10)\}_\varepsilon \rangle \\
& + e_A e_B e_C e_D {e'_P}^2 \langle \{a(10)p(01')p(1'0)\}_\varepsilon | \{c(10)c(01)a(10)\}_\varepsilon \rangle \\
& + e'_A e_B e_C e_D {e'_P}^2 \langle \{a(10)p(01')p(1'0)\}_\varepsilon | \{c(10)c(01)a(1'0)\}_\varepsilon \rangle \\
& + e'_A e'_B e_C e_D {e'_P}^2 \langle \{a(1'0)p(01')p(1'0)\}_\varepsilon | \{c(10)c(01)a(1'0)\}_\varepsilon \rangle \\
& + e'_A e_B e'_C e_D {e'_P}^2 \langle \{a(10)p(01')p(1'0)\}_\varepsilon | \{c(10)c(01')a(1'0)\}_\varepsilon \rangle \\
& + e'_A e_B e_C e'_D {e'_P}^2 \langle \{a(10)p(01')p(1'0)\}_\varepsilon | \{c(1'0)c(01)a(1'0)\}_\varepsilon \rangle \\
& + e'_A e'_B e'_C e_D {e'_P}^2 \langle \{a(1'0)p(01')p(1'0)\}_\varepsilon | \{c(10)c(01')a(1'0)\}_\varepsilon \rangle \\
& + e'_A e_B e'_C e'_D {e'_P}^2 \langle \{a(10)p(01')p(1'0)\}_\varepsilon | \{c(1'0)c(01')a(1'0)\}_\varepsilon \rangle \\
& + e'_A e'_B e_C e'_D {e'_P}^2 \langle \{a(1'0)p(01')p(1'0)\}_\varepsilon | \{c(1'0)c(01)a(1'0)\}_\varepsilon \rangle \\
& + e'_A e'_B e'_C e'_D {e'_P}^2 \langle \{a(1'0)p(01')p(1'0)\}_\varepsilon | \{c(1'0)c(01')a(1'0)\}_\varepsilon \rangle \\
& + e_A e'_B e_C e_D {e'_P}^2 \langle \{a(1'0)p(01')p(1'0)\}_\varepsilon | \{c(10)c(01)a(10)\}_\varepsilon \rangle \\
& + e_A e_B e'_C e_D {e'_P}^2 \langle \{a(10)p(01')p(1'0)\}_\varepsilon | \{c(10)c(01')a(10)\}_\varepsilon \rangle \\
& + e_A e_B e_C e'_D {e'_P}^2 \langle \{a(10)p(01')p(1'0)\}_\varepsilon | \{c(1'0)c(01)a(10)\}_\varepsilon \rangle \\
& + e_A e'_B e'_C e_D {e'_P}^2 \langle \{a(1'0)p(01')p(1'0)\}_\varepsilon | \{c(10)c(01')a(10)\}_\varepsilon \rangle \\
& + e_A e_B e'_C e'_D {e'_P}^2 \langle \{a(10)p(01')p(1'0)\}_\varepsilon | \{c(1'0)c(01')a(10)\}_\varepsilon \rangle \\
& + e_A e'_B e_C e'_D {e'_P}^2 \langle \{a(1'0)p(01')p(1'0)\}_\varepsilon | \{c(1'0)c(01)a(10)\}_\varepsilon \rangle \\
& + e_A e'_B e'_C e'_D {e'_P}^2 \langle \{a(1'0)p(01')p(1'0)\}_\varepsilon | \{c(1'0)c(01')a(10)\}_\varepsilon \rangle \qquad \text{(C.10)}
\end{aligned}
$$